\RequirePackage{lineno} 

\documentclass[twocolumn,preprintnumbers,amsmath,amssymb,nofootinbib]{revtex4}

\usepackage{graphicx}
\usepackage{dcolumn}
\usepackage{bm}
 
\usepackage{epstopdf}

\def\bea{\begin{eqnarray}}
\def\eea{\end{eqnarray}}

\def\pp{\mbox{$p$-$p$}}

\def\pbpb{\mbox{Pb-Pb}}
\def\ppb{\mbox{$p$-Pb}}

\def\pn{\mbox{$p$-N}}
\def\aa{\mbox{A-A}}
\def\nn{\mbox{N-N}}

\def\ee{\mbox{$e^+$-$e^-$}}

\def\pt{$p_t$}
\def\mt{$m_t$}
\def\yt{$y_t$}

\def\nch{$n_{ch}$}
\def\mmpt{$\bar p_t$}

\begin{document} 

\setlength{\pdfpagewidth}{8.5in}
\setlength{\pdfpageheight}{11in}

\setpagewiselinenumbers
\modulolinenumbers[5]

\addtolength{\footnotesep}{-10mm}\

\preprint{version 2.0\textsl{}}

\title{Systematic analysis of (multi)strange hadron $\bf p_t$ spectra from\\ small collision systems  at the large hadron collider
}

\author{Thomas A.\ Trainor}\affiliation{University of Washington, Seattle, WA 98195}


\date{\today}

\begin{abstract}

Small collision systems, e.g.\ $p$-$p$ and $p$-Pb collisions, comprise a potential reference for more-central A-A collisions with regard to production (or not) of a thermalized quark-gluon plasma (QGP). Small systems with low particle densities should evolve according to simple QCD mechanisms including projectile-nucleon dissociation and dijet production. But it is now claimed that QGP may appear  even in \pp\ collisions based on apparent evidence for radial flow from shape evolution of $p_t$ spectra and from variation of total yields for strange and multistrange hadrons relative to statistical models. The present study confronts such arguments with a detailed analysis of $p_t$ spectra for strange and multistrange hadrons from 5 TeV $p$-Pb collisions and 13 TeV $p$-$p$ collisions via a two-component model (TCM) of hadron production. Based on previous analysis of lighter hadrons the TCM accurately predicts spectra for Cascade and Omega hadrons.  Significant results include multistrange hadron spectra dominated by jet fragments, variation of strange-hadron abundances exaggerated by certain plot formats and spectrum extrapolations, and detailed relations between ensemble-mean $\bar p_t$ variation with event charge density and small shifts of jet fragment distributions on $p_t$. Within a TCM context $p$-$p$ and $p$-Pb collision systems with comparable jet contributions are found to be equivalent within data uncertainties. Attribution of certain data features to radial flow is doubtful.
\end{abstract}

\maketitle

\section{Introduction}

This article describes analysis of \pt\ spectra for strange and multistrange hadrons from 5 TeV \ppb\ collisions and 13 TeV \pp\ collisions via a two-component (soft+hard) model (TCM) of identified-particle (PID) spectra. One motivation is investigation of claims that evolution of yields for (multi)strange hadrons with charge multiplicity \nch\ or charge density $\bar \rho_0$ demonstrates that for higher \nch\ strangeness production comparable to that observed in central \pbpb\ collisions is observed. That result in turn has been interpreted to suggest that production of a thermalized, flowing quark-gluon plasma (QGP) may be achieved also in smaller collision systems albeit for higher charge densities. The importance of such conclusions has been emphasized: ``...the field of relativistic heavy ion physics is in the midst of a revolution...driven by the experimental observation of flow-like features in the collisions of small hadronic systems''~\cite{nagle}.

So-called strangeness enhancement was proposed as a signature of QGP formation in more-central \aa\ collisions~\cite{mullerqgp}, and that signature has been claimed for more-central heavy ion collisions at the RHIC and LHC~\cite{auausss,pbpbsss}. Reference~\cite{alippss} notes that  ``In central heavy-ion collisions, the yields of strange hadrons turn out to be consistent with the expectation from a grand-canonical ensemble, i.e.\ the production of strange hadrons is compatible with thermal equilibrium....''  Apparent enhancement in more-central \aa\ collisions is contrasted with strangeness production in small systems: ``On the other hand, the strange hadron yields in elementary collisions are suppressed with respect to the predictions of the (grand-canonical) thermal models.''
Reference~\cite{alippbss} refers to canonical suppression in similar terms: ``In smaller collision systems..., in particular proton-proton (pp) collisions, the relative abundance of multi-strange baryons is lower with respect to A-A collisions.... This led to the interpretation that strangeness enhancement is observed in \aa\ collisions. ... This means that strange hadrons are produced with a lower relative abundance in small systems, an effect known as canonical suppression.'' 

However, regarding previously-assumed contrasts in strangeness production between \aa\ and small collision systems Ref.~\cite{alippss} asserts that ``...strangeness enhancement is no longer considered an unambiguous signature for deconfinement [QGP formation].'' It notes that ``The \mbox{ALICE} Collaboration recently reported an enhancement in the relative production of (multi-) strange hadrons...in \pp\ collisions...and in \ppb\  collisions.... In the case of \ppb\ collisions, the yields of strange hadrons relative to pions reach values close to those observed in \pbpb\ collisions {\em at full equilibrium} [emphasis added].'' And Ref.~\cite{alippbss} interprets results from 5 TeV \ppb\ collisions to conclude that ``...lifting of strangeness suppression with system size has been observed with measurements in a single collision system. Hyperon to pion ratios are shown to increase with multiplicity in \ppb\ collisions from the values measured in pp to those observed in Pb-Pb. ...the behavior is qualitatively consistent with the lifting of canonical suppression with increasing multiplicity.'' Rather than admitting such results as challenging interpretations of \aa\ data the current trend has been to place small collision systems on a continuum with \aa\ regarding QGP formation. Such arguments may be strongly questioned~\cite{nature}.

The present study addresses the structure of PID \pt\ spectra for (multi)strange hadrons from 13 TeV \pp\ collisions reported in Ref.~\cite{alippss} and from 5 TeV \ppb\ collisions reported in Ref.~\cite{alippbss}. A context for this study is provided by previous TCM analysis of PID spectrum data for lower-mass hadrons from Ref.~\cite{aliceppbpid} reported in Refs.~\cite{ppbpid,pidpart1,pidpart2} and from Ref.~\cite{alicepppid} reported in Ref.~\cite{pppid}. This study addresses the evolution with collision-event charged-particle multiplicity \nch\ of differential PID \pt\ spectra, integrated yields and  ensemble-mean \mmpt\ for strange hadrons in comparison to nonstrange hadrons. The analysis is intended to extract {\em all statistically significant information} from particle data regarding those data features. In particular, contributions from minimum-bias (MB) jet production~\cite{mbdijets} are accurately identified as a determining factor.

A further context comes from Ref.~\cite{transport} wherein it is determined that PID spectrum evolution with \nch\ corresponds to transport of hadron species from soft (nonjet) to hard (jet-related) spectrum components while conserving {\em total} species abundances approximately consistent with statistical-model predictions. The degree of transport depends linearly on hadron mass permitting quantitative {\em predictions} of spectrum evolution for multistrange hadron transport that are confirmed by the present study.

Variations of abundance {\em fractions} vs charge multiplicity \nch\ for several hadron species are compared for 5 TeV \ppb\ collisions and 13 TeV \pp\ collisions. Variations for those systems are quite similar when compared on the basis of a TCM hard/soft ratio but {\em not} when compared on the basis of charge density $\bar \rho_0$. Local variations at low vs high \pt\ are found to be very different, suggesting the importance of jet production. The relation of such variations to strangeness {\em per se}, and the {\em significance} of measured variations compared to data uncertainties, are examined -- factors with critical relevance in regard to ``strangeness enhancement'' as an indicator for equilibration via particle rescattering and QGP formation.
  
This article is arranged as follows:
Section~\ref{spectrumtcm} introduces a PID TCM for small collision systems and reviews previous results for lighter hadrons from \pp\ and \ppb\ collisions.
Section~\ref{smallspec} presents TCM analysis of PID spectra for neutral kaons, Lambdas, Cascades and Omegas from 5 TeV \ppb\ collisions and 13 TeV \pp\ collisions.
Section~\ref{integrated} presents analysis of yields of (multi)strange hadrons integrated over limited \pt\ intervals at high \pt\ and also total yields.
Section~\ref{mptsec} presents TCM analysis of PID \mmpt\ data for  (multi)strange hadrons including the direct relation to jet-related hard-component properties.
Section~\ref{sys} discusses systematic uncertainties. 
Sections~\ref{disc} and~\ref{summ} present discussion and summary. 

\section{PID TCM for small systems}  \label{spectrumtcm}

In this section PID TCMs for 13 TeV \pp\ and 5 TeV \ppb\ collisions are introduced, referring to previous analyses of PID spectra for lower-mass hadrons~\cite{pppid,ppbpid,pidpart1,pidpart2}. Parameter values for PID model functions and for species fractions of soft $\bar \rho_s = n_s/\Delta \eta$ and hard $\bar \rho_h = n_h/\Delta \eta$ charge densities within acceptance $\Delta \eta$ from those previous independent studies are presented unchanged as a test of consistency. That material then provides a reference for analysis of PID spectra for strange and multistrange hadrons in the present study.

\subsection{PID spectrum TCM definition}   \label{pidspec}

Given a \pt\ spectrum TCM for unidentified-hadron spectra~\cite{ppprd,tomnewppspec} a corresponding TCM for identified hadrons can be generated by assuming that each hadron species $i$ comprises certain {\em fractions} of soft and hard TCM components denoted by $z_{si}(n_s)$ and $z_{hi}(n_s)$. The PID \pt\ spectrum TCM can then be written as
\bea \label{pidspectcm}
\bar \rho_{0i}(p_t,n_s) &\approx&  z_{si}(n_s) \bar \rho_{s} \hat S_{0i}(p_t) +   z_{hi}(n_s) \bar \rho_{h} \hat H_{0i}(p_t),
\eea
where for \pp\ collisions $\bar \rho_h \approx \alpha(\sqrt{s}) \bar \rho_s^2$~\cite{tomnewppspec} and $\bar \rho_0 = \bar \rho_{s} + \bar \rho_h$ is the measured event-class nonPID charge density. $\bar \rho_s$ is then obtained from measured  $\bar \rho_0$ as the root of $\bar \rho_0 = \bar \rho_{s} + \alpha \bar \rho_s^2$. Unit-normal model functions $\hat S_{0i}(m_t)$ and $\hat H_{0i}(y_t)$ are determined for each hadron species, but similarity to unidentified-hadron models is expected. 

The data soft component for a specific hadron species $i$ (except pions, with a resonance contribution) is well described by a L\'evy distribution as a density on $m_{ti}  = \sqrt{p_t^2 + m_i^2}$. The unit-integral soft-component model is 
\bea \label{s00}
\hat S_{0i}(m_{ti}) &=& \frac{A_i}{[1 + (m_{ti} - m_i) / n_i T_i]^{n_i}},
\eea
where $m_{ti}$ is the transverse mass for hadrons $i$ of mass $m_i$, $n_i$ is the L\'evy exponent, $T_i$ is the slope parameter and coefficient $A_i$ is determined by the unit-integral condition. As defined the soft-component model is a density on \pt\ or \mt\ which may be plotted vs pion $y_{t\pi} \equiv \ln((p_t + m_{t\pi})/m_\pi)$. The unit-integral hard-component model is defined as a Gaussian density on pion \yt\ with exponential (on $y_t$) or power-law (on $p_t$) tail at higher \yt\
\bea \label{h00}
\hat H_{0}(y_t) &\approx & B \exp\left\{ - \frac{(y_t - \bar y_t)^2}{2 \sigma^2_{y_t}}\right\}~~~\text{near mode $\bar y_t$}
\\ \nonumber
&\propto &  \exp(- q y_t)~~~\text{for higher $y_t$ -- the tail},
\eea
where the transition from Gaussian to exponential is determined by slope matching (see the macro in Ref.~\cite{hardspec}). Coefficient $B$ is determined by the unit-integral condition. This hard-component model density on $y_{t\pi}$ may be transformed to a density on \pt\ or \mt\  via  Jacobian $y_{t\pi} / m_{t\pi} p_t $.  The $\hat H_0$ tail then varies on \pt\ as power law $1/p_t^{q + 2.2}$ reflecting an underlying jet spectrum~\cite{fragevo}.

The PID spectrum TCM for \ppb\ collisions has the same structure as Eq.~(\ref{pidspectcm}) but $\bar \rho_s \rightarrow (N_{part}/2) \bar \rho_{sNN}$ and $\bar \rho_h \rightarrow N_{bin} \bar \rho_{hNN}$, where $N_{part}$ is the number of participant nucleons N, $N_{bin}$ ($= N_{part}-1$ for \ppb) is the number of \nn\ (in this case \pn) binary collisions, and densities $\bar \rho_{xNN}$ are averages over all participant pairs. For assumed \pn\ {\em linear superposition} within \ppb\ collisions $\bar \rho_{hNN} \approx \alpha \bar \rho_{sNN}^2$ as for \pp\ collisions. Hard/soft ratio $\bar \rho_h / \bar \rho_s \equiv x(n_s) \nu(n_s)$, with $x \approx \alpha \rho_{sNN}$ ($\approx \alpha \bar \rho_s$ for \pp\ collisions) and $\nu = 2 N_{bin} / N_{part}$ (= 1 for \pp\ collisions), is of central importance. Parameter estimation requires accurate determination of \ppb\ centrality or collision geometry vs measured charge density $\bar \rho_0$ as reported in Ref.~\cite{tomglauber}, which also notes that determination of \ppb\ geometry via a classical Glauber model produces very inaccurate results~\cite{tomexclude}. Table~\ref{rppbdata} presents centrality parameter values (unprimed values) for 5 TeV \ppb\ collisions.

\begin{table}[h]
	\caption{TCM fractional cross section $\sigma / \sigma_0$ (bin centers) and charge density $\bar \rho_0$, \nn\ soft component $\bar \rho_{sNN}$ and TCM hard/soft ratio $x(n_s)$ used for 5 TeV \ppb\ PID spectrum analysis~\cite{ppbpid}. $\sigma' / \sigma_0$ and $N_{bin}'$ values are from Table~2 of Ref.~\cite{aliceglauber}. Other parameter values are from Ref.~\cite{tomglauber}.
	}
	\label{rppbdata}
	\begin{center}
		\begin{tabular}{|c|c|c|c|c|c|c|c|c|} \hline
			$n$ &   $\sigma' / \sigma_0$ & $N_{bin}'$  &  $\sigma / \sigma_0$     & $N_{bin}$  & $\nu$ & $\bar \rho_0$ & $\bar \rho_{sNN}$ & $x(n_s)$ \\ \hline
			1	   &      0.025  & 14.7  & 0.15   & 3.20   & 1.52 & 44.6 & 16.6  & 0.188 \\ \hline
			2	 &  0.075  & 13.0 & 0.24    & 2.59   & 1.43 & 35.9 &15.9  & 0.180 \\ \hline
			3	 &  0.15  & 11.7 & 0.37 & 2.16  &  1.37 & 30.0  & 15.2  & 0.172 \\ \hline
			4	 &  0.30 & 9.36 & 0.58  & 1.70   & 1.26  & 23.0  & 14.1  & 0.159  \\ \hline
			5	 &  0.50  & 6.42 &0.80    & 1.31   & 1.13 & 15.8 &   12.1 & 0.137  \\ \hline
			6	 &  0.70 & 3.81 & 0.95   & 1.07   & 1.03  & 9.7  &  8.7 & 0.098 \\ \hline
			7	 & 0.90  & 1.94 & 0.99  & 1.00  & 1.00  &  4.4  & 4.2 &0.047  \\ \hline
		\end{tabular}
	\end{center}
\end{table}

\subsection{$\bf \hat S_{0i}$ and $\bf \hat H_{0i}$ model-function parameters}

Table~\ref{pidparams} shows 5 TeV \ppb\ TCM model parameters for hard component $\hat H_0(y_t)$ (first three) and soft component $\hat S_0(m_t)$ (last two) reported in Ref.~\cite{pidpart1}. Hard-component parameters vary significantly with hadron species and centrality: Gaussian width $\sigma_{y_t}$ is greater for mesons than for baryons while exponent $q$ is substantially greater for baryons than for mesons, and modes $\bar y_t$ for baryons shift to higher \yt\ with increasing centrality while modes for mesons (from \ppb\ collisions) shift to lower \yt. Soft-component model parameter $T$ is independent of collision energy but increases substantially with hadron mass (145 MeV for pions, $\approx$ 200 MeV for higher-mass hadrons). L\'evy exponent $n$ and hard-component exponent $q$ have substantial systematic energy dependence~\cite{alicetomspec}. 

\begin{table}[h]
	\caption{TCM model parameters for identified hadrons from 5 TeV \ppb\ collisions from Table~VI of Ref~\cite{pidpart1}: hard-component parameters $(\bar y_t,\sigma_{y_t},q)$ and soft-component parameters $(T,n)$. These values apply to the {\em most-central} \ppb\ event class. Numbers without uncertainties are adopted from a comparable hadron species with greater accuracy. 
	}
	\label{pidparams}
	\begin{center}
		\begin{tabular}{|c|c|c|c|c|c|} \hline
			& $\bar y_t$ & $\sigma_{y_t}$ & $q$ & $T$ (MeV) &  $n$  \\ \hline
			$ \pi^\pm $     &  $2.46\pm0.01$ & $0.57\pm0.01$ & $4.1\pm1$ & $145\pm3$ & $8.5\pm0.5$ \\ \hline
			$K^\pm$    & $2.65$  & $0.57$ & $4.1$ & $200$ & $14$ \\ \hline
			$K_s^0$          &  $2.65\pm0.01$ & $0.57\pm0.01$ & $4.1\pm0.1$ & $200\pm5$ & $14\pm2$ \\ \hline
			$p$        & $2.99\pm0.01$  & $0.47$ & $5.0$ & $210\pm10$ & $14\pm4$ \\ \hline
			$\Lambda$       & $2.99\pm0.01$  & $0.47\pm0.01$ & $5.0\pm0.05$  & $210$ & $14$ \\ \hline	
		\end{tabular}
	\end{center}
\end{table}

Table~\ref{engparamsyy} presents 13 TeV \pp\ TCM model parameters from Table V of Ref.~\cite{pppid}. Uncertainties for proton parameters $\bar y_t$ and $\sigma_{y_t}$ indicate the range of variation of those parameters for the variable-TCM hard component. Corresponding ensemble-mean \pt\ values $\bar p_{tsi}$ and $\bar p_{thi}(n_s)$ for soft and hard spectrum components corresponding to these model parameters are discussed in Sec.~\ref{mptsec}.

\begin{table}[h]
	\caption{TCM model parameters for identified hadrons from 13 TeV \pp\ collisions from Table V of Ref.~\cite{pppid}: hard-component parameters $(\bar y_t,\sigma_{y_t},q)$ and soft-component parameters $(T,n)$. The first two proton values are {\em averages over event classes}. Detailed variations are described below.
	}
	\label{engparamsyy}
	\begin{center}
		\begin{tabular}{|c|c|c|c|c|c|} \hline
			& $\bar y_t$ & $\sigma_{y_t}$ & $q$ & $T$ (MeV) &  $n$  \\ \hline
			$ \pi^\pm $     &  $2.46 \pm 0.02$ & $0.60 \pm 0.02$ & $3.7\pm 0.1$ & $145\pm 2$ & $8.0\pm 0.3$ \\ \hline
			$K^\pm$    & $2.68 \pm 0.02$  & $0.60\pm 0.02$ & $3.7\pm 0.1$ & $200\pm 5$ & $14\pm 2$ \\ \hline
			$p$        & $2.90\pm 0.05$  & $0.50\pm 0.03$ & $4.6\pm 0.2$ & $210\pm 5$ & $14\pm 2$ \\ \hline
		\end{tabular}
	\end{center}
\end{table}

The entries for \ppb\ collisions in Table~\ref{pidparams} define a {\em fixed} TCM reference independent of centrality that describes the {\em most central} event class (wherein $\bar y_t \approx 3.0$ for baryons)~\cite{pidpart1} wheres entries for \pp\ collisions in Table~\ref{engparamsyy} define a fixed TCM describing an {\em average over event classes}, with proton uncertainties reflecting variation of its hard component. In Refs.~\cite{pidpart2,pppid} variation of some hard-component model parameters is determined so as to describe each event class within statistical uncertainties (e.g.\ see Fig.~4 of Ref.~\cite{pidpart2}). Required variations are linear on hard/soft ratio $x(n_s) \nu(n_s)$: with increasing $x\nu$ hard-component modes $\bar y_t(n_s)$ shift to higher \yt\ for baryons while, {\em for \ppb\ spectra only}, hard-component widths above the mode decrease for mesons. In the present study the meson width evolution is reported in terms of equivalent mode shifts per $\bar y_t(n_s)$.

\subsection{TCM PID hadron fractions $\bf z_{si}(n_s)$ and $\bf z_{hi}(n_s)$}

Hadron species fractions $z_{si}(n_s)$ and $z_{hi}(n_s)$ are required to implement the TCM model denoted by Eq.~(\ref{pidspectcm}). Fractions may be inferred as measured quantities directly from PID spectrum data as reported in Ref.~\cite{pidpart1}, or individual fractions $z_{si}(n_s)$ and $z_{hi}(n_s)$ may be predicted from model parameters $z_{0i}(n_s)\equiv \bar \rho_{0i}(n_s) /  \bar \rho_0(n_s)$ and $\tilde z_i(n_s) \equiv z_{hi}(n_s)/ z_{si}(n_s)$ (both derived from {\em previously measured} $z_{si}(n_s)$ and $z_{hi}(n_s)$) 
via relations
\bea \label{zsix}
z_{si}(n_s) &=& \frac{1 + x(n_s) \nu(n_s)}{1 + \tilde z_i(n_s)x(n_s) \nu(n_s)} z_{0i}(n_s)
\\ \nonumber
z_{hi}(n_s) &=& \tilde z_i(n_s)z_{si}(n_s).
\eea
 5 TeV \ppb\ geometry (centrality) parameters $x(n_s)$ and $\nu(n_s)$ are determined in Refs.~\cite{alicetommpt,tommpt,tomglauber} based on ensemble-mean \mmpt\ data (see Table~\ref{rppbdata}). For \pp\ collisions $\nu \equiv 2 N_{bin} / N_{part} \rightarrow 1$. In what follows previous results for lower-mass hadrons are presented as an introduction to the PID TCM procedure. In Sec.~\ref{smallspec} the method is newly applied to (multi)strange hadrons.

Table~\ref{otherparamsx} shows PID parameters $z_{0i}$ and $\tilde z_i$ for five hadron species, determined from PID spectrum data for 5 TeV \ppb\ collisions, as reported in Ref.~\cite{pidpart1}. While $z_0$ was found to be independent of \ppb\ centrality within uncertainties the \ppb\ $\tilde z_i(n_s)$ exhibit significant centrality dependence as shown in Fig.~8 of Ref.~\cite{pidpart1}. Measurements of individual centrality trends for $z_{si}(n_s)$ and $z_{hi}(n_s)$ based on spectrum analysis are presented in Sec.~IV of Ref.~\cite{pidpart1}.   The $\tilde z_i = z_{hi}/z_{si}$ values included in Table~\ref{otherparamsx} represent averages over \ppb\ centrality. 

\begin{table}[h]
	\caption{\label{otherparamsx}
		TCM model parameters for identified hadrons from 5 TeV \ppb\ collisions in Ref.~\cite{pidpart1}. Numbers without uncertainties are adopted from a comparable hadron species with greater accuracy. Values for $\tilde z_i = z_{hi} / z_{si}$ are averages over \ppb\ centrality. Parameters $ \bar p_{ts}$ and $\bar p_{th}$ are determined by model functions $\hat S_0(y_t)$ and $\hat H_0(y_t)$ defined by Table~\ref{pidparams}.  Values for $\bar p_{thi}$ are event-class averages from Ref.~\cite{pidpart2}.
	}
	\begin{center}
		\begin{tabular}{|c|c|c|c|c|} \hline
			&   $z_{0i}$    &  $\tilde z_i$ &   $ \bar p_{tsi}$ (GeV/c)  & $ \bar p_{thi}$ (GeV/c)  \\ \hline
			$ \pi^\pm$        &   $0.82\pm0.01$  & $0.85\pm0.05$  & $0.40\pm0.02$ &    $1.15\pm0.03$  \\ \hline
			$K^\pm $   &  $ 0.128\pm0.002$   &  $2.7\pm0.2$ &  $0.60$&  $1.34$   \\ \hline
			$K_s^0$        &  $0.064\pm0.002$ &  $2.7\pm0.2$ &  $0.60\pm0.02$ &   $1.34\pm0.03$  \\ \hline
			$p $        & $ 0.065\pm0.002$    &  $5.6\pm0.2$ &  $0.73\pm0.02$&   $1.57\pm0.03$   \\ \hline
			$\Lambda $        &  $0.034\pm0.002$    & $6.5\pm0.5$ &   $0.76\pm0.02$ &    $1.65\pm0.03$ \\ \hline	
		\end{tabular}
	\end{center}
\end{table}

Table~\ref{otherparamsy} shows results for three hadron species from 13 TeV \pp\ collisions that may be compared with results for 5 TeV \ppb\ described above. With the exception of pion $\tilde z_i$ the $z_{xi}$ values determined for the two collision systems are consistent within data uncertainties. The significant change for pion $\tilde z_i$ is related to proton-pion crosstalk in $dE/dx$ analysis as explained in Ref.~\cite{pppid} Sec.~IX B. The significant difference between $\bar p_{thi}$ values for kaons from \ppb\ vs \pp\ collisions (1.34 vs 1.46 respectively) is explained in Sec.~\ref{directcomp} below.

\begin{table}[h]
	\caption{TCM species fraction coefficients $z_{0i}$ and $\tilde z_i$ for identified hadrons from 13 TeV \pp\ collisions from Table VI of Ref.~\cite{pppid}. The soft- and hard-component ensemble-means $\bar p_{tsi}$ and $\bar p_{thi}$ correspond to model parameters in Table~\ref{engparamsyy} and have been employed in a follow-up study~\cite{transport}. The large uncertainty for proton $\bar p_{thi}$ (averages over event class) corresponds to variation of proton hard-component mode $\bar y_t$.
	}
	\label{otherparamsy}
	\begin{center}
		\begin{tabular}{|c|c|c|c|c|} \hline
			&   $z_{0i}$    &  $\tilde z_i$ &   $ \bar p_{tsi}$ (GeV/c)  & $ \bar p_{thi}$ (GeV/c)  \\ \hline
			$ \pi^\pm$        &   $0.80\pm0.01$  & $0.60\pm0.05$  & $0.40\pm0.02$ &    $1.15\pm0.03$  \\ \hline
			$K^\pm $   &  $ 0.130\pm0.005$   &  $2.60\pm0.05$ &  $0.60\pm0.02$&  $1.46\pm0.03$   \\ \hline
			$p $        & $ 0.070\pm0.005$    &  $5.60\pm0.05$ &  $0.74\pm0.02$&   $1.55\pm0.10$   \\ \hline
		\end{tabular}
	\end{center}
\end{table}

In previous studies mean values of TCM parameters are reported in tables as above while any significant evolution with \nch\ is indicated within the text or by in-line equations. In the present study display equations are defined for relevant TCM variables. For instance, hard/soft fractional-coefficient ratio $\tilde z_i(n_s)$ is described by
\bea \label{tildez}
\tilde z_i(n_s) &=& \tilde z_i^* + \delta \tilde z_i^* x(n_s)\nu(n_s),
\eea
with parameters denoted in tables by starred quantities. 

\begin{figure}[h]
	\includegraphics[width=3.3in]{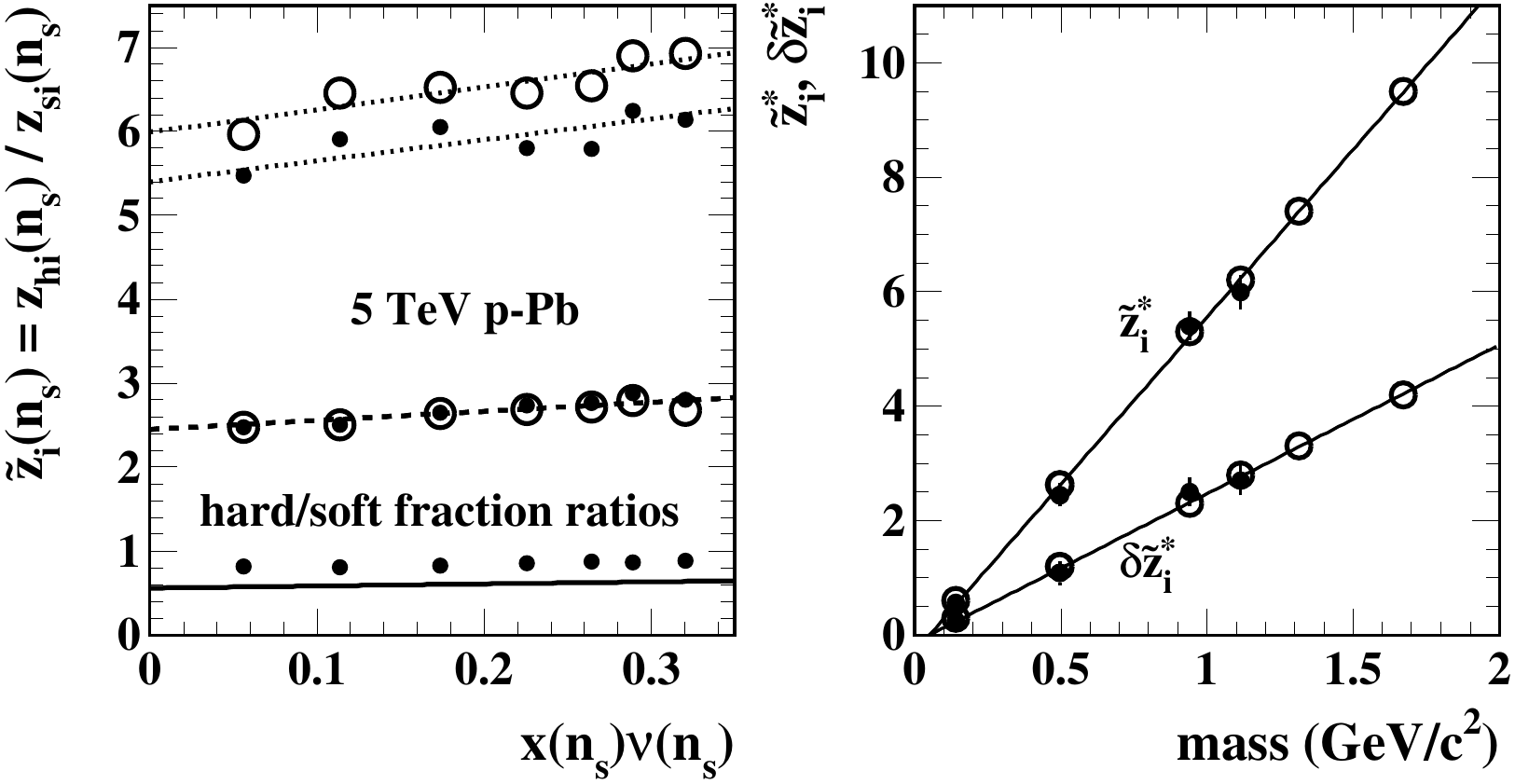}
	\caption{\label{tildezparams}
		Left: Ratios $\tilde z_{i}(n_s) = z_{hi}/z_{si}$ inferred from $z_{si}$ and $z_{hi}$ entries reported in Ref.~\cite{pidpart1} for charged (solid dots) and neutral (open circles) hadrons. The lines are linear parametrizations $\tilde z_i = \tilde z_i^* + \delta \tilde z_i^* x(n_s) \nu(n_s)$ that describe the ratio data: solid, dashed and dotted for pions, kaons and baryons respectively.
		Right: Coefficients $\tilde z_i^*$ and $\delta \tilde z_i^*$ for linear descriptions of ratio data in the left panel plotted vs hadron mass. The lines represent proportionality to hadron mass. The solid dots are values inferred in Ref.~\cite{pidpart1}. Open circles provide linear extrapolations on hadron mass for $\Xi$ and $\Omega$ in the present study.
	}  
\end{figure}

Figure~\ref{tildezparams} (left) shows ratios $\tilde z_{i}(n_s) = z_{hi}(n_s)/z_{si}(n_s)$ (points) inferred from $z_{si}(n_s)$ and $z_{hi}(n_s)$ data reported in Ref.~\cite{pidpart1} for charged hadrons (solid dots) and neutral hadrons (open circles), from pions at the bottom to Lambdas at the top. Dashed and dotted lines corresponding to kaons and baryons are best fits by eye to the points with Eq.~(\ref{tildez}). The pion line ($\approx 0.6$) corresponds to strict charge conservation as noted in Ref.~\cite{pppid} whereas the pion data points ($\approx 0.85$) from Ref.~\cite{pidpart1} may include crosstalk from misidentified protons. The resulting parameters are presented as solid dots in the right panel.

Figure~\ref{tildezparams} (right) shows  $\tilde z_i^*$ and $\delta \tilde z_i^*$ (solid dots) for  Eq.~(\ref{tildez}) plotted vs hadron mass. As noted in Refs.~\cite{pppid,transport} starred $\tilde z_i(n_s)$ parameters vary simply proportional to hadron mass and show no significant sensitivity to baryon identity or strangeness content. $\tilde z_i^*$ and $ \delta \tilde z_i^*$ values for $\Xi$ and $\Omega$ (rightmost open circles) are thus obtained for this study by extrapolation of the linear mass dependence. 

To generate the TCM curves for PID spectra in Figs.~\ref{ppbpiddata} and \ref{piddata} via TCM Eq.~(\ref{pidspectcm}) the parametrization $\tilde z_{i}(n_s)$ in Eq.~(\ref{tildez}) is applied to  Eqs.~(\ref{zsix}) to generate values $z_{si}(n_s)$ and $z_{hi}(n_s)$ that also require values for fractional abundance coefficients $z_{0i}$. The $z_{0i}$ were assumed independent of charge multiplicity for lighter hadrons~\cite{ppbpid,pidpart1,pidpart2,pppid}. 
However, significant variation of $z_{0i}$ with event class is observed for $\Xi$ and $\Omega$ in \pp\ and \ppb\ collisions and is incorporated in variable $z_{0i}(n_s)$ defined by Eq.~\ref{z0trend} (Sec.~\ref{variable}).

\section{Strange and multistrange TCM} \label{smallspec}

This section presents a comprehensive TCM analysis of PID spectra for strange and multistrange hadrons based on PID spectrum data from Ref.~\cite{alippss} for 13 TeV \pp\ collisions and from Ref.~\cite{alippbss} for 5 TeV \ppb\ collisions. Spectra for the present study are presented as densities on \pt\ plotted vs pion rapidity $y_{t\pi}$ with pion mass assumed.  $\hat S_{0i}(m_{ti})$ for species $i$ is defined by Eq.~(\ref{s00}). $\hat H_{0}(y_{t\pi})$ in Eq.~(\ref{h00}) is defined as a density on $y_{t\pi}$ where it has a simple form and is then converted to $\hat H_{0}(p_t)$ via the Jacobian factor $y_{t\pi} / m_{t\pi} p_t $. In general, plotting spectra as densities on \pt\ against logarithmic variable \yt\ permits superior visual access to important low-\pt\ structure where the {\em majority of jet fragments appears}. A further motivation is comparison of spectrum hard components interpreted to arise from a common underlying jet spectrum on \pt\ \cite{hardspec,fragevo}, in which case $y_{t\pi}$ serves simply as a logarithmic measure of hadron \pt\ with well-defined zero. 

The PID TCM for (multi)strange-hadron spectra is based in part on results from Refs.~\cite{ppbpid,pidpart1,pidpart2} (for  $K_\text{S}^0$ and $\Lambda$) and on the present study (for $\Xi$ and $\Omega$). Tables~\ref{pidparamz} and \ref{pidparamx} present hard-component model $\hat H_{0i}(y_t)$ parameter values with variable mode position $\bar y_t(n_s)$ defined by
\bea \label{ytbar}
\bar y_t(n_s) &=& \bar y_t^* + \delta \bar y_t^* x(n_s)\nu(n_s).
\eea
The soft-component parameter values are $T = 200$ MeV and $n = 14$ for the four species. For heavier hadrons there is reduced sensitivity to L\'evy exponent $n$. Note that in Ref.~\cite{pidpart2} kaon hard-component widths above the mode were observed to decrease. In the present study that evolution is expressed in terms of an equivalent mode shift to lower \yt. $\delta \bar y_t^*$ for neutral kaons is then {\em negative}. 
Starred parameters below relate to Eqs.~(\ref{tildez}), (\ref{ytbar}), (\ref{z0trend}) and (\ref{mpth}) that define variable-TCM parameters $\bar y_{ti}(n_s)$, $\tilde z_i(n_s)$, $z_{0i}(n_s)$ and $\bar p_{thi}(n_s)$ respectively. Solid curves in Figs.~\ref{ppbpiddata} and \ref{piddata} are the variable TCM with those definitions implemented.

\subsection{5 $\bf TeV$ $\bf p$-Pb PID Spectrum data}

The 5 TeV \ppb\ PID spectrum data used for this analysis were reported in Ref.~\cite{aliceppbpid} (for $K_\text{S}^0$ and $\Lambda$ data, 25 million  events) and Ref.~\cite{alippbss} (for $\Xi$ and $\Omega$ data, 100 million events). Events are sorted into seven multiplicity classes based on charge accumulated within a V0A or V0M detector combination~\cite{aliceppbpid}. Mean charge densities $dN_{ch} / d\eta \rightarrow \bar \rho_0$  as integrated within $|\eta| < 0.5$ or angular acceptance $\Delta \eta = 1$ are 45, 36.2, 30.5, 23.2, 16.1, 9.8 and 4.3 for event class $n \in [1,7]$ respectively. 

\begin{figure}[h]
	\includegraphics[width=1.62in]{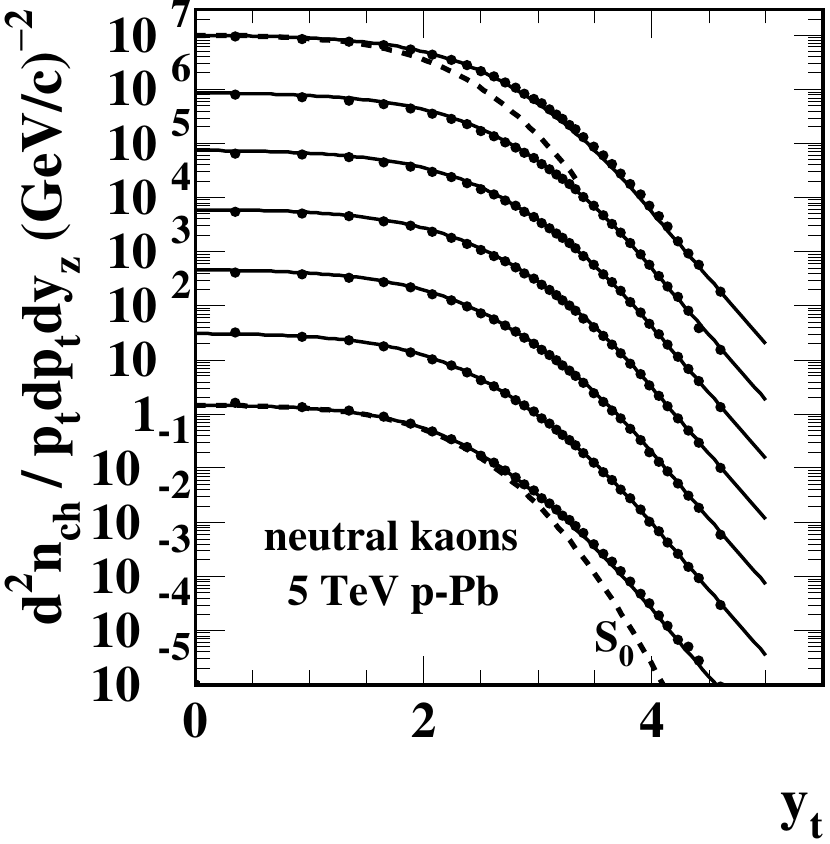}
	\includegraphics[width=1.65in]{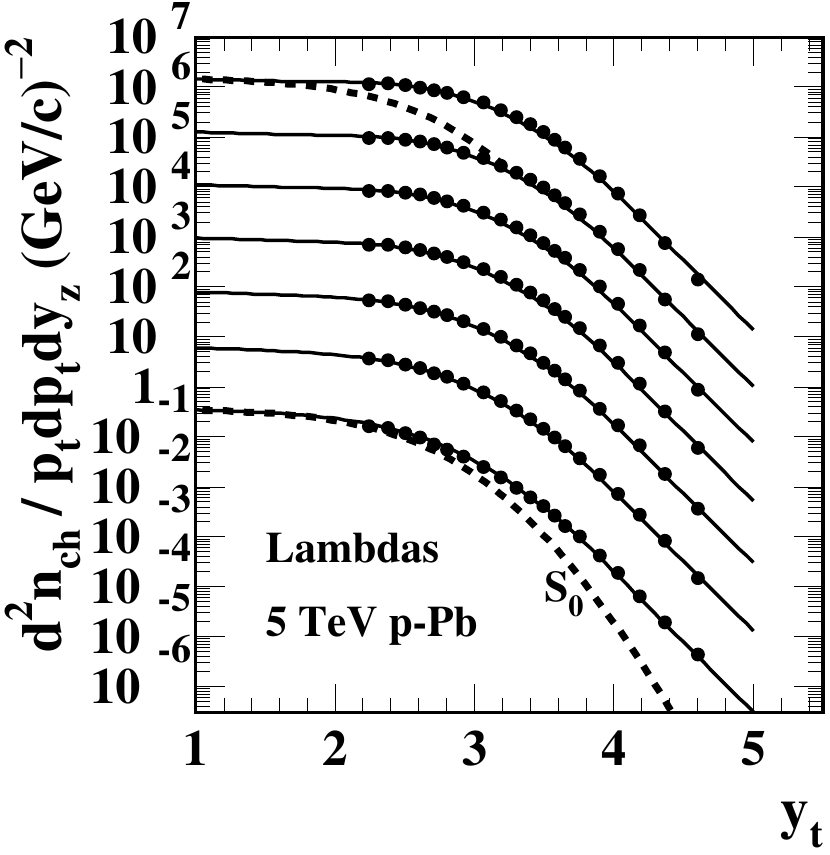}
	\put(-145,105) {\bf (a)}
	\put(-23,105) {\bf (b)}
	\\
	\includegraphics[width=1.65in]{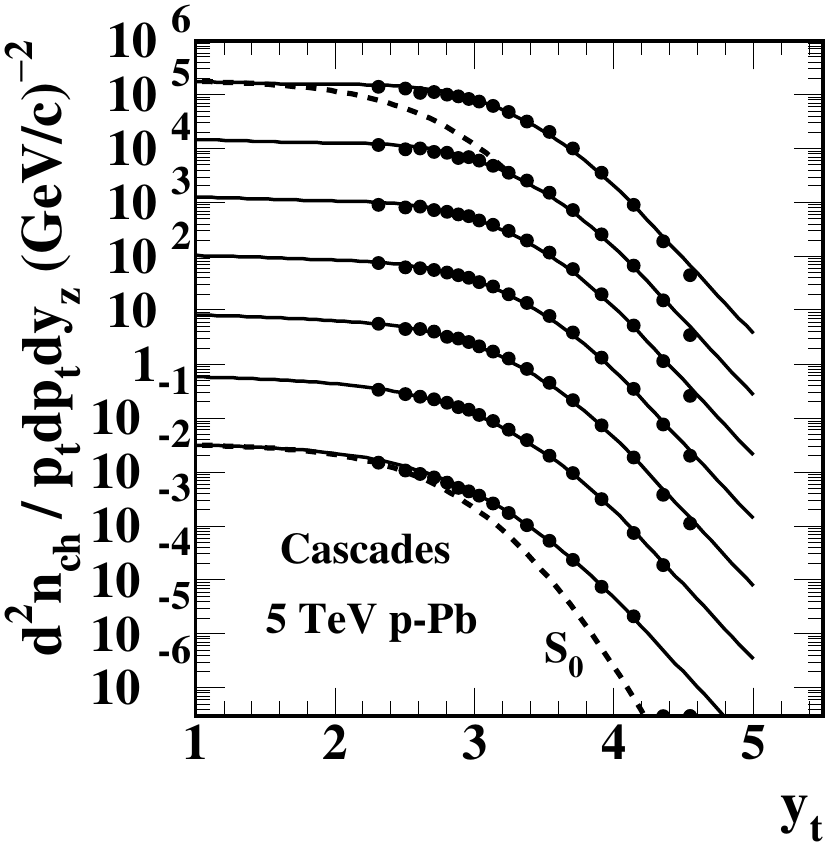}
	\includegraphics[width=1.65in]{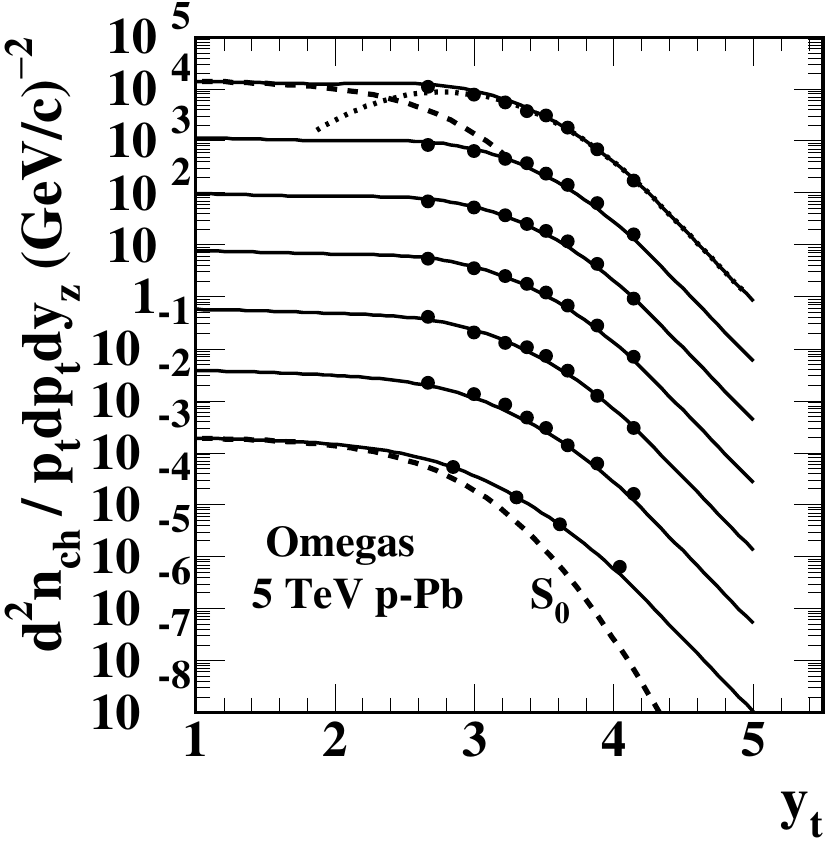}
	\put(-145,105) {\bf (c)}
	\put(-23,105) {\bf (d)}
	\\
	\caption{\label{ppbpiddata}
		\pt\ spectrum data (points) for strange hadrons from 5 TeV \ppb\ collisions:
		(a) Neutral kaons and
		(b) Lambdas from Ref.~\cite{aliceppbpid},
		(c) Cascades and
		(d) Omegas from Ref.~\cite{alippbss}.
		Solid curves represent the TCM. Dashed curves represent TCM soft components in the form $z_{si}(n_s) \bar \rho_s \hat S_0(p_t)$. The dotted curve in (d) is   the TCM hard component in the form $z_{hi}(n_s) \bar \rho_h \hat H_0(p_t)$.
	} 
\end{figure}

Figure~\ref{ppbpiddata} shows PID \pt\ spectra (points) for neutral kaons (a), Lambdas (b), Cascades (c) and Omegas (d) plotted vs pion rapidity \yt. The solid curves are TCM spectra determined as described above. The dashed curves are soft-component models in the form $z_{si}(n_s) \bar \rho_s \hat S_{0i}(p_t)$. The dotted curve in (d) is the Omega hard-component model in the form $z_{hi}(n_s) \bar \rho_h \hat H_{0i}(p_t)$. The latter serves to demonstrate that a large fraction of Omegas from central \ppb\ collisions is jet fragments.

Table~\ref{pidparamz} shows hard-component parameters for 5 TeV \ppb\ collisions. Parameters $\bar y_t^*$ and $\delta \bar y_t^*$ are used to define variable mode $\bar y_t(n_s)$ per Eq.~(\ref{ytbar}). Soft-component slope parameters $T = 200$ MeV and $n = 14$ are held fixed at those values for more-massive hadrons. 

\begin{table}[h]
	\caption{TCM hard-component model parameters $(\bar y_t,\sigma_{y_t},q)$ for strange hadrons from 5 TeV \ppb\ collisions. The first two columns define $\bar y_t(n_s)$ via Eq.~(\ref{ytbar}).
	} \label{pidparamz}
	\begin{center}
		\begin{tabular}{|c|c|c|c|c|} \hline
			& $\bar y_t^*$ & $\delta \bar y_t^*$ & $\sigma_{y_t}$ & $q$    \\ \hline
			$K_\text{S}^0$          &  $2.80\pm0.01$ &$-0.60\pm0.05$ & $0.58\pm0.01$ & $4.1\pm0.1$   \\ \hline
			$\Lambda$       & $2.80\pm0.03$ & $0.40\pm0.05$ & $0.50\pm0.01$ & $5.0\pm0.1$   \\ \hline	
			$ \Xi $     &  $2.98\pm0.05$ & $0.20\pm0.05$ & $0.50\pm0.02$ & $5.0\pm0.2$   \\ \hline
			$\Omega$    & $3.05\pm0.05$  & $0.40\pm0.05$ & $0.50\pm0.05$ & $5.0\pm0.5$   \\ \hline
		\end{tabular}
	\end{center}
\end{table}

Table~\ref{coefparams} below shows fractional-coefficient parameters for 5 TeV \ppb\ collisions. Parameters $\tilde z_i^*$ and $\delta \tilde z_i^*$ are used to define variable hard/soft ratio parameter $\tilde z_i(n_s)$ per Eq.~(\ref{tildez}). Variable $z_{0i}(n_s)$ is defined by Eq.~\ref{z0trend} (Sec.~\ref{variable}). $z_{0i}(n_s)$ and $\tilde z_i(n_s)$ are then combined to define fractional coefficients $z_{si}(n_s)$ and $z_{hi}(n_s)$ per Eqs.~(\ref{zsix}). That combination plus Eq.~(\ref{pidspectcm}) determines the solid curves in Fig.~\ref{ppbpiddata}.  

\subsection{13 $\bf TeV$ $\bf p$-$\bf p$ PID Spectrum data} \label{alicedata}

The 13 TeV \pp\ PID spectrum data used for this analysis were reported in Ref.~\cite{alippss}. Spectrum data were obtained from 50 million \pp\ collision events satisfying an INEL $> 0$ minimum-bias trigger (at least one charged particle within $|\eta| < 1$). Collision events were sorted into ten multiplicity classes based on charge accumulated within a V0 detector (V0M amplitude). Mean charge densities $dN_{ch} / d\eta \rightarrow \bar \rho_0$  as integrated within $|\eta| < 0.5$ or angular acceptance $\Delta \eta = 1$ are 25.8, 19.8 16.1, 13.8, 12.1 10.1 8.1, 6.5, 4.6 and 2.5 for $n \in [1,10]$ respectively. Note that a factor ten increase in $\bar \rho_0$ implies approximately {\em 100-fold increase in jet production} per the fundamental relation $\bar \rho_h \approx \alpha \bar \rho_s^2$ for \pp\ collisions~\cite{ppprd,pppid}.

\begin{figure}[h]
	\includegraphics[width=1.61in]{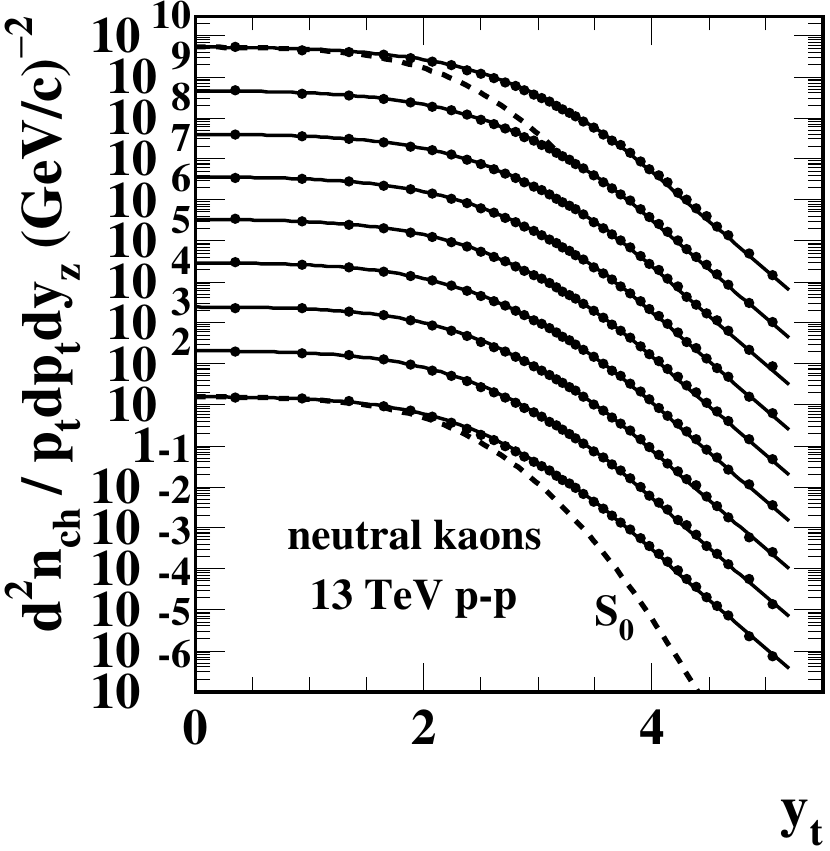}
	\includegraphics[width=1.65in]{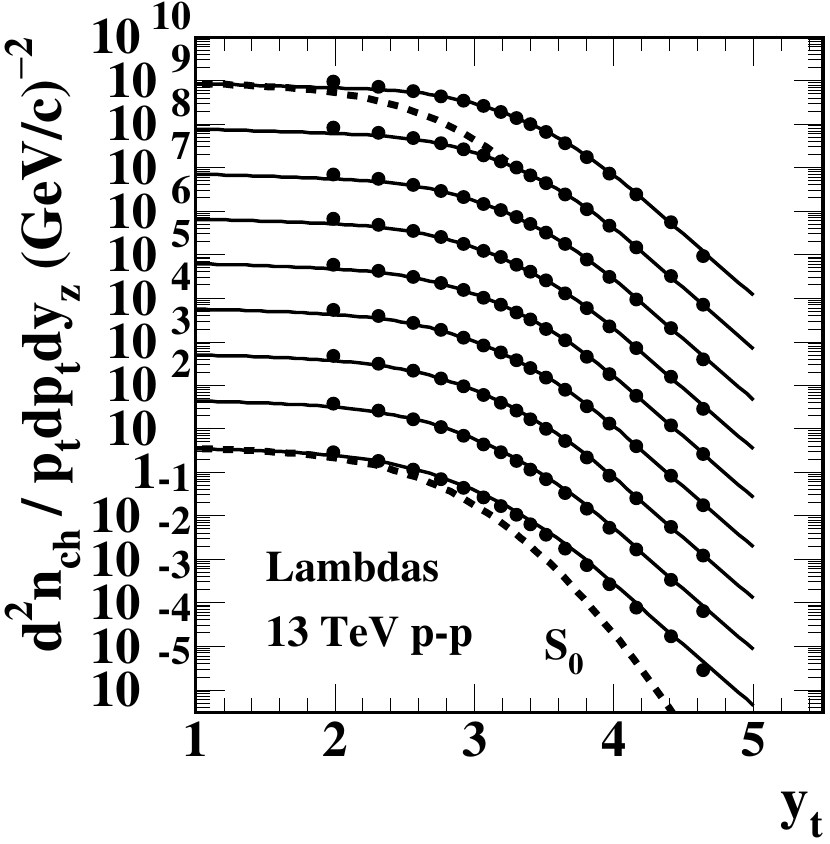}
	\put(-145,105) {\bf (a)}
	\put(-23,105) {\bf (b)}
	\\
	\includegraphics[width=1.65in]{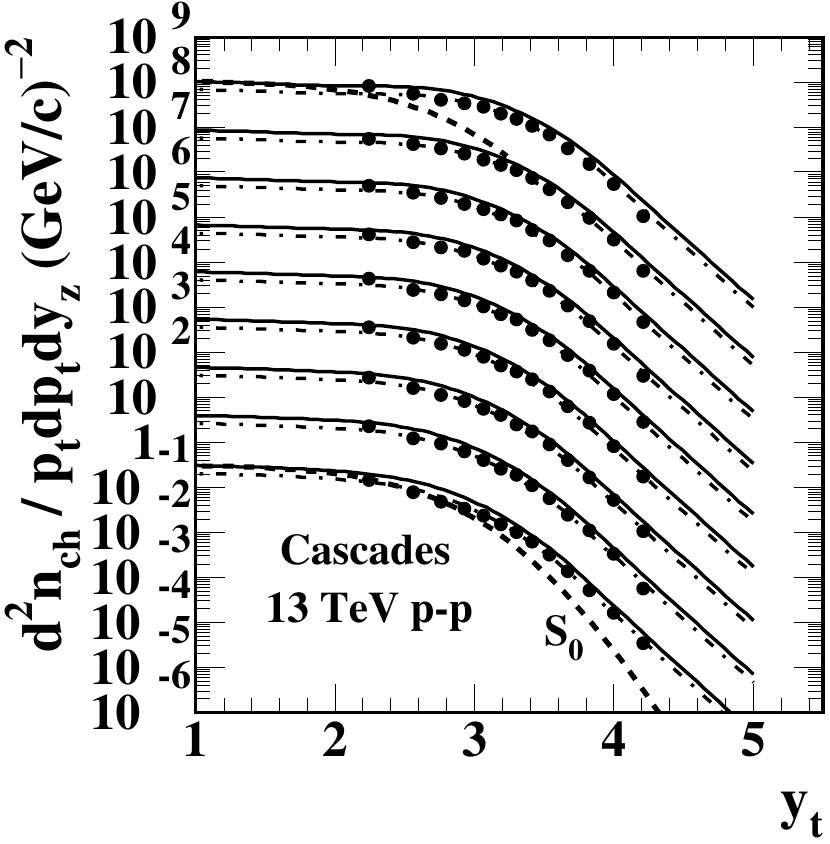}
	\includegraphics[width=1.65in]{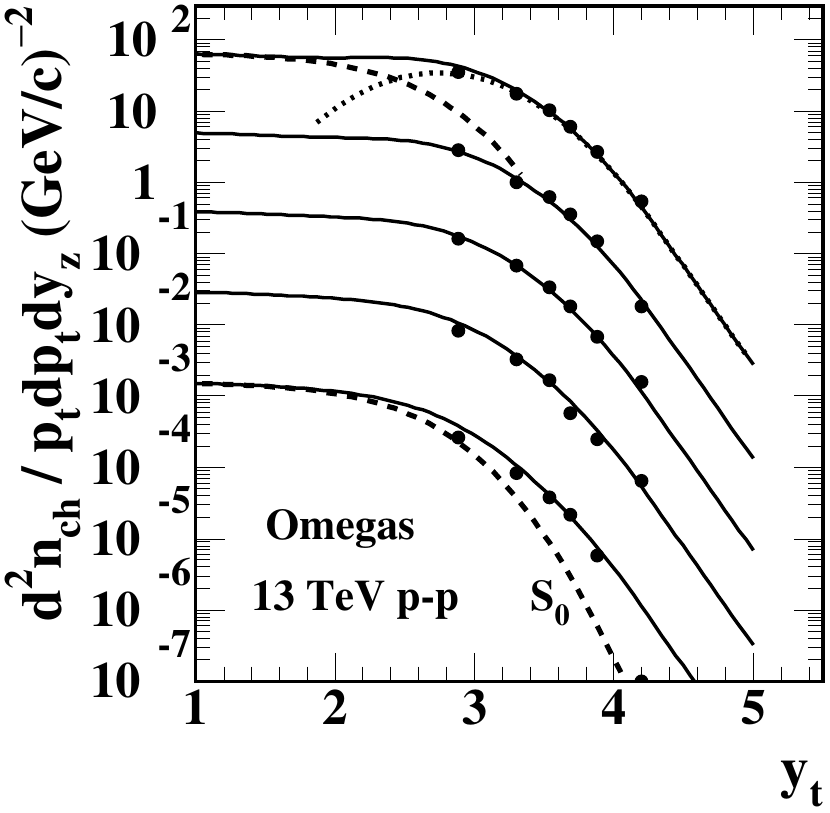}
	\put(-145,105) {\bf (c)}
	\put(-23,105) {\bf (d)}
	\\
	\caption{\label{piddata}
		\pt\ spectrum data (points) for strange hadrons:
		(a) Neutral kaons,
		(b) Lambdas,
		(c) Cascades and
		(d) Omegas from Ref.~\cite{alippss}.
		Solid curves represent the PID spectrum TCM. The dash-dotted curves in (c) are the TCM curves reduced by factor 2/3. The dashed curves are the TCM soft component in the form $z_{si}(n_s) \bar \rho_s \hat S_0(y_t)$.  The dotted curve in (d) is   the TCM hard component in the form $z_{hi}(n_s) \bar \rho_h \hat H_0(p_t)$.
	}  
\end{figure}

Figure~\ref{piddata} shows \pp\ PID spectrum data (points) as densities on \pt\ vs transverse rapidity \yt\ with pion mass assumed. Published spectra have been divided by published $p_t$ values relative to the presentation in Fig.~2 of Ref.~\cite{alippss}. Especially for Cascades note that there is no other rescaling. The spectra are multiplied by powers of 10 according to $10^{n-1}$ where $n \in [1,10]$ is the centrality class index and $n = 10$ is here most central, following the practice in Ref.~\cite{alippss}. Elsewhere in this paper event class $n = 1$ is most central (see $\bar \rho_0$ values listed above). The solid curves are full TCM parametrizations.  The Cascade data spectra  fall systematically below the TCM predictions. The dash-dotted curves are TCM predictions scaled down by factor 2/3. The dashed curves are TCM soft components in the form $ z_{si }\bar \rho_s \hat S_0(y_t)$. The dotted curve in (d) is Omega hard-component model $z_{hi}(n_s) \bar \rho_h \hat H_{0i}(p_t)$ indicating that almost all detected Omegas are jet fragments.

The plotting format in Figs.~\ref{ppbpiddata} and \ref{piddata} is desirable for two reasons: 
(a) Soft component $\hat S_0(y_t)$ that (for example) describes low-\pt\ data within their statistical uncertainties for \ppb\ and \pp\ collisions~\cite{pidpart1,pidpart2} closely approximates a Boltzmann exponential on \mt\ at lower \mt\ and thus follows an $A - m_i \cosh(y_t)/T$ trend in this semilog format on \yt\ (where $m_i$ is the mass for hadron species $i$ and A is some constant).
(b) Hard component $\hat H_0(y_t)$ that describes high-\pt\ data within statistical uncertainties for \ppb\ and \pp\ collisions follows an exponential $\exp(-q\, y_t)$ at higher \yt\ equivalent to a power law on \mt\ or \pt\ that manifests in this format as a straight line for $y_t > 4$. 

In contrast, the plotting format chosen for Fig.~2 of Ref.~\cite{alippss} shows spectra in the form $d^2 N_{ch} / dp_t dy_z$ that is missing a factor $1/p_t$ appropriate for momentum in the transverse plane and thus does not approximate a Boltzmann exponential for low \mt. The data are presented on linear \pt\ rather than logarithmic \yt, and data binning is also defined on \pt\ requiring dramatic variation of bin widths to control statistical uncertainties. On linear \pt\ there is no hint of the simple data trends evident on \yt. 

Table~\ref{pidparamx} shows hard-component parameters for 13 TeV \pp\ collisions. Parameters $\bar y_t^*$ and $\delta \bar y_t^*$ are used to define variable mode $\bar y_t(n_s)$ per Eq.~(\ref{ytbar}). Soft-component slope parameter $T = 200$ MeV and $n = 14$ are held fixed at those values for more-massive hadrons (as for \ppb). 

\begin{table}[h]
	\caption{TCM hard-component model parameters $(\bar y_t(n_s),\sigma_{y_t},q)$ for strange hadrons from 13 TeV \pp\ collisions. The first two columns define $\bar y_t(n_s)$ via Eq.~(\ref{ytbar}).
	}
	\label{pidparamx}
	\begin{center}
		\begin{tabular}{|c|c|c|c|c|} \hline
			& $\bar y_t^*$ & $\delta \bar y_t^*$ & $\sigma_{y_t}$ & $q$    \\ \hline
			$K_\text{S}^0$          &  $2.70\pm0.01$ &$0.00\pm0.05$ & $0.58\pm0.01$ & $3.8\pm0.1$   \\ \hline
			$\Lambda$       & $2.80\pm0.03$ & $0.70\pm0.05$ & $0.50\pm0.01$ & $4.6\pm0.1$   \\ \hline	
			$ \Xi $     &  $2.75\pm0.05$ & $0.90\pm0.05$ & $0.46^\dagger\pm0.01$ & $4.6\pm0.2$   \\ \hline
$\Omega$    & $3.00\pm0.05$  & $0.55\pm0.05$ & $0.50\pm0.05$ & $4.6\pm0.5$   \\ \hline
		\end{tabular}
	\end{center}
$^\dagger$Cascades $\sigma_{y_ti} = 0.46$ is required by data.
\end{table}

Table~\ref{coefparams2} below shows fractional-coefficient parameters for 13 TeV \pp\ collisions. Parameters $\tilde z_i^*$ and $\delta \tilde z_i^*$ are used to define variable hard/soft ratio parameter $\tilde z_i(n_s)$ per Eq.~(\ref{tildez}). Variable $z_{0i}(n_s)$ is defined by Eq.~\ref{z0trend} (Sec.~\ref{variable}). $z_{0i}(n_s)$ and $\tilde z_i(n_s)$ are then combined to define fractional coefficients $z_{si}(n_s)$ and $z_{hi}(n_s)$ per Eqs.~(\ref{zsix}). That combination plus Eq.~(\ref{pidspectcm}) determines the solid curves in Fig.~\ref{piddata}.  

\subsection{Data description quality}

In previous TCM analyses of PID spectra from 5 TeV \ppb\ collisions~\cite{ppbpid,pidpart1,pidpart2} and 13 TeV \pp\ collisions~\cite{pppid} data description quality was studied in detail using the Z-score statistic~\cite{zscore}. It was thereby established that spectrum data are described by the TCM within their statistical uncertainties. In the present analysis emphasizing multistrange baryons statistical uncertainties are substantially larger. The description quality is then apparent simply by visual comparison of data and TCM in Figs.~\ref{ppbpiddata} and \ref{piddata}.
The most recent modification to the variable TCM is Eq.~\ref{z0trend} that provides for variable abundances $z_{0i}(n_s)$ for Cascades and Omegas per Tables~\ref{coefparams} and \ref{coefparams2} based on yields from Refs.~\cite{alippbss,alippss} (see Sec.~\ref{totyields}).

\section{PID integrated strange yields} \label{integrated}

This section responds to presentations of total integrated PID yields $\bar \rho_{0i}$ from 5 TeV \ppb\ collisions and 13 TeV \pp\ collisions reported in Refs.~\cite{alippss} and \cite{alippbss} respectively. Variations of total yields, various combinations thereof and integrals over restricted high-\pt\ intervals as defined by Eq.~(\ref{ratint}) are considered relative to increasing total charge density $\bar \rho_0$. The relation of yield data to canonical suppression of strangeness is considered.

\subsection{PID integrated total yields} \label{totyields}

Figures~8-11, 13 and 14 of Ref.~\cite{alippss} relate to integrated total yields $\bar \rho_{0i}$ for 13 TeV \pp\ PID spectra in various combinations or ratios. Figures 8, 9 and 11 show $\bar \rho_{0i}$ vs total charge density $\bar \rho_{0}$. Figure 10 shows $\bar \rho_{0i}$ vs $\bar \rho_{0j}$ for various species combinations $(i,j)$. Figure 13 shows ratios $\bar \rho_{0i}/\bar \rho_{0j}$  for various combinations $(i,j)$, and Fig.~14 repeats Fig.~8 in comparison with Monte Carlos. Given the basic result from Fig.~8 of Ref.~\cite{alippss} the yield trends in the other plot formats  are easily generated.  Based on its Fig.~11, a log-log version of its Fig.~8, Ref.~\cite{alippss} asserts that ``As can be seen in Fig.~11, the yields of strange hadrons increase with the charged particle multiplicity following a power law behaviour...,'' the same for 7 and 13 TeV.  A straight line on a log-log plot does represent a power law, but in this case the power is trivially $\approx 1$ (modulo variations in slope as revealed in Fig.~\ref{ratios} below).

The TCM of Eq.~(\ref{pidspectcm}) in combination with Eq.~(\ref{zsix}) previously assumed $\bar \rho_{0i} = z_{0i} \bar \rho_{0}$ with $z_{0i}$  independent of event class~\cite{pidpart1}. In the present study ratios $ z_{0i}(n_s)  = \bar \rho_{0i} / \bar \rho_{0}$ plotted vs $\bar \rho_0$ or $x\nu$  may represent all the information carried by PID spectrum integrals $\bar \rho_{0i}$ and are addressed differentially to determine species abundance variation.

Figure~\ref{ratios} (left) compares \pp\ data (solid dots) and \ppb\ data (open circles) for integrated yields $\bar \rho_{0i}$ in the form of ratios $\bar \rho_{0i} / \bar \rho_{0} \approx z_{0i}(n_s)$. Some \pp\ data are derived from $K^\pm/\pi$ and $p/\pi$ ratios appearing in Fig.~5 of Ref.~\cite{alicepppid}.
$K_\text{S}^0$, $\Lambda$, $\Xi$ and $\Omega$ total yields appear in Fig.~8 of Ref.~\cite{alippss}.
The \ppb\ data include total yields for $\pi$, $K^\pm$, $K_\text{S}^0$, $p$ and $\Lambda$ from Ref.~\cite{aliceppbpid}
and $\Xi$ and $\Omega$ total yields from Table~3 of Ref.~\cite{alippbss}.
The \pp\ $X/\pi$ ratios are converted to $z_{0i}$ estimates by multiplying the ratios by fixed $z_{0i}^* \approx 0.8$ for pions. In addition, they are corrected for pion $dE/dx$ bias by factor 1/0.85 as discussed below. The error bars show systematic errors uncorrelated across event classes. 
The solid lines are $z_{0i}^*$ values from Tables~\ref{coefparams} and \ref{coefparams2}. The hatched bands represent NSD values $\bar \rho_{0\text{NSD}} \approx 5$ for 5 TeV \ppb\ collisions and $\approx 6.5$ for 13 TeV \pp\ collisions. 

\begin{figure}[h]
	\includegraphics[width=3.3in]{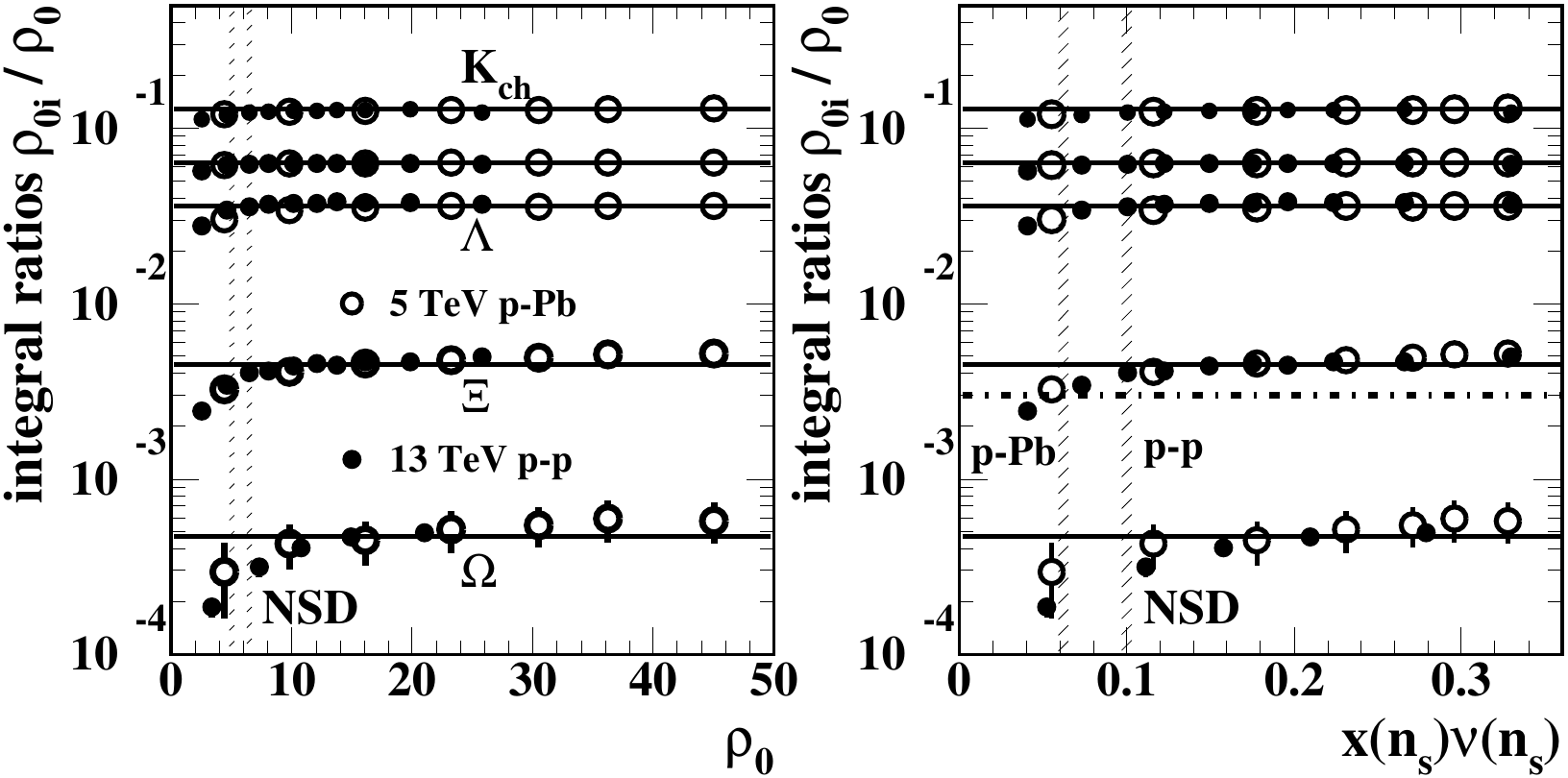}
	\caption{\label{ratios}
		Left: PID yield ratios $z_{0i}(n_s) \equiv \bar \rho_{0i} / \bar \rho_0$ vs $\bar \rho_0$ based on 13 TeV \pp\ data in Fig.~8 of Ref.~\cite{alippss} (solid dots) and 5 TeV \ppb\ data in Table~4 of Ref.~\cite{alippbss} (open circles) which then represent variable quantity  $z_{0i}(n_s)$ in the TCM context. The \ppb\ Omega errors have been doubled to make them visible. The solid lines are fixed values $z_{0i}^*$ as reported in the present study. The hadron species, from the top, are charged kaons, neutral kaons, Lambdas, Cascades and Omegas.
		Right: Data in the left panel plotted vs TCM hard/soft ratio $x(n_s) \nu(n_s)$ demonstrating the near equivalence of \pp\ and \ppb\ collision systems for quantity $z_{0i}(n_s)$ plotted vs that ratio. The dash-dotted line corresponds to \pp\ Cascade spectra in Fig.~\ref{piddata} (c). 
	} 
\end{figure}

Figure~\ref{ratios} (right) shows results in the left panel replotted vs hard/soft ratio $x(n_s) \nu(n_s)$ (with $\nu = 1$ for \pp). In that plot format the near equivalence of \ppb\ and \pp\ data is apparent.  A similar equivalence between \pp\ and \ppb\ systems is revealed for ensemble-mean \mmpt\ data in Sec.~\ref{mptsec}. \ppb\ Omega and Cascade errors are doubled to make them visible. The dash-dotted line corresponds to Cascade spectra in Fig.~\ref{piddata} (c) (dash-dotted)  consistent with $z_{0i}^* \approx 0.0030$ but contradicts published total yields.

\subsection{Variable fractional coefficients $\bf z_{0i}(n_s)$} \label{variable}

Significant variation of parameters $z_{0i}(n_s)$ for Cascade and Omega baryons is obvious in Fig.~\ref{ratios}. In order to incorporate that feature into a PID TCM the yield data should be examined more differentially.
Variable $z_{0i}(n_s)$ is  expressed in terms of TCM hard/soft ratio $x \nu$ as 
\bea \label{z0trend}
z_{0i}(n_s) \approx z_{0i}^* [1 + \delta z_{0i}^*(x\nu - 0.20)].
\eea
The value 0.20 is a consequence of choosing $z_{0i}^*$ values that correspond to spectra for central event class 5 (for \pp) or 4 (for \ppb), in which case the crossover in Fig.~\ref{ratios} (right) occurs near $x\nu \approx 0.2$.  $z_{0i}^*$ and $\delta z_{0i}^*$ values appear in Tables~\ref{coefparams} and \ref{coefparams2}. 

\begin{figure}[h]
	\includegraphics[width=3.3in]{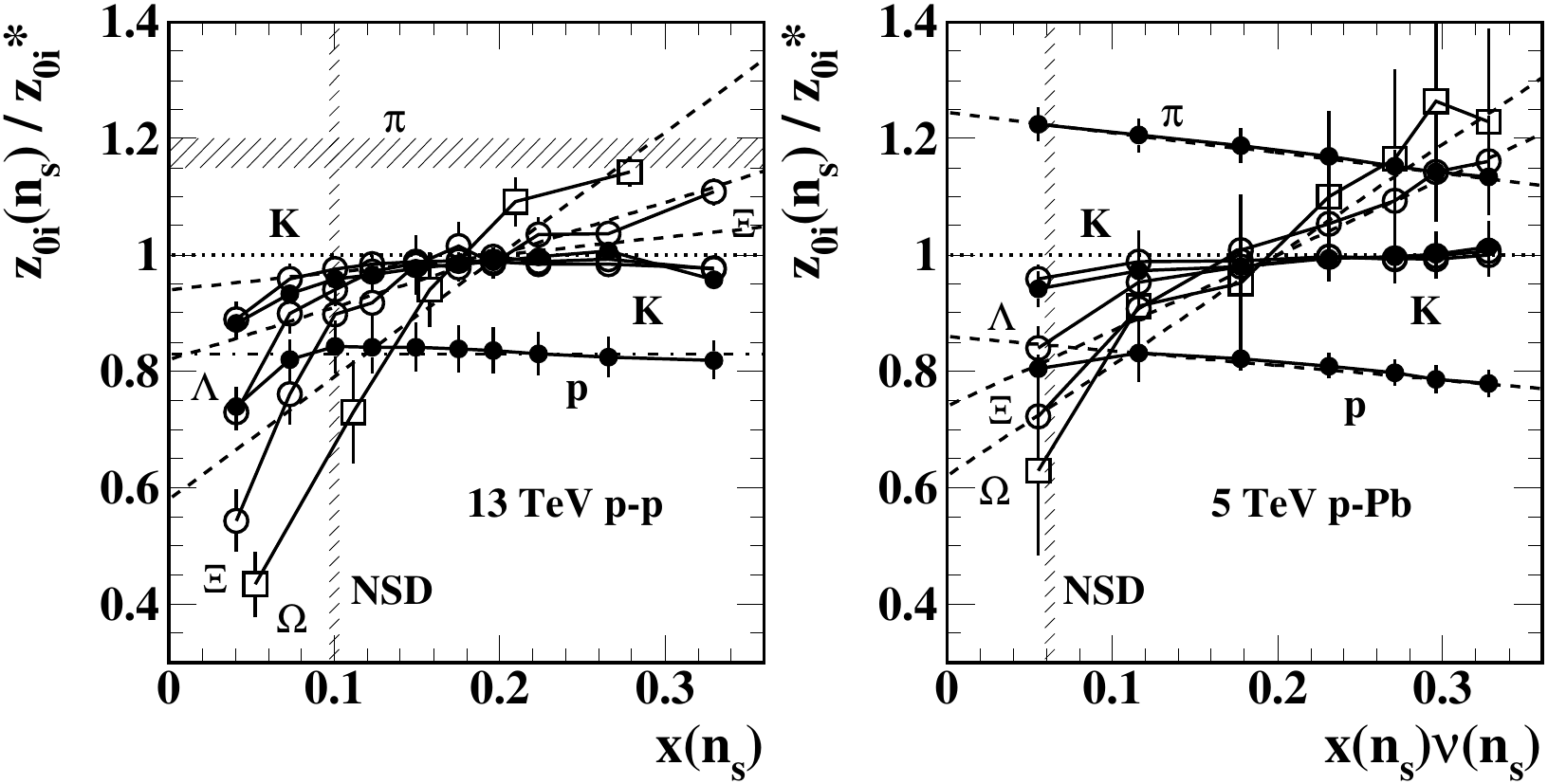}
	\caption{\label{ratiosx}
		Rescaled hadron fraction coefficients in the form $z_{0i}(n_s) / z_{0i}^*$ for 13 TeV \pp\ collisions (left) and 5 TeV \ppb\ collisions (right) and for charged hadrons (solid dots) and neutral hadrons (open circles, except open boxes for Omegas). The vertical hatched bands denote hard/soft ratio values corresponding to NSD collisions. The horizontal hatched bands estimate the biased $z_{0i}(n_s) / z_{0i}^* \approx 1.17$ value noted for \pp\ pions. The dash-dotted lines are the average value of the proton inefficiency correction function of  Eq.~(10) in Ref.~\cite{pppid}.
	} 
\end{figure}

Figure~\ref{ratiosx} presents a more differential analysis of coefficients $z_{0i}(n_s)$ showing ratios $z_{0i}(n_s) / z_{0i}^*$ vs hard/soft ratio $x(n_s)$ for \pp\ collisions (left) and vs $x(n_s)\nu(n_s)$ for \ppb\ collisions (right) and for charged hadrons (solid dots) and neutral hadrons (open circles, or open squares for Omegas) where the curves guide the eye. In the right panel it is notable that the {\em published} yields for pions and protons relative to statistical-model expectations differ systematically from reference value 1 to an extent consistent with bias arising from $dE/dx$ PID crosstalk as discussed in Refs.~\cite{pidpart1,pppid}. The dash-dotted lines in Fig.~\ref{ratiosx} represents the {\em average} value 0.82 of proton inefficiency function $\epsilon_p(p$-$p)$ defined by Eq.~(10) of Ref.~\cite{pppid}. 

All hadron species decrease substantially relative to reference value 1 for \nch\ {\em below} the NSD value (e.g.\ $\bar \rho_0 \approx 6.5 \Rightarrow x \approx 0.1$ for \pp\ collisions, vertical hatched band). No attempt is made within the TCM to describe such {\em local} low-\nch\ trends. The {\em dashed lines} in the two panels describe a more-gradual variation approximately linear vs hard/soft ratio $x$ or $x\nu$. Slopes of the dashed lines are then incorporated (as $\delta z_{0i}^*$) into Eq.~\ref{z0trend}. The dashed lines for Cascades and Omegas pass through 1.0 for $x\nu = 0.2$ while the hatched band for pions passes through 1.17 and dash-dotted line for protons through 0.81 at  that point.

Tables~\ref{coefparams} and \ref{coefparams2} summarize fractional-coefficient parameters based on TCM analysis of PID spectra in the present study plus $X/\pi$ ratios and/or integrated PID yields from Refs.~\cite{aliceppbpid,alicepppid,alippbss,alippss}. 
The $z_{0i}^*$ values are determined such that data ratio trends approximate 1 near $x\nu = 0.20$ in Fig.~\ref{ratiosx}. 
The exceptions are for pion and proton yields where the $z_{0i}^*$ values as presented in Table~\ref{otherparamsy} are fixed at 0.80 and 0.070 respectively from Ref.~\cite{pppid} {\em based on proton spectra corrected for $dE/dx$ bias and on pion $z_{si}(n_s)$ values determined by charge conservation}.

\begin{table}[h]
	\caption{\label{coefparams}
			TCM fractional-coefficient parameters for identified hadrons from 5 TeV \ppb\ collisions as reported in Refs.~\cite{aliceppbpid,alippbss}.  These parameters are relevant to Eqs.~(\ref{zsix}).  The first two columns define $z_{0i}(n_s)$ via Eq.~(\ref{z0trend}). The last two columns define $\tilde z_i(n_s)$ via Eq.~(\ref{tildez}). The effects of pion and proton bias are discussed in the text.
	}
	\begin{center}
		\begin{tabular}{|c|c|c|c|c|} \hline
			&   $100z_{0i}^*$ & $\delta z_{0i}^*$ &  $\tilde z_i^*$ &   $ \delta \tilde z_i^*$   \\ \hline
			$ \pi^\pm$        &   $80\pm1$  & $-0.35\pm0.1$  & $0.60\pm0.05$ &    $0.3\pm0.05$  \\ \hline
$K^\pm $   &  $ 13.0\pm0.5$   &  $0.0\pm0.1$ &  $2.63\pm0.2$&  $1.20\pm0.03$   \\ \hline
$K_s^0$        &   $6.4\pm0.2$ & $0.0\pm0.1$ &  $2.63\pm0.2$ &  $1.20\pm0.2$   \\ \hline
$p $        & $ 7.0\pm0.5$    &  $-0.25\pm0.1$ &  $5.3\pm0.2$&   $2.3\pm0.10$   \\ \hline
			$\Lambda $        &   $3.6\pm0.1$ & $0.0\pm0.2$ & $6.2\pm0.5$ &   $2.8\pm0.4$  \\ \hline	
			$ \Xi$        &    $0.45\pm0.03$  & $1.0\pm0.2$ & $7.4\pm0.5$  & $3.3\pm0.5$   \\ \hline
			$\Omega $   &   $ 0.047\pm0.005$   & $2.0\pm0.2$ &  $9.5\pm0.5$ &  $4.2\pm0.8$   \\ \hline
		\end{tabular}
	\end{center}
\end{table}

\begin{table}[h]
	\caption{\label{coefparams2}
		TCM fractional-coefficient parameters for identified hadrons from 13 TeV \pp\ collisions as reported in Refs.~\cite{alicepppid,alippss}. The dagger for Cascades $z_{0i}^*$ recalls the issue for Cascade spectra in Fig.~\ref{piddata} (c) that seem consistent with $z_{0i}^* \approx 0.0030$ which value however contradicts corresponding yield data as shown in Fig.~\ref{ratios} (right).  The last two columns define $\tilde z_i(n_s)$ via Eq.~(\ref{tildez}).
	}
	\begin{center}
		\begin{tabular}{|c|c|c|c|c|} \hline
			&   $100z_{0i}^*$ & $\delta z_{0i}^*$ &  $\tilde z_i^*$ &   $ \delta \tilde z_i^*$   \\ \hline
			$ \pi^\pm$        &   $80\pm1$  &  --  & $0.60\pm0.05$ &    $0.3\pm0.05$  \\ \hline
$K^\pm $   &  $ 13.0\pm0.5$   &  $0.3\pm0.1$ &  $2.63\pm0.2$&  $1.20\pm0.03$   \\ \hline
			$K_s^0$        &   $6.4\pm0.2$ & $0.0\pm0.1$ &  $2.63\pm0.2$ &  $1.20\pm0.2$   \\ \hline
$p $        & $ 7.0\pm0.5$    &  -- &  $5.3\pm0.2$&   $2.3\pm0.10$   \\ \hline
			$\Lambda $        &   $3.8\pm0.1$ & $0.0\pm0.2$ & $6.2\pm0.5$ &   $2.8\pm0.4$  \\ \hline	
			$ \Xi$        &    $0.45\pm0.03^\dagger$  & $1.0\pm0.2$ & $7.4\pm0.5$  & $3.3\pm0.5$   \\ \hline
			$\Omega $   &   $ 0.047\pm0.005$   & $2.0\pm0.2$ &  $9.5\pm0.5$ &  $4.2\pm0.8$   \\ \hline
		\end{tabular}
	\end{center}
	$^\dagger z_{0i}^*=0.0030$ from Cascades spectra disagrees with yields.
\end{table}

Note that the same $\tilde z_i(n_s)$ (right two columns) are applied to 5 TeV \ppb\ collisions and 13 TeV \pp\ collisions with no change in parameter values.
Hard/soft ratio $\tilde z_i^*$, measuring the jet contribution to species $i$, varies from $\approx 0.6$ for pions to $\approx 9.5$ for Omegas reflecting the mass dependence in Fig.~\ref{tildezparams} (right). The spectra for heavier hadrons are thus dominated by MB jet fragments for larger event \nch\ while still preserving integrated yields that remain close to statistical-model predictions due to conservative transport from soft to hard components as reported in Ref.~\cite{transport}. Also note  that $z_{0i}^* \approx 0.0030$ is required for Cascade spectra from 13 TeV \pp\ collisions (see Fig.~\ref{piddata}) whereas for 5 TeV \ppb\ collisions it is 0.0045. The difference is well outside systematic uncertainties. However, the integrated Cascade {\em yields} for 13 TeV \pp\ collisions from Ref.~\cite{alippss} are consistent with the 0.0045 value for \ppb\ collisions. See the \pp\ Cascade yield data points (solid dots) in Fig.~\ref{ratios}.
In Eq.~(\ref{z0trend}) hard/soft ratio $x\nu$ varies over interval 0.3 implying that for either \pp\ or \ppb\ collisions $z_{0i}(n_s)$ varies by 30\% for Cascades and 60\% for Omegas. Either variation is less than two $\sigma$.

\subsection{PID integrated high-$\bf p_t$ yields} \label{highpt}

This subsection relates specifically to Fig.~6 of Ref.~\cite{alippss}. Referring to spectra in ratio to a MB reference spectrum as in its Figs.~2-4 it is acknowledged that the ratios ``...reach a plateau for $p_T \geq 4$ GeV/c'' corresponding to the power-law trends obvious in the plot format of Sec.~\ref{smallspec} above. In its Fig.~6, spectrum integrals over $p_t > 4$ GeV/c are plotted vs charge density $\bar \rho_0$. The quantities are ``self-normalized,'' that is, divided by their average over event classes. It is noted that ``...the high-$p_T$ yields of strange hadrons increase  faster than the charged particle multiplicity. ...the data also hint at the increase being non-linear. ...the...yields of baryons are higher than those of [neutral kaons].'' It is of interest to contrast those observations with information derived from the TCM.

\begin{figure}[h]
	\includegraphics[width=3.3in]{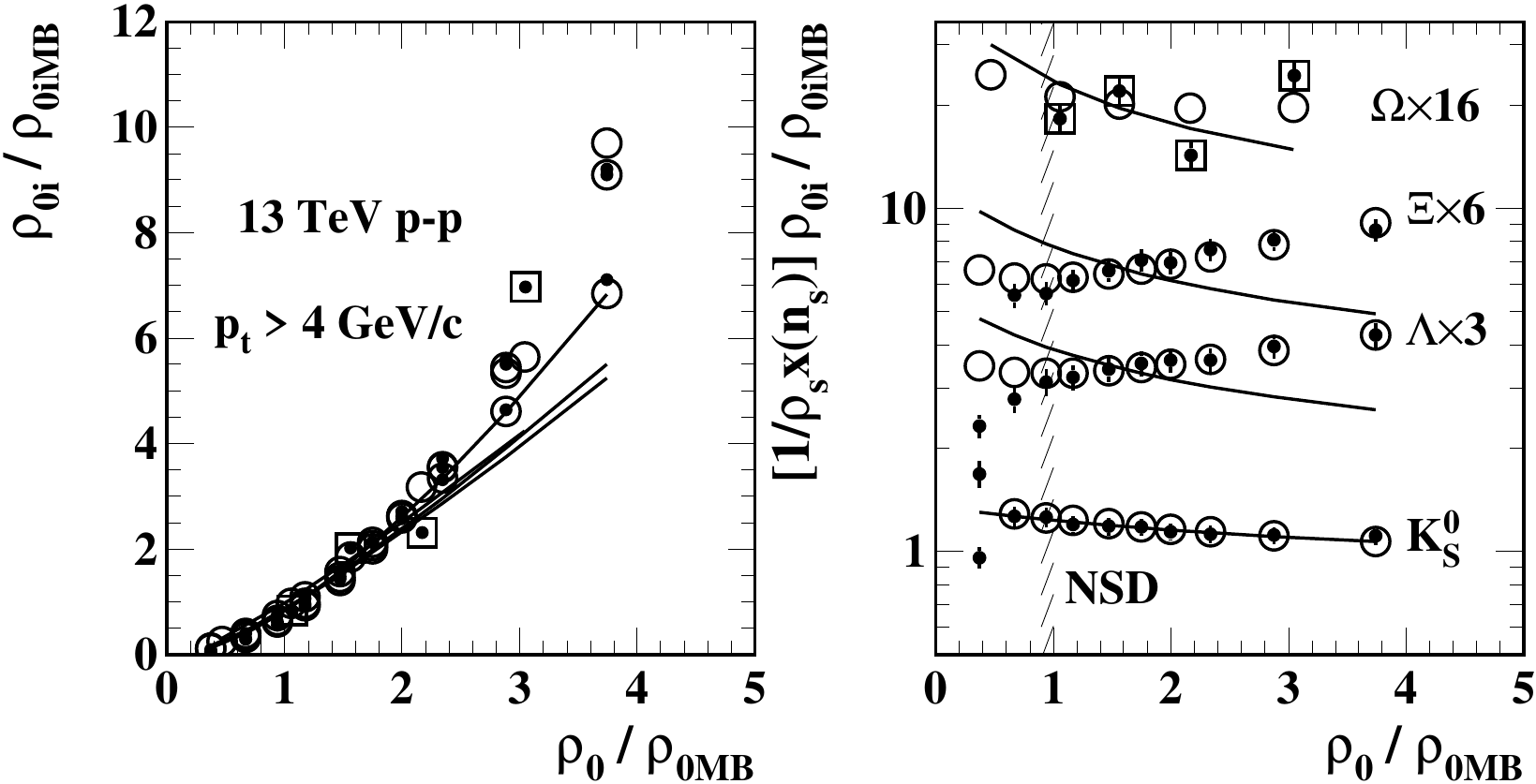}
	\put(-140,55) {\bf (a)}
	\put(-21,55) {\bf (b)}
\\
	\includegraphics[width=1.65in]{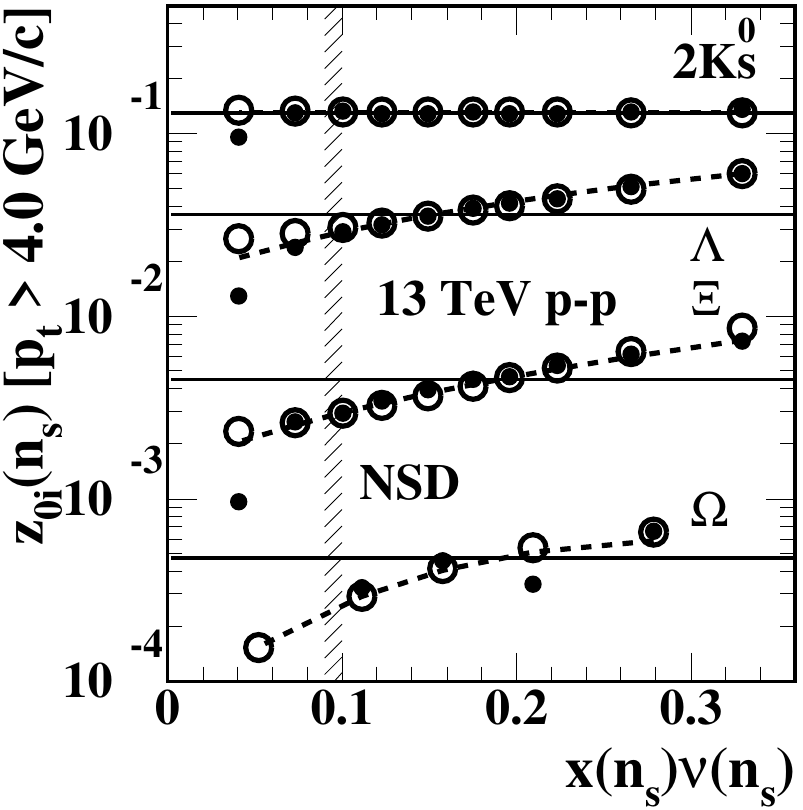}
	\includegraphics[width=1.62in]{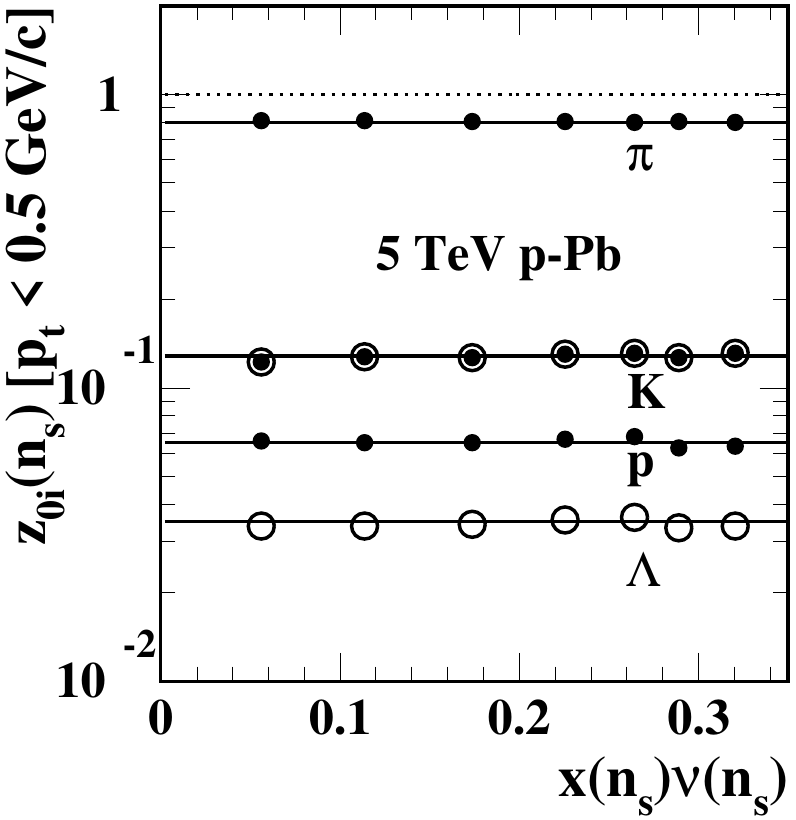}
	\put(-140,54) {\bf (c)}
	\put(-20,82) {\bf (d)}
	\caption{\label{highptx}
		(a) High-\pt\ PID spectrum integrals ($p_t > 4$ GeV/c) from 13 TeV \pp\ collisions (solid dots) for four hadron species as presented in Fig.~6 (top row) of Ref.~\cite{alippss}. Open squares identify $\Omega$ data. Open circles and curves are TCM results.
		(b) Data and curves in the left panel rescaled by $\bar \rho_h = x(n_s) \bar \rho_s$. Solid dots are data from Ref.~\cite{alippss}. Open circles are derived from TCM spectra (curves) in Fig.~\ref{piddata}. Solid curves are TCM trends with no hard-component mode shift on \yt.
		(c) Contents $I_h / \bar \rho_h$ of panel (b) transformed according to Eq.~(\ref{panelc}).
		(d) A similar procedure applied to spectrum integrals over \pt\ interval [0,0.5] GeV/c showing no significant $z_{0i}$ variation.
	}  
\end{figure}

Figure~\ref{highptx} (a) shows PID yield data (solid dots) for 13 TeV \pp\ collisions from Ref.~\cite{alippss} corresponding to its Fig.~6 (upper left). PID spectra in Fig.~\ref{piddata} above (solid dots) are integrated over high-\pt\ intervals ($> 4$ GeV/c, except $>3.8$ GeV/c for Omegas). Integrated yields are then rescaled by the same integral applied to a minimum-bias spectrum for the corresponding hadron species.  A similar rescaling is applied to total charge density $\bar \rho_0$ on the $x$ axis, where $\bar \rho_{0\text{MB}} \approx 6.9$. The open squares highlight the Omega data points. Note in Fig.~\ref{piddata}  that there are only two Omega spectrum data points above 3.8 GeV/c for four event classes and only one for the lowest \nch\ (explaining a missing Omega point in Fig.~\ref{highptx} (b)). The open circles result from the same procedure applied to TCM spectra (solid curves in Fig.~\ref{piddata}). The solid curves in (a) are explained below.  Panel (a) does not provide visual access to detailed trends. That figure represents much more information than is immediately apparent.

Given  structure of Eq.~(\ref{pidspectcm}) the integrals have the form
\bea \label{ratint}
I_h(n_s;4,\infty)&\equiv& z_{hi}(n_s) \bar \rho_{h} \int_\text{4 GeV/c}^\infty p_t dp_t \hat H_{0i}(p_t,n_s).~~~
\eea
The data increase in panel (a) relative to linear $\bar \rho_0$ is effectively dominated by factor $\bar \rho_h \propto \bar \rho_s^2 \sim \bar \rho_0^2$ in Eq.~(\ref{ratint}), that quadratic relation being a {\em signature feature of jet production} in \pp\ collisions~\cite{jetspec2}. The \pt\ lower bound on the integral in Eq.~(\ref{ratint}) is well above  the hard-component mode (also excluding the data soft component) and thus effectively represents an integral of the high-\pt\ tail of the jet fragment distribution particularly sensitive to mode or width variations of the peaked hard component.

Figure~\ref{highptx} (b) shows an alternative plotting format based on the TCM. Given the structure of Eq.~(\ref{ratint}) data and curves in panel (a) may be divided by density $\bar \rho_h$ to obtain more-detailed information. If data hard components $\hat H_{0i}(p_t)$ were independent of event class,  integrals in Eq.~(\ref{ratint}) should all be the same.  Variation with $\bar \rho_0$ would then be determined solely by PID coefficients $z_{hi}(n_s)$ defined by Eqs.~(\ref{zsix}) (solid curves in panel (b)). But as reported in Refs.~\cite{pidpart2,pppid} spectrum hard components $\hat H_{0i}(p_t,n_s)$ vary systematically with event class, meson functions (for \ppb\ collisions) shifting to {\em lower} \yt\ while baryon functions shift to higher \yt, with corresponding substantial variation of  integrals in Eq.~(\ref{ratint}). Trends for high-\pt\ integrals in  (b) should then correspond to trends for PID ensemble-mean $\bar p_{thi}$ in Fig.~\ref{ybarcomp} below.  The error bars represent uncorrelated systematic uncertainties.

Rescaling of TCM elements for Fig.~\ref{highptx} (b) is performed as follows. For each hadron species the TCM integrals (open circles, derived from solid curves in Fig.~\ref{piddata}) are rescaled to agree {\em on average} with the ALICE data (solid dots) as they appear in panel (b).  TCM hard-component modes $\bar y_t$ are then temporarily fixed at event class 5 (or 3 for Omegas) and TCM curves $z_{hi}(n_s)$ for each hadron species (solid curves) are rescaled to coincide with the resulting TCM integrals. {\em Variable} TCM hard components are then restored to produce final panel (b) and TCM points and curves are reverse transformed  to panel (a). 

Panel (c) provides further simplification. The solid curves in panel (b) $\propto z_{hi}(n_s)$ (including factor $z_{0i}(n_s)$) serve as a reference for the integrated-spectrum trends (solid and open points). An estimator for  $z_{0i}(n_s)$ can then be obtained from  Eq.~(\ref{ratint}) via Eq.~(\ref{zsix}) as 
\bea \label{panelc}
z_{0i}(n_s) \sim \frac{1 + \tilde z(n_s) x \nu}{\tilde z(n_s) (1 + x\nu)} I_h(n_s;4,\infty) / \bar \rho_h
\eea
in which fixed values would emerge if the integrals in Eq.~(\ref{ratint}) were independent of event class. Since data (solid points) in panel (a) were rescaled with inaccessible factors~\cite{alippss} the results in panel (c) have been rescaled to agree on average with the $z^*_{0i}$ values in Fig.~\ref{ratios}.

Panel (c) reveals that integration of the high-\pt\ portion of spectra {\em with intent to evaluate}  $z_{0i}(n_s)$ produces a misleading result because of evolution of the spectrum hard component (e.g.\ mode shifts on \yt) with changing event conditions. The issue  is {\em relative} variation on $x\nu$. There is no significant variation for \pp\ mesons $K^0_\text{S}$ but {\em equal substantial} variations for three baryon species. In particular, there is no apparent dependence on strangeness {\em per se}. The next question is how these results at higher \pt\ relate to the full \pt\ integrals in Fig.~\ref{ratios} (right).

Panel (d) shows the same procedure applied to the low-\pt\ [0,0.5] GeV/c portion of spectra. The inferred $z_{0i}(n_s)$ trends are then independent of event class (i.e.\ $n_s$) consistent with observed stability of the data soft component. The results in panels (c) and (d) may be compared with those in Fig.~\ref{ratios} (right). The variations in that figure result from a linear combination of no variation at lower \pt\ (d) and strong variation at higher \pt\ (c, a jet contribution). The combination then depends on how ``jetty'' a given hadron species is, but that is determined by $\tilde z_i(n_s)$. From Fig.~\ref{tildezparams} (right) $\tilde z_i(n_s)$ is 0.6 for pions vs 9.5 for Omegas. Close examination of Fig.~\ref{piddata} (d) shows that for most-central data every Omega detected is a jet fragment (lies on the dotted hard-component curve). Thus, relative to Omegas the large variations for baryons in panel (c) are strongly reduced for Cascades and negligible for Lambdas in Fig.~\ref{ratios}. 
The point of this exercise is that hidden in Fig.~6 of Ref.~\cite{alippss} is evidence that variation of fraction $z_{0i}(n_s)$ in Fig.~\ref{ratios}, interpreted to indicate strangeness enhancement, arises from small variations of jet production unrelated to strangeness {\em per se}.

\subsection{Omega baryons and jet production} \label{omegajets}

\begin{figure}[h]
	\includegraphics[width=1.66in]{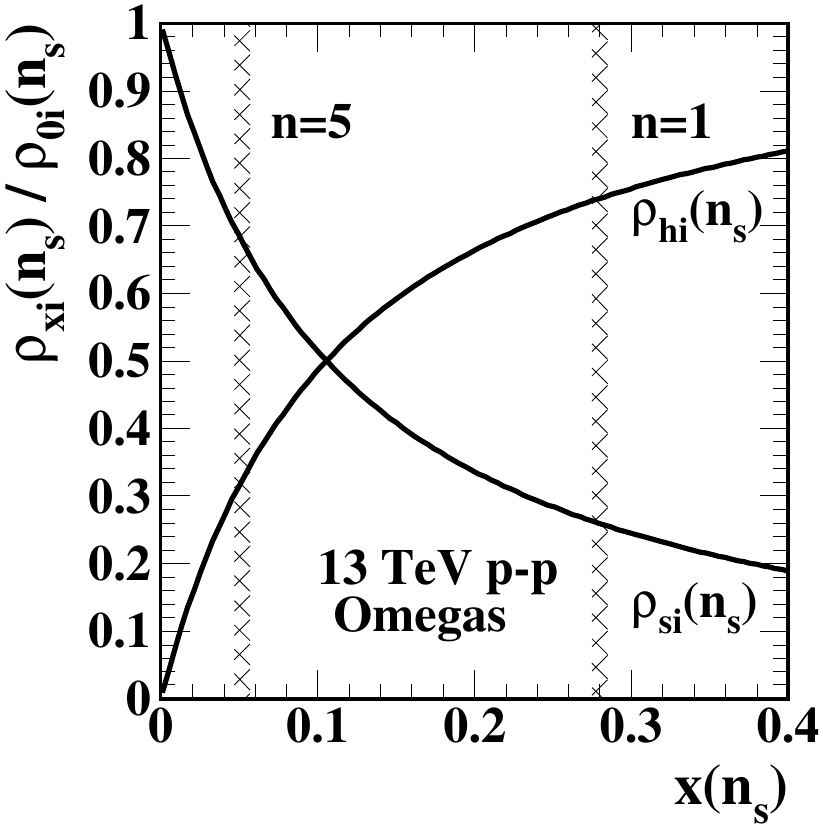}
	\includegraphics[width=1.64in]{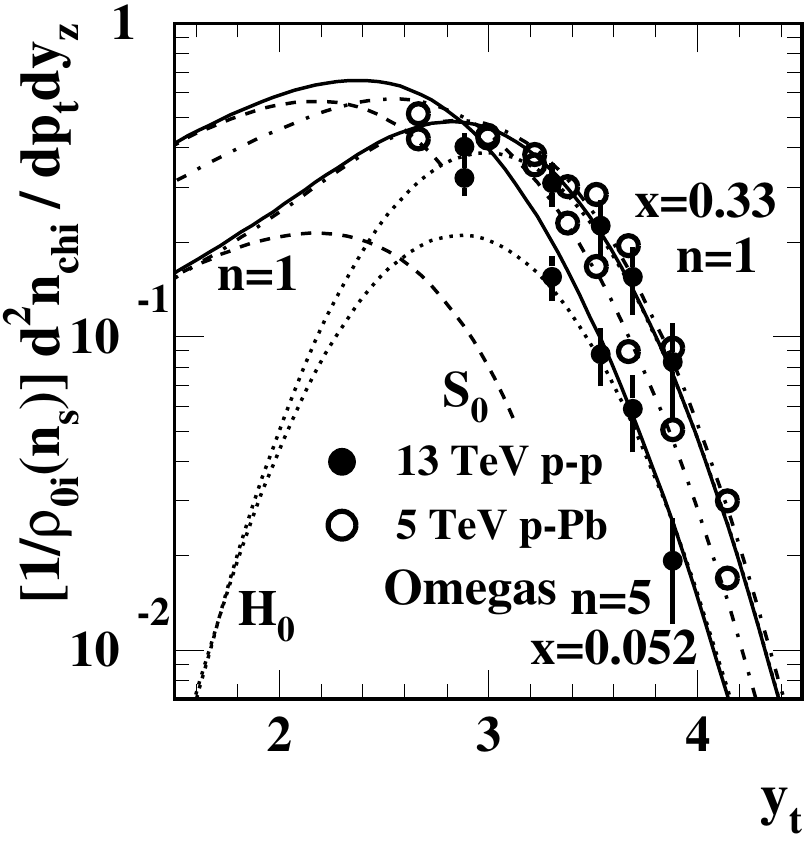}
	\caption{\label{omegarats}
		Left: Hard- and soft-component yield ratios $\bar \rho_{si} / \bar \rho_{0i}$ and $\bar \rho_{hi} / \bar \rho_{0i}$ for 13 TeV \pp\ collisions as reported in Ref.~\cite{transport}, including here a prediction (curves) for Omega baryons based on extrapolation of a mass trend for lower-mass hadrons. The hatched bands correspond to hard/soft ratio $x(n_s)$ values for Omega spectra from \pp\ event classes 1 (high \nch) and 5 (low \nch). The curves are labeled with the numerators on the $y$ axis.
		Right: PID spectra for Omega baryons for event classes 1 and 5 from 13 TeV \pp\ collisions (solid points) and for classes 2 and 6 from 5 TeV \ppb\ collisions (open circles). The solid curves are \pp\ TCM spectra with {\em fixed values} $z_{0i}^*$ all rescaled as indicated in the $y$-axis label. The dashed and dotted curves are corresponding soft and hard \pp\ TCM model functions. The dash-dotted curves are the \ppb\ TCM spectra.
	}  
\end{figure}

Figure~\ref{omegarats} (left) shows results reported in Ref.~\cite{transport}. Hadron numbers within jet (hard) and nonjet (soft) components of the final state are closely correlated so as to preserve the {\em total} hadron number or fraction for each species consistent with statistical-model predictions. The rate of jet production indicated by hard/soft coefficient ratio $\tilde z_i$ for each species (relative to total charge density $\bar \rho_{0}$) is simply proportional to hadron mass {\em irrespective of baryon identity or strangeness}. The curves correspond to ratios $\bar \rho_{si}/\bar \rho_{0i}$ and $\bar \rho_{hi}/\bar \rho_{0i}$ for Omega baryons and K mesons (to illustrate mass dependence) predicted by Ref.~\cite{transport} and depend only on parameter $\tilde z_i(n_s)$ simply proportional to hadron mass $m_i$. This plot illustrates how low-\pt\ behavior [$\bar \rho_{si}(n_s) = z_{si}(n_s)\bar \rho_s$] predicts high-\pt\ behavior [$\bar \rho_{hi}(n_s) =z_{hi}(n_s)\bar \rho_h$], for example enabling correction of proton inefficiencies in Refs.~\cite{pidpart1,pppid}.

The hatched bands correspond to values of hard/soft ratio $x(n_s)$ for 13 TeV \pp\ event classes 1  (0.33, high \nch) and 5 (0.052, low \nch). For event class 5 more  than two-thirds of Omegas are nonjet (projectile proton fragments) while for event class 1 {\em three quarters of Omegas are jet fragments}, thus representing a dramatic change in hadron production mechanisms while still maintaining approximately the {\em same} Omega fraction $z_{0i}^* \approx 0.00047$.

Figure~\ref{omegarats} (right) shows Omega spectra (solid dots) for \pp\ event classes 1 and 5 in the form $[1/\bar \rho_{0i}(n_s)]\, d^2n_{chi} / dp_t dy_z$ without the usual additional factor $1/p_t$, better indicating what fraction of particles lies within a given \pt\ interval (in this instance data spectra are not compared with a Boltzmann distribution on \mt). The $z_{0i}(n_s)$ values for $\bar \rho_{0i}$ are obtained from Eq.~(\ref{z0trend}) with parameter values from Table~\ref{coefparams2}. The error bars are total systematic uncertainties published with the data. 

The solid curves are TCM Omega spectra corresponding to Fig.~\ref{piddata} (d) rescaled in the same way. In that format the TCM spectra each integrate on \pt\ to 1, and in ratio the model curves are independent of $z_{0i}(n_s)$. The dashed curves are TCM soft components, and the dotted curves are corresponding hard components -- jet fragment distributions. Note the substantial mode shift to higher \yt\ for $n = 5 \rightarrow 1$. Corresponding to the left panel, for event class 1 the integrated hard component is three times the integrated soft component and the data points lie essentially entirely on the hard component. Thus, jets dominate Omegas from \pp\ collisions for event class $n = 1$.

The open circles and dash-dotted curves show 5 TeV \ppb\ Omega data and TCM for event classes $n = 2$ and 6. For $n=2$ the \ppb\ data and TCM are statistically equivalent to $n = 1$ \pp\ data and TCM. However, the $n = 6$ \ppb\ data correspond to $x\nu \approx 0.12$ in contrast to \pp\ $n = 5$ with $x\nu \approx 0.052$. (The $n=7$ \ppb\ spectrum has only four data points.) Thus, \ppb\ and \pp\ data and TCM differ substantially for the lower-\nch\ event class. Nevertheless, parameter values in Tables~\ref{coefparams} and~\ref{coefparams2}  describe \pp\ and \ppb\ Omega spectra within uncertainties.  

Taken together,  Omega data demonstrate that when compared on hard/soft ratio $x(n_s)\nu(n_s)$ ({\em not} on charge density $\bar \rho_0$) \pp\ and \ppb\ collision systems are equivalent. The TCM for Omega spectra is based in part on {\em predictions} from Ref.~\cite{transport} wherein it is determined that fractional coefficient $z_{0i}$ controls integrated yields from entire hadron spectra, {\em including jet contributions} up to high \pt, while the TCM accurately describes dramatic changes in differential spectrum shapes in terms of two fragmentation processes. A challenge for advocates of thermal models and canonical suppression (of strangeness) in small collision systems, claiming that for increasing ``centrality'' a volume increases and suppression is relaxed, is how a simple thermal model for Omegas may accommodate a heterogeneous system {\em dominated by jet production}.

\subsection{Canonical suppression and $\bf z_{0i}(n_s)$}

Deviations of parameters $z_{0i}(n_s)$ from fixed values $z_{0i}^*$ are discussed in Refs.~\cite{alippbss,alippss} in connection with canonical suppression of strangeness abundance.  Referring to claims that suppression or enhancement of specific hadron species with varying \nch\ depends strongly on strangeness content the following observations based on Fig.~\ref{ratiosx} are relevant: For \pp\ collisions {\em all} hadron species (except possibly pions) exhibit substantial reduction below the NSD $\bar \rho_0$ value relative to the reference. Hadrons that {\em elsewhere} exhibit significant variation of $z_{0i}(n_s)$ with \nch\ are Cascades and Omegas for \ppb\ and \pp\ collisions. 

Figure~4 of Ref.~\cite{alippbss} shows ratios $\bar \rho_{0i}/\bar \rho_{0j}$ vs $\bar \rho_0$ for $\Xi / \pi$ and $\Omega / \pi$, where the $\bar \rho_{0}$ (x) axis is logarithmic while the ratio ($y$) axis is linear. One can argue that such a linear vs log plot format exaggerates $2\sigma$ data variations and emphasizes deviations {\em below NSD charge densities}. Note that if the relation $y = kx$ is plotted as $y$ vs $\ln(x)$ the {\em apparent} local slope is $dy/d\ln x = kx$, not a constant value.  In contrast, Fig.~\ref{ratiosx} above is linear vs linear with reference value 1. An observation is made that the ratios increase significantly with charge density: ``The relative increase is more pronounced for [Omegas] than for [Cascades] ... indicating that strangeness content may control the rate of increase with multiplicity.'' What we might expect for intermediate \nch\ values, given Fig.~\ref{ratios} above, is ratio values $z_{0i} / z_{0j} = 0.0045 / 0.8 = 0.0056$ for  $\Xi / \pi$ and $0.00047 / 0.8 \approx  0.0006$ for $\Omega / \pi$. Those ratios correspond to the lines ($z_{0i}^*$ values) in Fig.~\ref{ratios} compared to pion $z_{0i} \approx 0.80$ with negligible variation~\cite{pppid}. Close examination of Fig.~4 in Ref.~\cite{alippbss} suggests that assumed pion $z_{0i}$ is $\approx 0.95$, whereas the value determined in Ref.~\cite{pppid} for 13 TeV \pp\ collisions is 0.80$\pm 0.01$ consistent with a statistical-model prediction~\cite{statmodel}. That result supports inference of $dE/dx$ bias for pions associated with Fig.~\ref{ratiosx} above. It is notable that the \pbpb\ result for Cascades is consistent with the higher-\nch\ result for \ppb\ whereas the \pbpb\ result for Omegas is 50\% higher than for \ppb. That difference is quite relevant to Fig.~5 of Ref.~\cite{alippbss}. Species ratios with 30\% uncertainties and 50\% unexplained inconsistencies should not be determining.

Figure~13 of Ref.~\cite{alippss} shows $X/K_\text{S}^0$ ratios for Lambdas, Cascades and Omegas from 13 TeV \pp\ collisions in which the ratios increase significantly with increasing charge density $\bar \rho_0$. Again, the horizontal axis is logarithmic while the vertical axis is linear, exaggerating the variation at lower \nch. The data are interpreted as follows: ``In the context of a canonical thermal model..., an increase of the relative strangeness abundance depending on the strange quark content can be interpreted as a consequence of a change in the system volume, called canonical suppression.'' Based on entries in Table~\ref{coefparams2} expected ratio values are 0.30 for Lambdas, 0.070 for Cascades and 0.0070 for Omegas. Data are consistent  with those values for larger \nch. Note that the more significant deviations from fixed values are confined to \nch\ values {\em below the NSD charge density} 6.5 for 13 TeV \pp\ collisions. 

Figure~\ref{highptx} presents the greatest difficulty regarding claims  for canonical suppression of strangeness. Strong variation of {\em inferred} $z_{0i}(n_s)$ at high \pt\ for Lambdas, Cascades and Omegas in panel (c) is clearly correlated with properties of the spectrum hard component representing a jet fragment distribution as shown in Fig.~\ref{ybarcomp} below. In contrast, $z_{0i}(n_s)$ at low \pt\ in panel (d) shows no significant variation. Figure~\ref{ratios} (right) then shows an average over \pt\ strongly correlated with hadron-mass-dependent jet production.  Association of modest variations of species abundances or species/species ratios with {canonical suppression} of strangeness thus seems unjustified. Such claims confront a context wherein differential spectrum analysis reveals copious jet production, especially for more-massive hadrons as in Fig.~\ref{omegarats}. Reference~\cite{transport} summarizes extensive TCM analysis of PID spectra from \pp\ and \ppb\ collisions and notes the implication that despite heterogeneous collision systems with jet and nonjet components {\em total} PID yields are consistent with some statistical models. That combination may have major implications for interpretation of \aa\ data.

\section{$\bf \bar p_t$ trends for strange hadrons} \label{mptsec}

This section presents a TCM analysis of PID ensemble-mean \mmpt\ data for strange and multistrange hadrons based on PID spectrum data from Refs.~\cite{aliceppbpid,alippbss,didiersss} for 5 TeV \ppb\ collisions and from Refs.~\cite{alicepppid,alippss} for 13 TeV \pp\ collisions. Conventional interpretations of PID  \mmpt\ data in terms of flows seem unconvincing compared to consequences of jet production in high-energy nuclear collisions. In this section \mmpt\ data are explored within a PID TCM context. \mmpt\ trends for \pp\ and \ppb\ collision systems vs hard/soft ratio $x\nu$ correspond closely and relate simply to two basic fragmentation processes. In particular, manifestations of systematic shifts of hard-component (jet fragment) distributions reveal substantial differences between mesons and baryons but show no correlation with strangeness.

\subsection{PID TCM for ensemble-mean $\bf \bar p_t$}

Given the PID spectrum TCM in Eq.~(\ref{pidspectcm}) the corresponding ensemble-mean {\em total} \pt\ for identified hadrons of species $i$, integrated over some angular acceptance $\Delta \eta$ that includes integrated charge $n_{chi}$, is
\bea \label{ptintid}
\bar P_{ti} 
&=&  n_{si}(n_s) \bar p_{tsi} +  n_{hi}(n_s) \bar p_{thi}(n_s).
\eea 
where $n_{xi}(n_s) = z_{xi}(n_s)n_x$.
The TCM for conventional ensemble-mean $\bar p_{ti}$  is then
\bea \label{pampttcmid}
\frac{\bar P_{ti}}{n_{chi}}  &\equiv&    \bar p_{ti}   \approx \frac{\bar p_{tsi} + \tilde z_i(n_s) x(n_s)\nu(n_s) \, \bar p_{thi}(n_s)}{1 +  \tilde z_i(n_s)x(n_s)\nu(n_s)}.
\eea
As an alternative, a \mmpt\ expression based on the TCM  has the particularly simple form
\bea \label{pampttcmpid}
\frac{\bar P_{ti}}{n_{si}} &=& \frac{n_{chi}}{n_{si}}\bar p_{ti} \approx \bar p_{tsi} + \tilde z_i(n_s) x(n_s)\nu(n_s) \, \bar p_{thi}(n_s).~~~
\eea
Factors $\tilde z_i(n_s) x(n_s)\nu(n_s)$ are defined above and soft- and hard-component mean values $\bar p_{tsi}$ and $\bar p_{thi}(n_s)$ are obtained from TCM models $\hat S_{0i}(y_t)$ and $\hat H_{0i}(y_t,n_s)$, where $\hat H_{0i}(y_t,n_s)$ may be slowly varying with $n_s$~\cite{pidpart2,pppid}. In what follows \mmpt\ data for lower-mass hadrons introduce the TCM application via previous analysis. The method is then newly applied to (multi)strange hadrons.

\subsection{Ensemble-mean $\bf \bar p_t$ for lower-mass hadrons}

\begin{figure}[h]
	\includegraphics[width=1.65in]{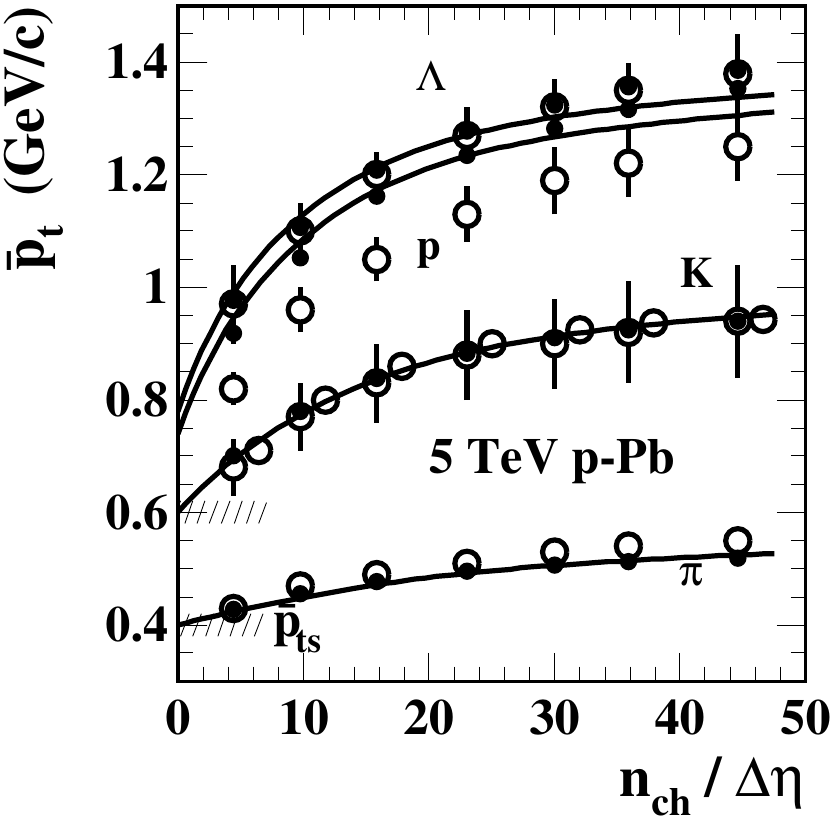}
	\includegraphics[width=1.65in]{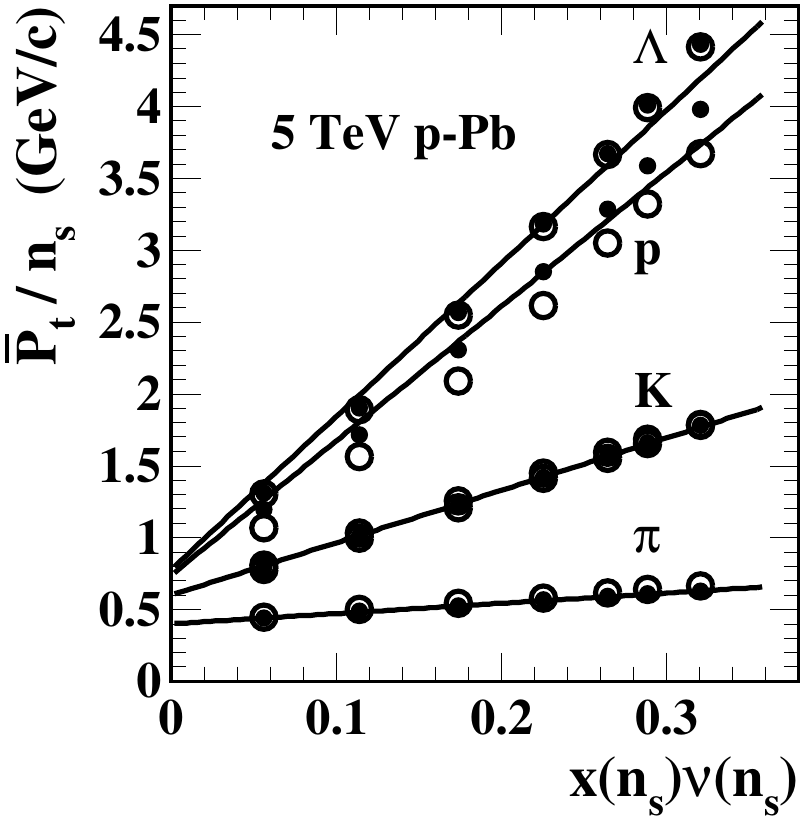}
	\put(-143,85) {\bf (a)}
	\put(-19,80) {\bf (b)}\\
	\includegraphics[width=3.3in]{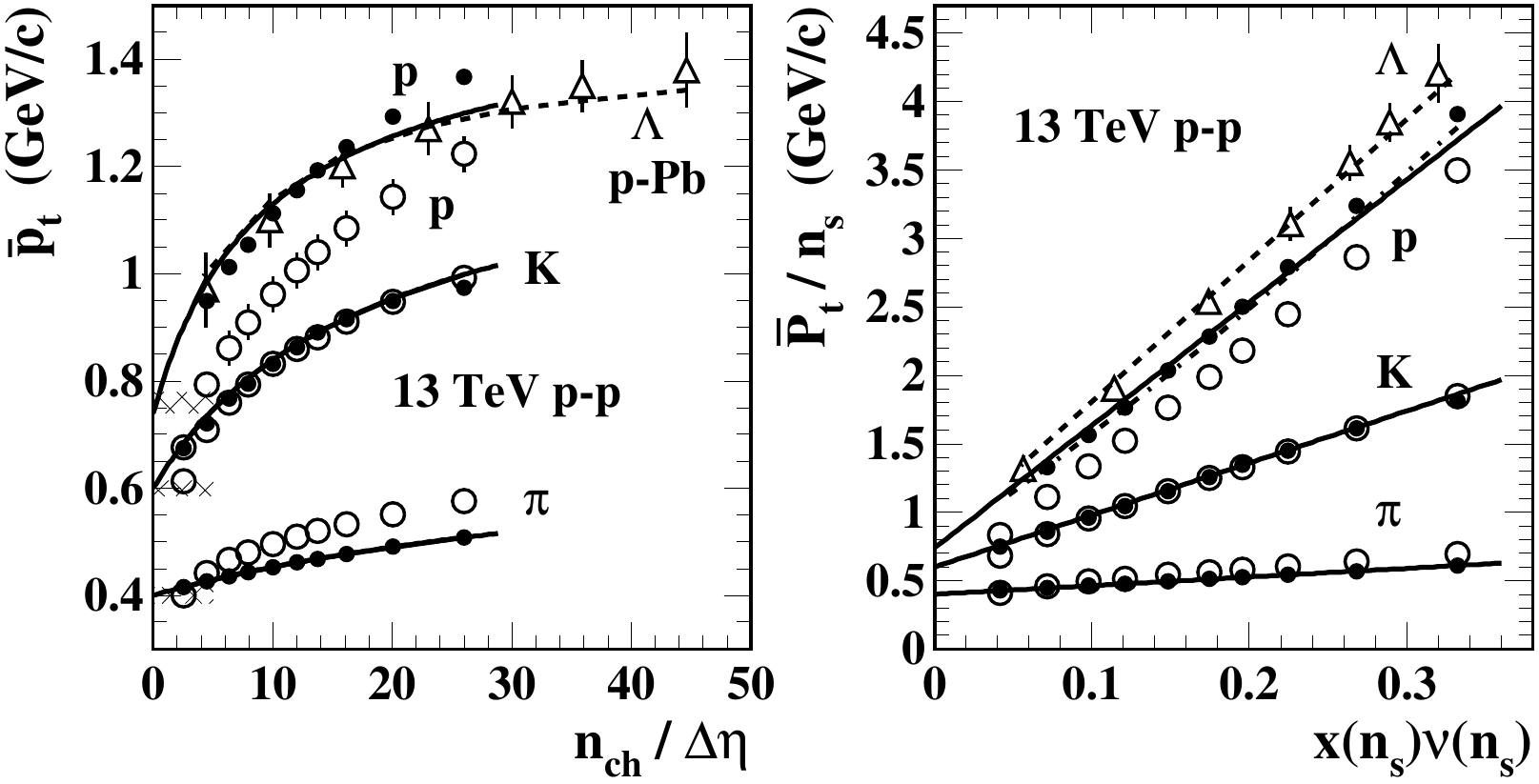}
	\put(-141,80) {\bf (c)}
	\put(-20,75) {\bf (d)}
	\caption{\label{mpttrends}
		Left: \mmpt\ data (open circles) for 5 TeV \ppb\ collisions (a) from Ref.~\cite{aliceppbpid} and for 13 TeV \pp\ collisions (c) from Ref.~\cite{alicepppid}. Solid dots are full TCM evaluated with \nch\ on data points. The curves are Eq.~(\ref{pampttcmid}) with {\em fixed} $\bar p_{thi}(n_s)$ for event class 4.		
		Right: Results in the left panels replotted in the format of Eq.~(\ref{pampttcmpid}) revealing approximately linear trends on $x(n_s)\nu(n_s)$. In this TCM format the equivalence of two collision systems differing in size and collision energy is apparent.
	} 
\end{figure}

Figure~\ref{mpttrends} shows \mmpt\ data (open circles) for 
5 TeV \ppb\ collisions from Ref.~\cite{aliceppbpid} (a) and for 13 TeV \pp\ collisions from Ref.~\cite{alicepppid} (c) plotted in the format of Eq.~(\ref{pampttcmid}) (as in  Ref.~\cite{alicepppid}). The solid dots are derived from TCM spectrum models (including for {\em corrected} proton spectra). The open triangles in (c) are Lambda data for 5 TeV \ppb\ collisions as reported in Ref.~\cite{aliceppbpid} for comparison. The solid curves are Eq.~(\ref{pampttcmid}) with fixed $\bar p_{tsi}$ and with fixed $\bar p_{thi}(n_s)$ for event class 4. The pion value $\bar p_{tsi}\approx 0.40$ GeV/c includes a resonance contribution which reduces the $\bar p_{tsi}$ value from what is inferred from $\hat S_{0i}(m_t)$. For \pp\ data in (c) $ x\nu \rightarrow x$ in Eq.~(\ref{pampttcmid}). Published pion \mmpt\ data (lowest open points in (a, c) fall substantially {\em above} the TCM predictions (lowest solid curves), which may relate to  {\em pion-proton misidentification} as discussed in Ref.~\cite{pppid}.

Figure~\ref{mpttrends} (b,d) shows the same data and curves in the format of Eq.~(\ref{pampttcmpid}) with trends vs hard/soft (jet/nonjet) ratio $x(n_s)\nu(n_s)$ or $x(n_s)$ exhibiting approximately linear dependences. Note that in this format the Lambda data in (b,d) correspond closely to {\em corrected} proton data (solid dots). Also note in (b,d) that baryon (including corrected proton) \mmpt\ increases somewhat faster than the TCM linear reference (solid lines) due to hard-component shifts to higher \yt\ with increasing \nch\ as reported in Refs.~\cite{pidpart2,pppid}. The simplicity of this format illustrates the importance of a TCM reference for precision analysis of data trends. 

\subsection{Ensemble-mean $\bf \bar p_t$ for multistrange hadrons} \label{strangemmpt}

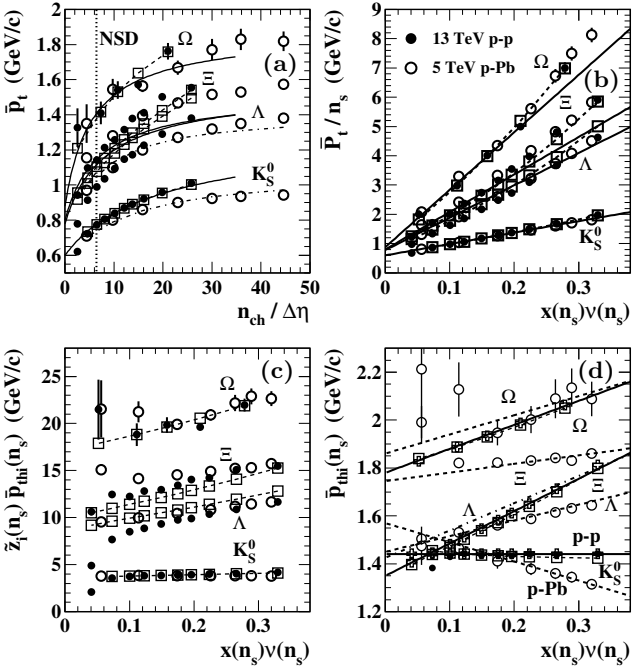
\begin{figure}[h]
	\includegraphics[width=3.3in]{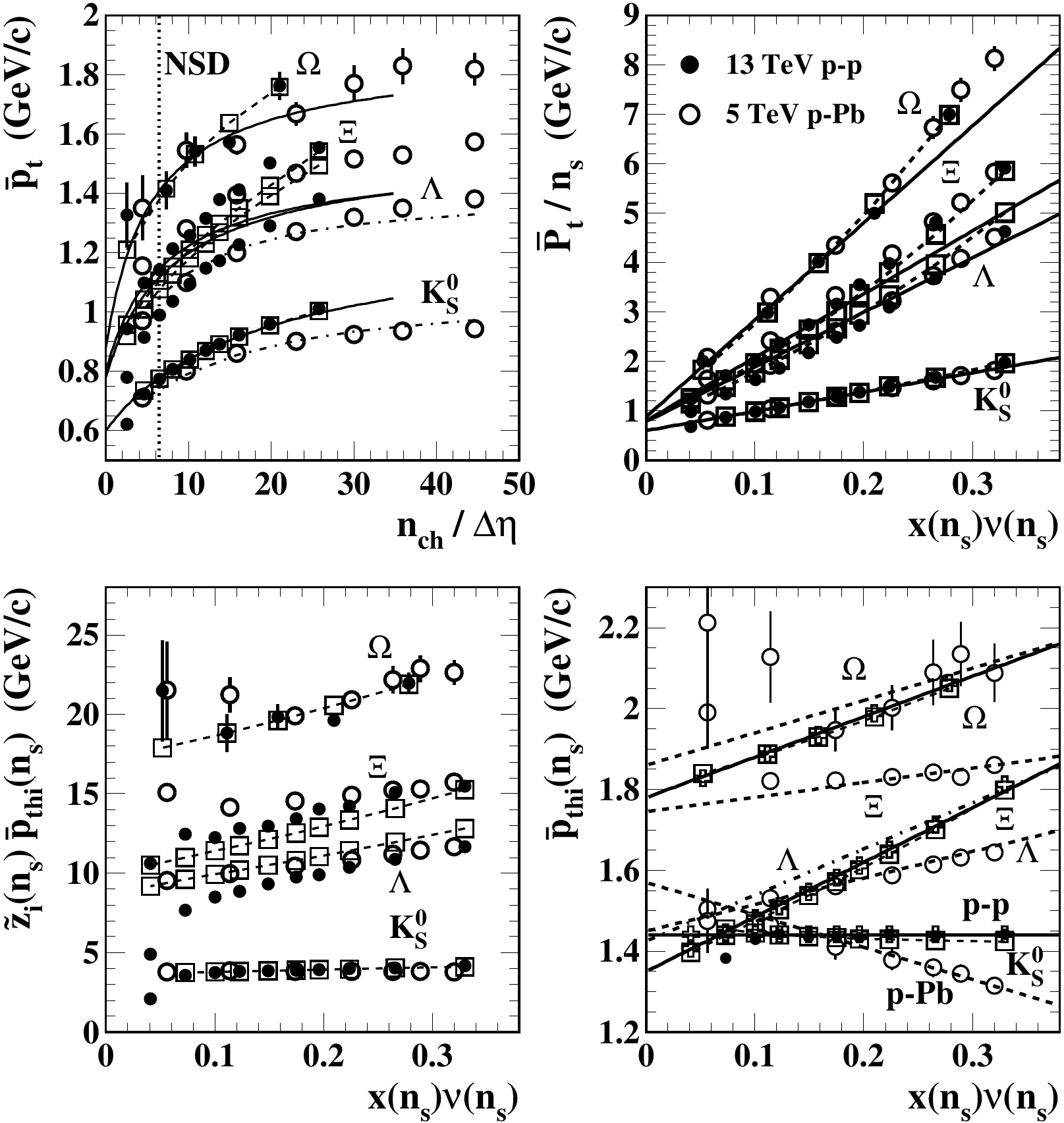}
	\put(-140,225) {\bf (a)}
	\put(-19,219) {\bf (b)}
	\put(-141,109) {\bf (c)}
	\put(-21,109) {\bf (d)}
	\caption{\label{mptss}
		(a) Ensemble-mean \mmpt\ data for 13 TeV \pp\ collisions from Fig.~5 of Ref.~\cite{alippss} (solid dots) and for 5 TeV \ppb\ collisions (open circles) from Ref.~\cite{aliceppbpid}. Open squares (connected by dashed curves) are \pp\ \mmpt\ values inferred from TCM curves in Fig.~\ref{piddata}. 
		Dash-dotted curves are \ppb\ \mmpt\ TCM trends from Refs.~\cite{ppbpid,pidpart2}. 
		(b) Data and curves from (a) corresponding to Eq.~(\ref{pampttcmid}) transformed to the TCM format of Eq.~(\ref{pampttcmpid}).
		(c) Quantity $\tilde z_i(n_s) \bar p_{thi}(n_s)$ derived from data and curves in (b) based on estimates of $\bar p_{tsi}$ and TCM hard/soft ratio $x(n_s)\nu(n_s)$.
		(d) Hard components $\bar p_{thi}(n_s)$ derived from data in (c) based on $\tilde z_i(n_s)$ trends described by Eq.~(\ref{tildez}). The open circles are $K_\text{S}^0$, $\Lambda$, $\Xi$ and $\Omega$ data from 5 TeV \ppb\ collisions. Open squares and dashed curves are derived from the \pp\ variable TCM. Open crosses are values derived directly from corresponding TCM hard-component model functions $\hat H_0(y_t,n_s)$. 	Solid dots correspond to 13 TeV \pp\ $K_\text{S}^0$ data. The dash-dotted line represents the TCM for 13 TeV \pp\ Lambdas.
	}  
\end{figure}

Figure~\ref{mptss} shows the \mmpt\ TCM applied to multistrange hadron data. The format in panels (a) and (b) of Fig.~\ref{mptss} correspond to left and right panels of Fig.~\ref{mpttrends}. Data for $K_\text{S}^0$, $\Lambda$, $\Xi$ and $\Omega$ from 13 TeV \pp\ collisions (solid dots in all panels) are as reported in Ref.~\cite{alippss}. The open circles from 5 TeV \ppb\ collisions are $K_\text{S}^0$ and $\Lambda$ data as reported in Ref.~\cite{aliceppbpid} and $\Xi$ and $\Omega$ data as reported in Ref.~\cite{didiersss}. Solid curves correspond to a \pp\ PID TCM with fixed hard components corresponding to event class 5 (except event class 3 for $\Omega$). Open squares (connected by dashed curves) are \mmpt\ values inferred from corresponding {\em variable} TCM PID spectra (solid curves in Fig.~\ref{piddata}).  

Figure~\ref{mptss} (a) presents a conventional \mmpt\ format per Eq.~(\ref{pampttcmid}). Panel (b) presents the same data and curves transformed to the format of Eq.~(\ref{pampttcmpid}) by multiplication with factor $1 + \tilde z_i(n_s) x(n_s) \nu(n_s)$ ($\nu \rightarrow 1$ for \pp\ data). In the format of Eq.~(\ref{pampttcmpid}) \pp\ and \ppb\ data are directly comparable, and the close correspondence between the two systems (note open circles vs solid dots) is evident whereas that is not the case in panel (a).

Panel (c) shows a further manipulation via Eq.~(\ref{pampttcmpid}) so that the product $\tilde z_i(n_s) \bar p_{thi}(n_s)$ is isolated given estimates of $\bar p_{tsi}$ from TCM model functions $\hat S_{0i}(y_t)$. For each hadron species the same transformation is applied to data and curves. The general increases vs $x(n_s) \nu(n_s)$ can be attributed to shifts of baryon hard components to higher \yt\ and {\em mass-dependent} increases of hard/soft ratios $\tilde z_i(n_s)$. The \pp\ $K_\text{S}^0$ data (solid dots) show good agreement with the TCM and seem to agree with \ppb\ data in that format. However, there appear to be significant deviations from the TCM for \pp\ $\Lambda$ and $\Xi$ data. 

Panel (d) isolates $\bar p_{thi}(n_s)$ via division of panel (c) by mass-dependent $\tilde z_i(n_s)$. For the TCM that is a consistency check. The open squares are derived from TCM spectra (solid curves in Fig.~\ref{piddata}) while the open crosses are derived directly from $\hat H_0(y_t)$ model functions. If TCM analysis were linear and self-consistent the two results should be essentially identical, and they are. The open circles are $\bar p_{thi}(n_s)$ values inferred by transforming 5 TeV \ppb\ $\bar p_{ti}$ data from panel (a) as described above, with statistical errors only. The solid dots correspond to 13 TeV \pp\ $K_\text{S}^0$ data from Ref.~\cite{alippss}. The dash-dotted line represents TCM results for Lambdas from 13 TeV \pp\ collisions. This is a highly differential  format with precision $\approx 30$ MeV/c or 2\% of $\bar p_{thi}$ averages over event classes.

Table~\ref{mptparamsx} shows soft $\bar p_{tsi}$ and hard $\bar p_{thi}(n_s)$ ensemble-mean \mmpt\ components for four hadron species derived from corresponding model functions with parameters from Tables~\ref{pidparamz} for \ppb\ data. Hard-component variations with event class are parametrized by the expression
\bea \label{mpth}
\bar p_{thi}(n_s) &=& \bar p_{thi}^* + \delta \bar p_{thi}^* x(n_s)\nu(n_s)
\eea
(with $\nu$ = 1 for \pp\ data). Table entries in the last two columns correspond to dashed lines in Fig.~\ref{mptss} (d) and Fig.~\ref{ybarcomp} (right) . The first columns in  these  two  tables are identical because for given energy non-pion $\bar p_{tsi}$ varies only with hadron mass.

\begin{table}[h]
	\caption{\label{mptparamsx}
		Ensemble-mean \mmpt\ TCM parameters for 5 TeV \ppb\ collisions. The $\bar p_{thi}^*$ and $\delta \bar p_{thi}^*$ entries correspond to the dashed lines in Fig.~\ref{mptss} (d) and \ref{ybarcomp} (right). Soft-component entries $\bar p_{tsi}$ are the same as for 13 TeV \pp\ collisions.
	}
	\begin{center}
		\begin{tabular}{|c|c|c|c|} \hline
			&   $\bar p_{tsi}$ (GeV/c)    &  $\bar p_{thi}^*$ (GeV/c)  &   $ \delta \bar p_{thi}^*$  (GeV/c)  \\ \hline
			$K_s^0$        &  $0.60\pm0.01$ &  $1.57\pm0.02$ &  $-0.80\pm0.05$   \\ \hline
			$\Lambda $        &  $0.78\pm0.02$    & $1.46\pm0.02$ &   $0.65\pm0.05$  \\ \hline	
			$ \Xi$        &   $0.80\pm0.03$  & $1.75\pm0.02$  & $0.36\pm0.05$   \\ \hline
			$\Omega $   &  $ 0.88\pm0.05$   &  $1.86\pm0.03$ &  $0.80\pm0.05$   \\ \hline
		\end{tabular}
	\end{center}
\end{table}

Table~\ref{mptparams} shows soft $\bar p_{tsi}$ and hard $\bar p_{thi}(n_s)$ ensemble-mean \mmpt\ TCM components for four hadron species derived from corresponding model functions with parameters from Table~\ref{pidparamx} for \pp\ data. Entries in the last two columns correspond to solid and dash-dotted (for Lambdas) lines in Fig.~\ref{mptss} (d) and solid lines in Fig.~\ref{ybarcomp} (right).

\begin{table}[h]
	\caption{\label{mptparams}
	Ensemble-mean \mmpt\ TCM parameters for 13 TeV \pp\ collisions. The $\bar p_{thi}^*$ and $\delta \bar p_{thi}^*$ entries correspond to the solid lines in Fig.~\ref{mptss} (d) and \ref{ybarcomp} (right). Soft-component entries $\bar p_{tsi}$ are obtained from models $\hat S_{0i}(p_t)$ that depend only on hadron mass as shown in Fig.~\ref{shapes} (left).
	}
	\begin{center}
		\begin{tabular}{|c|c|c|c|} \hline
			&   $\bar p_{tsi}$ (GeV/c)    &  $\bar p_{thi}^*$ (GeV/c)  &   $ \delta \bar p_{thi}^*$  (GeV/c)  \\ \hline
			$K_s^0$        &  $0.60\pm0.01$ &  $1.44\pm0.02$ &  $0.00\pm0.05$   \\ \hline
			$\Lambda $        &  $0.78\pm0.02$    & $1.42\pm0.02$ &   $1.14\pm0.05$  \\ \hline	
			$ \Xi$        &   $0.80\pm0.03$  & $1.35\pm0.02$  & $1.35\pm0.05$   \\ \hline
			$\Omega $   &  $ 0.88\pm0.05$   &  $1.78\pm0.03$ &  $1.00\pm0.05$   \\ \hline
		\end{tabular}
	\end{center}
\end{table}

Much of the information carried by PID \mmpt\ data is represented in panel (c): The combination $\tilde z_i(n_s) \bar p_{thi}(n_s)$, the slope for \mmpt\ trends in panel (b), increases by factor 30 with hadron mass: $\approx 0.7$ GeV/c for pions vs $\approx 20$ GeV/c for Omegas by the combination of $\tilde z_i(n_s)$ (simply proportional to hadron mass per Fig.~\ref{tildezparams}) and $\bar p_{thi}(n_s)$ (via spectrum hard components reflecting PID fragmentation functions~\cite{ppbpid}). Variation of $\tilde z_i(n_s)$ shows no sensitivity to baryon identity or strangeness content {\em per se} and, given panel (d), there is no clear systematic trend for $\bar p_{thi}(n_s)$ other than mesons vs baryons. 

\subsection{Direct comparison of hard-component trends} \label{directcomp}

As demonstrated in Fig.~\ref{ratios} (right) hadron species abundances are similar for \pp\ and \ppb\ collisions when compared vs hard/soft ratio $x(n_s)\nu(n_s)$. As demonstrated in Fig.~\ref{shapes} (left) below, soft components $\hat S_{0i}(y_t)$ vary simply with mass independent of collision system. Since \mmpt\ soft components $\bar p_{tsi}$ are derived from those model functions the same applies to them. Thus, interesting variations apply only to evolution of jet-related hard components $\hat H_{0i}(y_t,n_s)$. Especially for higher-mass hadrons with limited \pt\ coverage exponent $q$ is not determining; $q$ is fixed at 3.7 for kaons and 4.6 for baryons. Modes $\bar y_{ti}(n_s)$ and Gaussian widths $\sigma_{y_ti}$ are the critical parameters.

\begin{figure}[h]
	\includegraphics[width=3.3in]{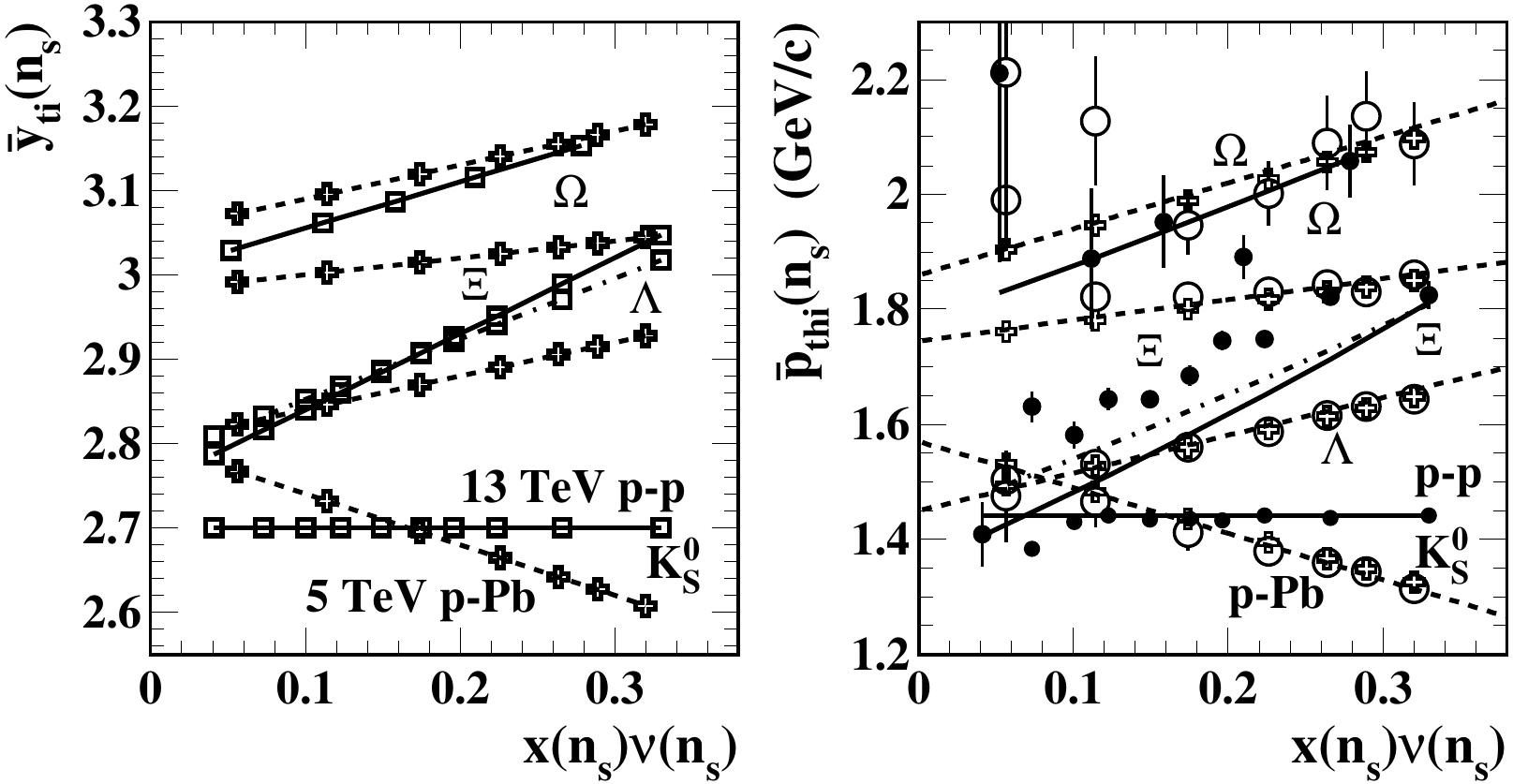}
	\caption{\label{ybarcomp}
		Left: Variable hard-component modes $\bar y_t(n_s)$ vs hard/soft ratio $x(n_s)\nu(n_s)$ for four hadron species and for 13 TeV \pp\ collisions (solid lines, open boxes) and 5 TeV \ppb\ collisions (dashed lines, open crosses). The trends are determined by $\bar y_t^*$ and $\delta \bar y_t^*$ parameters from Tables~\ref{pidparamz} for 5 TeV \ppb\ collisions and \ref{pidparamx} for 13 TeV \pp\ collisions.
		Right: Corresponding TCM $\bar p_{thi}(n_s)$ trends obtained from TCM models $\hat H_{0i}(y_t)$ (lines, open crosses) with the same plot conventions. Also shown are $\bar p_{thi}(n_s)$ trends obtained from PID \mmpt\ data as described for Fig.~\ref{mptss} (d) for \ppb\ collisions (open circles) and \pp\ collisions (solid dots). The dash-dotted line corresponds to the 13 TeV \pp\ Lambda TCM.
		}  
\end{figure}

Figure~\ref{ybarcomp} presents a comparison between TCM hard-component variable modes $\bar y_{ti}(n_s)$ (left) and \mmpt\ hard-components $\bar p_{thi}(n_s)$ (right). The TCM mode shifts are determined by Eq.~(\ref{ytbar}) with parameters $\bar y_{ti}^*$ and $\delta \bar y_{ti}^*$ in Tables~\ref{pidparamz} for 5 TeV \ppb\ collisions (dashed) and \ref{pidparamx} for 13 TeV \pp\ collisions (solid). The 13 TeV \pp\ Lambda mode trend is indicated by the dash-dotted line to distinguish from the \pp\ Cascade trend (solid).

Figure~\ref{ybarcomp} (right) shows corresponding $\bar p_{thi}(n_s)$ trends derived from published \mmpt\ data (solid dots and open circles for \pp\ and \ppb\ collisions respectively). Data are in this case plotted with  statistical uncertainties only. Solid (\pp) and dashed (\ppb) curves are Eq.~(\ref{mpth}) with parameter values from Tables~\ref{mptparamsx} and \ref{mptparams} respectively. The dash-dotted curve represents \pp\ Lambdas. The open crosses are derived directly from \ppb\ TCM hard components. 

Small $\bar y_{ti}(n_s)$ vs  $\bar p_{thi}(n_s)$ differences in relative position arise mainly from differences in TCM hard-component widths $\sigma_{y_ti}$ that also affect the resulting $\bar p_{thi}(n_s)$ trends. The $\bar p_{thi}(n_s)$ trends for  $K_\text{S}^0$ are higher relative to other species than the $\bar y_{ti}(n_s)$ trends on the left because $\sigma_{y_ti}$ for kaons is 0.58 (above the mode) compared to 0.50 for other species (except Cascades as noted below). The \pp\ TCM trend for Cascades $\bar p_{thi}(n_s)$ at right (solid line) is lower relative to other species compared to its $\bar y_{ti}(n_s)$ trend at left because {\em for \pp\ spectrum data only} the required Cascade width $\sigma_{y_ti}$ is 0.46 rather than 0.50 (see below).

$\bar p_{thi}(n_s)$ values derived from \mmpt\ data generally agree with TCM trends within statistical uncertainties, with the exception of \pp\ Cascades. (Lambda data for \pp\ are not plotted to minimize confusion among different symbols.) \pp\ cascade data in  the right panel (solid dots) lie systematically {\em above} the trend (solid line) derived from TCM spectra in Fig.~\ref{piddata}. As noted for $z_{0i}(n_s)$ trends, \pp\ Cascade {\em spectra} are unique in requiring $z_{0i}^* = 0.00030$ rather than 0.00047 and $\sigma_{y_ti}=0.46$ rather than 0.50. Yet {\em integral measures} derived from the same $\Xi$ spectra such as $z_{0i}(n_s)$ values in Fig.~\ref{ratios} (right)  conform to expectations from other species and to the TCM. 

Note that correspondence between the PID TCM and details of PID \mmpt\ data at the percent level, and correspondence between left and right panels in Fig.~\ref{ybarcomp}, relies on a TCM based on accurate determination of jet contributions to PID spectra and on accurate determination of \ppb\ centrality and hard/soft ratio $x(n_s)\nu(n_s)$~\cite{tomglauber}.
While \mmpt\ data may provide precision tests for certain issues, in general \mmpt\ data represent information {\em reduction} relative to differential spectra: spectrum structure can be used to interpret \mmpt\ data, but \mmpt\ data cannot be used to interpret spectrum structure.

Ensemble-mean \mmpt\ values in right columns of Tables~\ref{otherparamsx} and \ref{otherparamsy}, based on model-function parameters in Tables~\ref{pidparams} and \ref{engparamsyy}, were used in a study of hadron species transport from soft to hard component within small collision systems~\cite{transport}.  The apparent disagreement between $\bar p_{thi}$ value 1.34 for neutral kaons from \ppb\ collisions and 1.46 for charged kaons from \pp\ collisions arises from different evolution with \nch\ for two systems as illustrated in Fig.~\ref{ybarcomp} (right): The value for midcentral ($n = 4$) \ppb\ is 1.35 whereas the constant value for \pp\ is 1.44, agreeing within uncertainties with the entries in Tables~\ref{otherparamsx} and \ref{otherparamsy}.

\subsection{Direct comparison of (multi)strange spectra}

The trends in Fig.~\ref{ybarcomp} suggest making direct comparisons between data and TCM spectra for pairs of event classes from  \pp\ and \ppb\ collision data that should correspond most closely. Rescaled spectra in the form $\bar \rho_{0i} / \bar \rho_s$ are compared. Based on Eq.~(\ref{pidspectcm}) the rescaled spectra should depend only on hard/soft ratio $x(n_s)\nu(n_s)$. For spectrum equivalence quantities $z_{si}(n_s)$, $\tilde z_i(n_s)$ and $\hat H_{0i}(y_t,n_s)$ in Eq.~(\ref{pidspectcm}) should agree between two collision systems. For a given hadron species the first two quantities agree if $x(n_s)\nu(n_s)$ for the two systems agree. Agreement for $\hat H_{0i}(y_t,n_s)$ is then determined by the $\bar p_{thi}(n_s)$ trends or $\bar y_t(n_s)$ trends in Fig.~\ref{ybarcomp}. Thus, for each hadron species, event classes for \pp\ and \ppb\ collisions near {\em crossovers} of $\bar p_{thi}(n_s)$ trends on  $x(n_s)\nu(n_s)$ are selected. 

\begin{figure}[h]
	\includegraphics[width=3.3in]{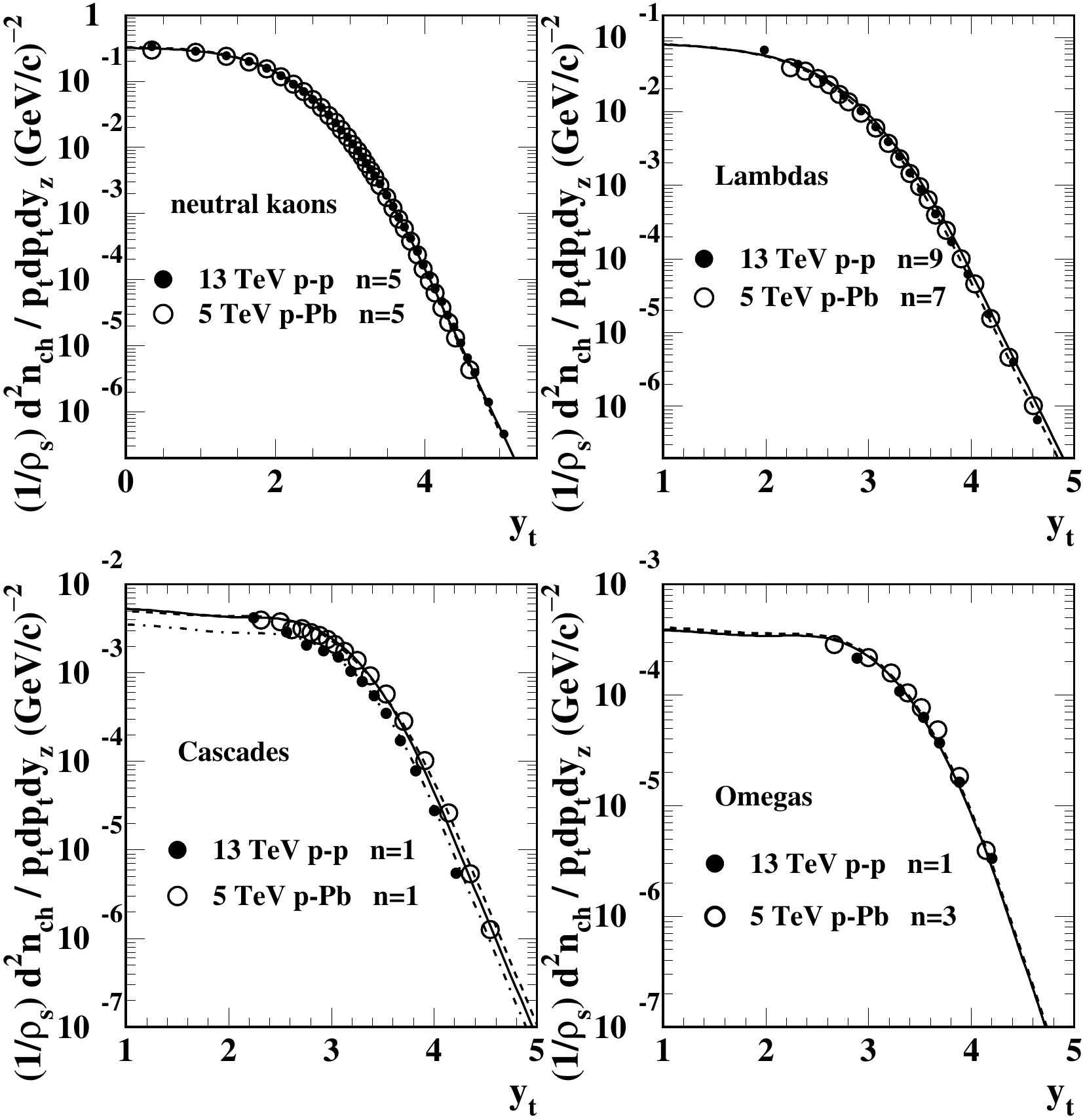}
	\put(-140,225) {\bf (a)}
	\put(-23,225) {\bf (b)}
	\put(-140,100) {\bf (c)}
	\put(-23,100) {\bf (d)}
	\caption{\label{comparehc}
		Comparisons of TCM (curves) and spectrum data (points) between selected event classes of \pp\ collisions from Ref.~\cite{alippss} (solid dots, solid curves) and \ppb\ collisions from Ref.~\cite{alippbss} (open circles, dashed curves) for
		(a) neutral kaons,
		(b) Lambdas,
		(c) Cascades and
		(d) Omegas. The paired event classes correspond to \pp\ vs \ppb\ crossovers of $\bar p_{thi}(n_s)$ trends in Fig.~\ref{ybarcomp} (right) that indicate equivalent hard components.
	} 
\end{figure}
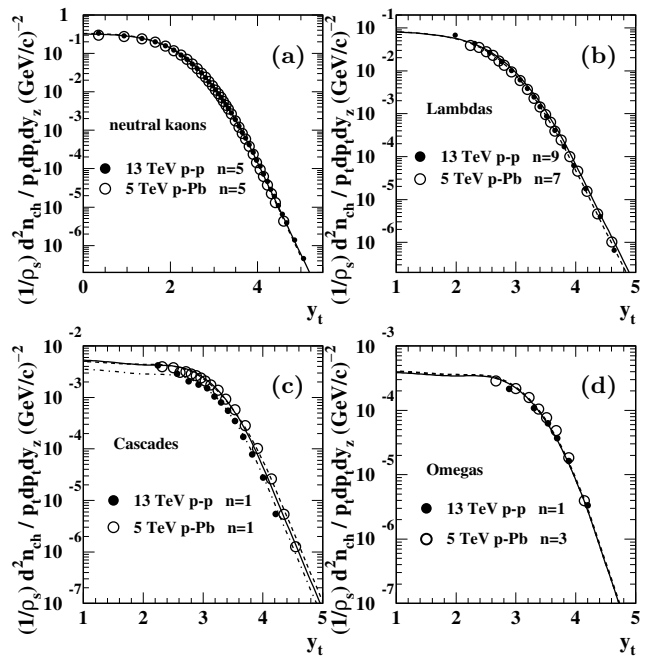

Figure~\ref{comparehc} (a) shows neutral kaon spectrum data and TCM for event class 5 from both \pp\ and \ppb\ collisions corresponding to the crossover of kaon trends for $\bar p_{thi}(n_s)$ near $x(n_s)\nu(n_s) \approx 0.16$ in Fig.~\ref{ybarcomp} (right). Data and TCM for \pp\ collisions are solid dots and solid curve wheres for \ppb\ collisions they are open circles and dashed curve. The data and curves are duplicated from Figs.~\ref{ppbpiddata} and \ref{piddata}.

Figure~\ref{comparehc} (b) shows a similar comparison for Lambdas corresponding to the  $\bar p_{thi}(n_s)$ crossover near $x(n_s)\nu(n_s) \approx 0.08$ in Fig.~\ref{ybarcomp} (right). Again,  the correspondence is within data uncertainties and confirms agreement of data and TCM trends.

Figure~\ref{comparehc} (c) shows direct comparison for Cascades. In this case issues raised previously are apparent: (a) The \pp\ cascade data spectrum falls below the \ppb\ spectrum by 2/3, corresponding to $z_{0i} = 0.0030$ for \pp\ vs 
0.0045 for \ppb. In addition, the \pp\ Cascade spectrum requires width $\sigma_{y_ti} \approx 0.46$ rather than 0.50 for other hadron species, evident by the disagreement between \ppb\ TCM (dashed) and \pp\ TCM (solid) at higher \yt. The width change has an effect on $\bar p_{thi}(n_s)$ because $\hat H_{0i}(y_t)$ is asymmetric, having a power-law high-\pt\ tail. The lowered dash-dotted TCM curve corresponds to $z_{0i} = 0.0030$

Figure~\ref{comparehc} (d) shows comparable results for Omegas showing consistency with general trends among hadron species and collision systems. Here event class $n=3$ for \ppb\ collisions (out of seven classes) is compared to event class 1 for \pp\ collisions (out of five classes) as most similar in their jet-related hard-component configuration. 

In panels (c) and (d) the curvature of the TCM trends actually  {\em changes sign} near \yt\ = 2 ($p_t \approx 0.5$ GeV/c). That distinctive variation results from the large hard/soft ratios for the higher-mass hadrons: The hard components, with modes near \yt\ = 3, push well above the soft components creating the minima near \yt\ = 2. See Fig.~10 (right) of Ref.~\cite{ppprd} for anticipation of such structure.

\section{Systematic uncertainties} \label{sys}

This analysis emphasizes detailed comparisons among collision systems and hadron species requiring consistency at the level of data random uncertainties across several studies. \ppb\ geometry determination as reported in Refs.~\cite{tommpt,tomglauber,ppbpid} must be compatible with jet production in \pp\ collisions~\cite{ppprd,alicetomspec,pppid}. The possibility for major systematic error in the \ppb\ geometry determination is demonstrated by the difference between primed and unprimed numbers in Table~\ref{rppbdata}. TCM hard/soft ratio $x(n_s)$ for \pp\ collisions or $x(n_s)\nu(n_s)$ for \ppb\ collisions is the basic TCM control parameter  that provides a common basis for differential comparison of  several systems and species. Examples include Figs.~\ref{ratios}, \ref{mptss}, \ref{ybarcomp} and \ref{comparehc}.

\subsection{TCM model function consistency}

For heavier hadrons and limited \pt\ acceptances TCM ``tail'' parameters $n$ and $q$ may be neglected. TCM soft component $\hat S_0(m_t)$ includes the collision-energy dependent L\'evy exponent $n$ \cite{alicetomspec} that however does not play a significant role for more-massive hadrons where, per Ref.~\cite{transport}, the spectra are dominated by jet (hard-component) contributions.  Within data \pt\ acceptances for the present study the TCM soft component is indistinguishable from a Boltzmann exponential on \mt. Slope parameter $T \approx 200$ MeV describes data for all hadron species except pions within data uncertainties. TCM hard component $\hat H_0(y_t)$ includes exponent $q$ which however has large uncertainties due to data \pt\ upper limits near 4 GeV/c, the transition point from Gaussian to exponential for $\hat H_0(y_t)$. The TCM for the present study is effectively a three-parameter model: $T_i$, $\bar y_{ti}(n_s)$ and $\sigma_{y_t i}$.

TCM hard component $\hat H_0(y_t,n_s)$ exhibits significant variation, with hard/soft ratio or soft multiplicity $n_s$, of Gaussian widths $\sigma_{{y_t}i}$ and modes $\bar y_{ti}$ that differs substantially between mesons and baryons~\cite{pidpart2,pppid}. For the present study effective variation with \nch\ is represented by variable modes $\bar y_{ti}(n_s)$ defined by Eq.~(\ref{ytbar}) and by starred parameter values in Tables~\ref{pidparamz} and \ref{pidparamx}. The results differ substantially between \pp\ and \ppb\ data.  

\subsection{Hard-component mode shifts and $\bf \bar p_{thi}(n_s)$}

Given published PID \mmpt\ data and Eq.~(\ref{pampttcmpid}) further \mmpt\ analysis relies on $\tilde z_i(n_s)$ that is determined only by hadron mass (see Fig.~\ref{tildezparams}), hard/soft ratio $x(n_s)\nu(n_s)$ that was previously determined by nonPID analysis~\cite{tommpt,tomglauber} and $\bar p_{tsi}$ that is determined only by hadron mass via $\hat S_{0i}(m_t)$. There is no adjustment to accommodate specific PID data. Resulting $\bar p_{thi}(n_s)$ trends in Figs.~\ref{mptss} (d) and \ref{ybarcomp} (right) can be compared with $\bar y_t(n_s)$ trends in Fig.~\ref{ybarcomp} (left) as a test of self-consistency of the TCM system.

In Ref.~\cite{pppid} the modes for \ppb\ protons {\em and} Lambdas shift according to $\bar y_t(n_s) \approx 2.86 + 0.40 x\nu$ that is compatible with {\em ad hoc} expression $\bar y_t(n_s) \approx 2.96 - 0.015(n - 4)$ with $n \in [1,7]$ the event class number as reported in Ref.~\cite{ppbpid}. The trend for protons from 13 TeV \pp\ collisions reported in Ref.~\cite{pppid} is $\bar y_t(n_s) \approx 2.81 + 0.40 x\nu$ compared to $2.80 + 0.40 x\nu$ for \ppb\ $\Lambda$ data from Table~\ref{pidparamz} of the present study compared to  $2.80 + 0.70 x\nu$ for \pp\ $\Lambda$ data from Table~\ref{pidparamx}. Significant differences between collision systems might indicate inconsistent analysis methods. However, the comparison in Fig.~\ref{ybarcomp} between $\bar y_t(n_s)$ trends on the left and $\bar p_{thi}(n_s)$ trends on the right (derived from published PID \mmpt\ data) confirms that the $\bar y_t(n_s)$ trends accurately reflect real data spectrum variation. The corresponding TCM and data $\bar p_{thi}(n_s)$ trends agree at the few percent level, i.e.\ within data statistical uncertainties, with the exception of \pp\ Cascade data.

In essence, most of the information in PID \mmpt\ data lies in the $\bar p_{thi}(n_s)$ trends that are predicted accurately by spectrum hard-component evolution with \nch. PID hard components are in turn consistent with (mass-dependent) PID fragmentation functions convoluted with a common underlying scattered-parton (gluon) energy spectrum~\cite{fragevo,jetspec2,ppbpid}. Some quantitative differences between mesons and baryons and between \ppb\ and \pp\ data may depend in part on event selection bias, e.g.\  via V0M yields.

\subsection{PID TCM hadron species fraction accuracy}

Accuracy of hadron species fractions $z_{si}(n_s)$ and $z_{hi}(n_s)$ determined by parameters $z_{0i}(n_s)$ and $\tilde z_i(n_s)$ via Eqs.~(\ref{zsix}) is a central issue for PID spectrum studies. That is an especially critical issue for (multi)strange hadron abundances -- their variation with charge multiplicity \nch\ in relation to claims of canonical suppression and thermalization in small collision systems. For lower-mass hadrons the $z_{0i}(n_s)$ were assumed fixed and the $\tilde z_i(n_s)$ were determined via low-\pt\ spectrum trends~\cite{ppbpid,pidpart1,pppid}. $dE/dx$ biases for protons and pions were estimated and, in the case of protons, the spectra were corrected~\cite{pidpart1}. The combined results led to discovery of soft $\rightarrow$ hard hadron transport conserving total species abundances~\cite{transport}. For higher-mass (multistrange) hadrons substantial variation of $z_{0i}(n_s)$ is reported. One may question whether the variation is dominated by systematic error or not and, if correct, are reported variations consistent between \pp\ and \ppb\ collision systems?

Tables~\ref{coefparams} and \ref{coefparams2} provide final comparison between collision systems. The $\tilde z_i^*$ and $\delta \tilde z_i^*$ values are simply proportional to hadron mass as in Fig.~\ref{tildezparams} (right) the same for both systems. The yields (measured by $z_{0i}(n_s)$) for protons and pions from \ppb\ collisions exhibit a substantial {\em decrease} with increasing \nch\ as represented by negative values for $\delta z_{0i}^*$ in Table~\ref{coefparams}. No such values appear in Table~\ref{coefparams2} because only $X/\pi$ {\em ratios} are reported in Ref.~\cite{alicepppid} for \pp\ collisions, not total yields as in Ref.~\cite{aliceppbpid} for \ppb\ collisions. The \pp\ charged-kaon yield data (solid dots) in Fig.~\ref{ratiosx} (obtained from $K^\pm/\pi$ {\em ratios} assuming {\em constant} $z_{0i} = 0.80$ for pions) increase significantly with \nch\ whereas the neutral-kaon data (open circles) corresponding to integrated yields~\cite{alippss} remain constant within uncertainties. However, the spectrum data for the two kaon species are statistically equivalent. Apparent increase for charged kaons may arise from a pion {\em decrease} in the ratio denominator. Different Lambda $z_{0i}^*$ values 0.036 and 0.038 are required by yield data in Fig.~\ref{ratiosx}. Within data uncertainties kaons and Lambdas as strange hadrons show {\em no  significant increase} with \nch. Cascade and Omega variation is then open to question.

The Cascade and Omega yields require substantial spectrum extrapolation from rather high \pt\ data lower bounds down to zero. For 13 TeV \pp\ collisions ``In order to compute $\langle p_T \rangle$ and the $p_T$-integrated production yields, the spectra are fitted with a Tsallis-L\'evy distribution to extrapolate in the unmeasured $p_T$ region. The systematic uncertainties on this extrapolation procedure are evaluated using other fit functions, as discussed in Sec.~4''~\cite{alippss}. The estimated extrapolation uncertainties for Omegas (for example) are 4\% for central and 12\% for peripheral event classes. For 5 TeV \ppb\ collisions ``The calculation of $p_T$-integrated yields can be performed by using data in the measured region and a parametrization-based extrapolation elsewhere. The Boltzmann-Gibbs Blast-Wave...model gives a good description of each $p_T$ spectrum and has been used as a tool for this extrapolation''~\cite{alippbss}. Estimated uncertainties for Omega yields are 8\% for central and 15\% for peripheral event classes.

Extrapolation of high-mass hadron spectra with a monolithic spectrum model may be strongly questioned. As pointed out in Fig.~\ref{comparehc} (d) for example the Omega {\em data} spectra for higher \nch\ are dominated by the jet contribution which requires its own unique model function $\hat H_0(y_t)$. Also see Fig.~\ref{omegarats} (right). Attempting to extrapolate data spectra with a {\em monolithic} model function (more appropriate for the data spectrum soft component alone) may then lead inevitably to substantial overestimation of the soft-component contribution for higher \nch.  Estimating extrapolation uncertainties by comparison with other monolithic model functions {\em having the same basic limitation} may greatly underestimate the true uncertainty.

\begin{figure}[h]
	\includegraphics[width=1.65in]{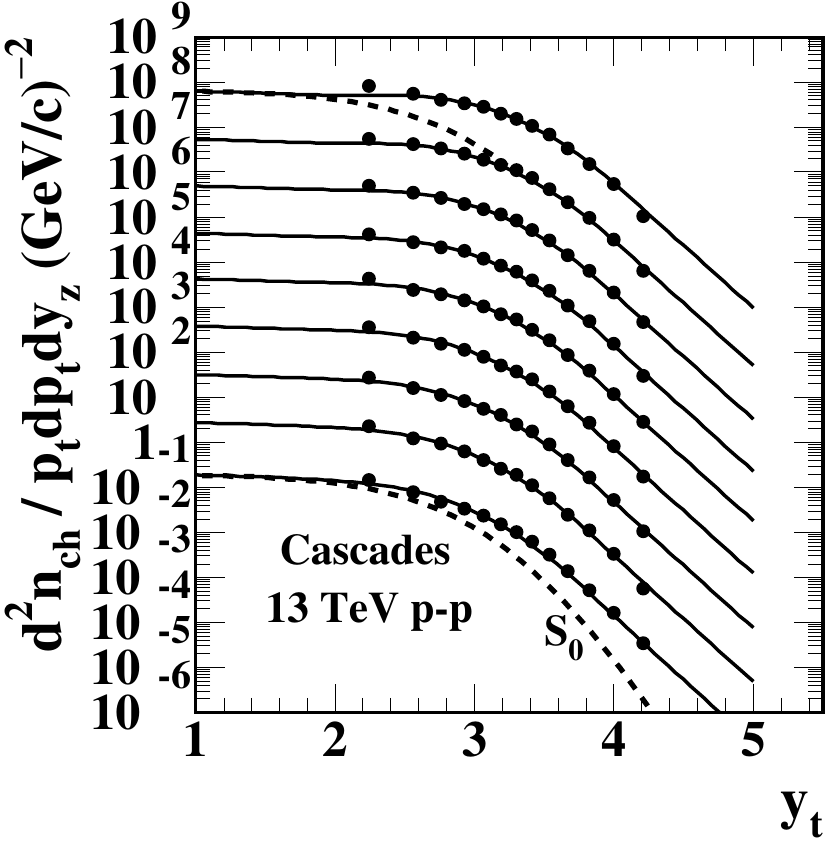}
	\includegraphics[width=1.65in]{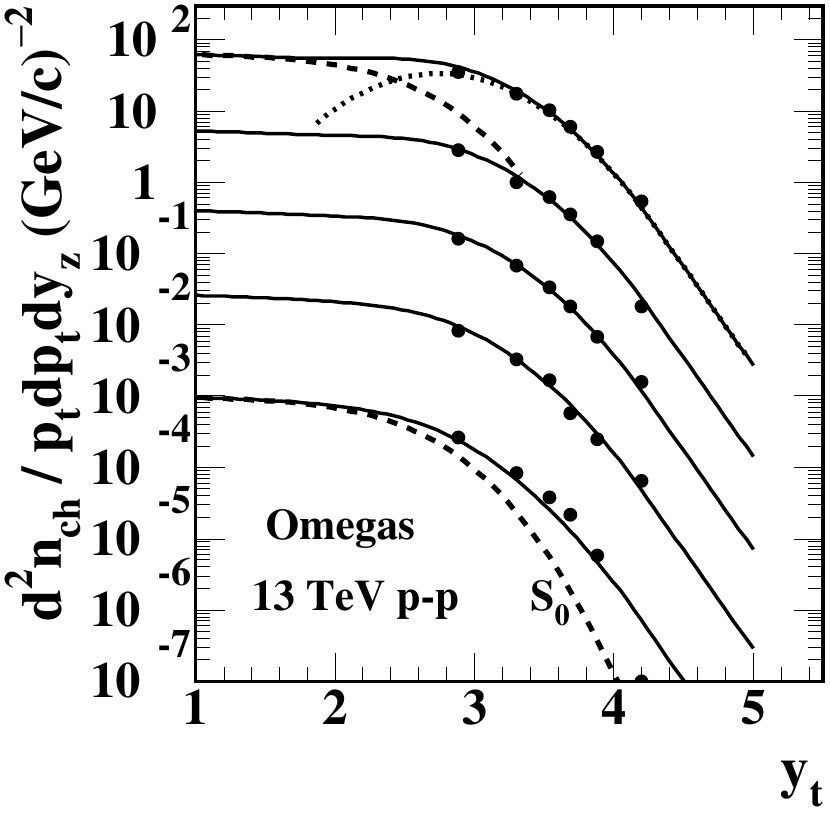}
	\caption{\label{axbad}
		Left: TCM (solid) for Cascade spectra (points) from 13 TeV \pp\ collisions using published yields (Cascade points in Fig.~\ref{ratiosx}, left) ($\times 2/3$) to obtain $z_{0i}(n_s)$ values rather than linear TCM trends per Table~\ref{coefparams2} (dashed lines in Fig.~\ref{ratiosx}).
		Right: Similar comparison for Omegas. The Omega data exclude the published yield for the most-peripheral Omega spectrum.
	} 
\end{figure}

Figure~\ref{axbad} illustrates extrapolation issues. Left and right panels are Fig.~\ref{piddata} Cascades (c) and Omegas (d) repeated but with $z_{0i}(n_s)$ values derived from published total yields as shown in Fig.~\ref{ratiosx} (left) open circles (Cascades) and open boxes (Omegas). The published values replace the TCM values obtained from Eq.~(\ref{z0trend}) and parameters from Table~\ref{coefparams2}. The greatest changes are for the most-peripheral event classes. For Cascades in the left panel the comparison is not definitive given the data, suggesting that estimated uncertainties are appropriate. However,  for Omegas at right the published yield for the lowest event class is definitely excluded by  spectrum data. Systematic yield underestimation for more-massive hadrons resulting from extrapolation via a monolithic model function may be related to the spectrum curvature changes noted in Fig.~\ref{comparehc} (c,d) arising from jet production.

To summarize: with the exception of suppression of all hadrons (except pions) below NSD \nch\ the lack of significant variation above that point for singly-strange hadrons (mesons or baryons) coupled with substantial uncertainty in the method of integrating total yields by extrapolation for Cascades and Omegas with their limited \pt\ acceptances calls into question support for claims of canonical suppression of strangeness based on total yield variation. Invoking ratios to pions (with the pion nontrivial variation), including yield suppression below NSD \nch, and using a linear $y$ vs logarithmic $x$ plot format tends to exaggerate increases to favor a specific hypothesis.

\section{Discussion} \label{disc}

This section considers (a) the quality of the TCM description of data and particularly evolution of PID spectrum shapes and their interpretation in terms of radial flow as opposed to minimum-bias jet production, (b) variation of integrated PID hadron yields in the context of strangeness enhancement claims, (c) trends for PID high-\pt\ yields in relation to jet-related spectrum hard components and (d) to what extent evolution of PID spectra for (multi)strange hadrons may be interpreted to indicate formation of a thermalized QGP in small systems.

\subsection{TCM PID spectrum evolution}

\begin{figure}[h]
	\includegraphics[height=1.65in]{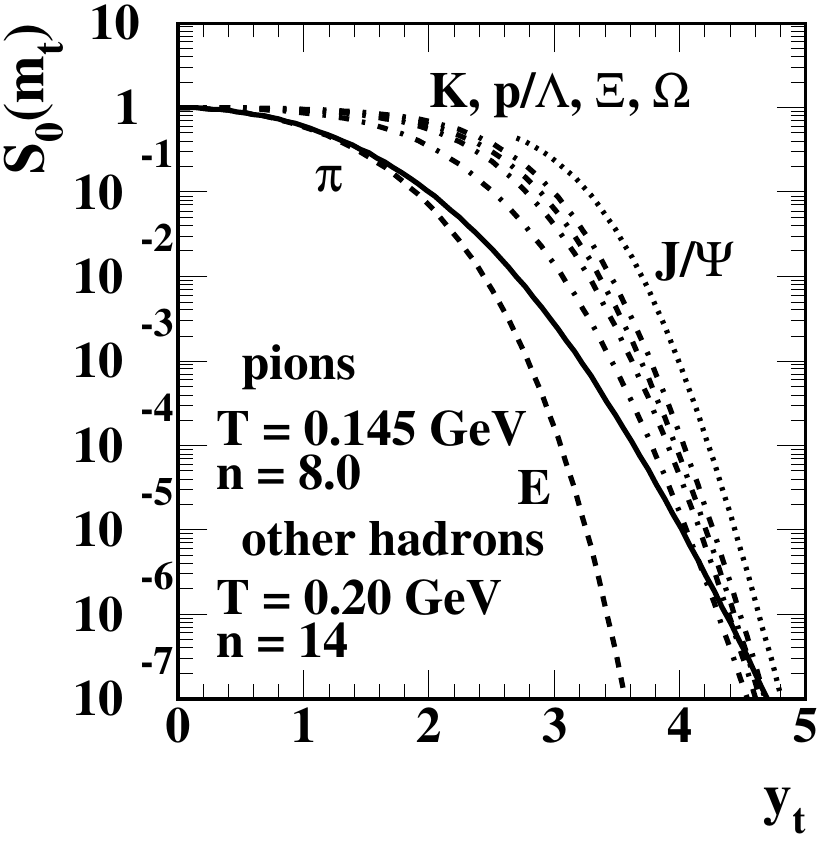}
	\includegraphics[height=1.65in]{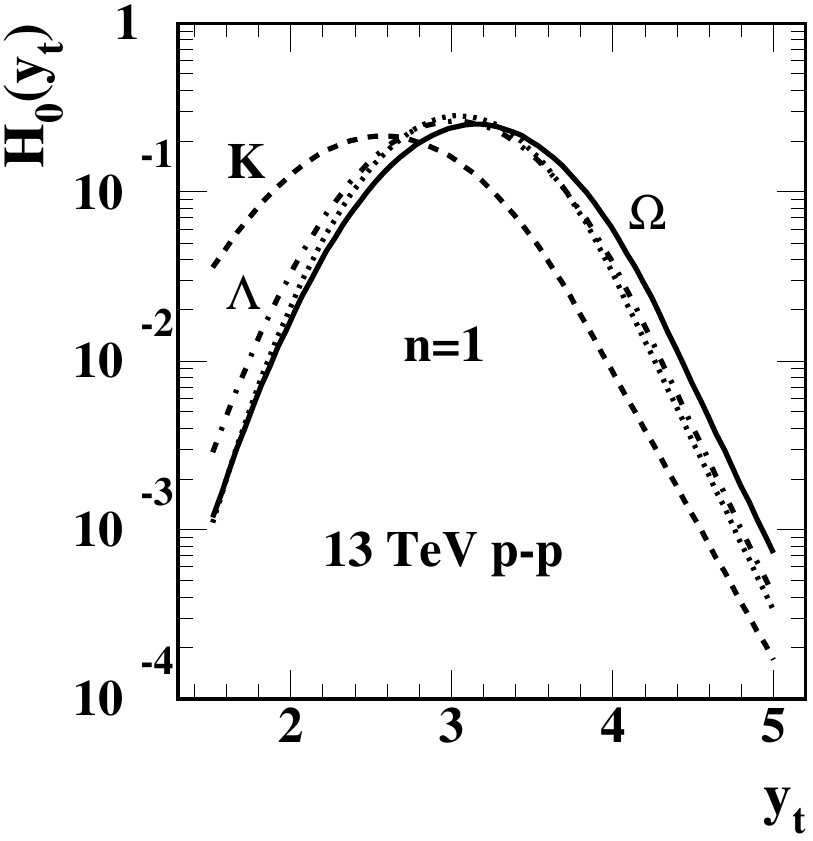}
	\caption{\label{shapes}
		Left: Soft-component model function $\hat S_{0i}(m_t)$ evolution with hadron mass. That trend does not vary with specific A-B collision conditions.
		Right: Hard-component model variation with hadron species for $n = 1$ (highest \nch) spectra from 13 TeV \pp\ collisions. In general, meson modes are lower on \yt\ and widths are greater than those for baryons.
	} 
\end{figure}

Figure~\ref{shapes} (left) shows TCM soft components $\hat S_0(m_t;T,n)$ (as densities on \mt) for pions (solid), Kaons, Lambdas, Cascades and Omegas (dash-dotted) and J/$\psi$ (dotted) for 13 TeV \pp\ collisions. The unit-normal functions are rescaled here to coincide at \yt\ = 0. The dashed curve is a Boltzmann exponential corresponding to pions. Slope parameter $T$ remains fixed at 145 MeV for pions and 200 MeV for more-massive hadrons. L\'evy exponent $n(\sqrt{s})$ has a simple logarithmic energy dependence~\cite{tomnmf}. At lower \pt\ (below 1 GeV/c or $y_t \approx 2.7$) the functions are approximately exponential and thus vary as $A - m_i\cosh(y_{ti})$ which explains the evident mass dependence of the shape. There is typically no  other dependence on the collision system beyond collision energy.

Figure~\ref{shapes} (right) shows TCM hard components $\hat H_0(y_t;\bar y_t, \sigma_{y_t},q)$ (as densities on \yt) for event class $n = 1$ (high \nch) (per TCM curves in Fig.~\ref{piddata}) that exhibit typical hard-component trends: meson peak modes $\bar y_t$ appear at lower \yt\ while widths $\sigma_{y_t}$ are larger compared to baryons. In contrast to TCM soft components hard components do vary significantly with collision conditions (e.g., event \nch) as specified in Tables~\ref{pidparamz} and \ref{pidparamx} and as manifested in $\bar p_{thi}$ trends in Fig.~\ref{ybarcomp} (right) corresponding to $\bar y_t(n_s)$ trends in the left panel. However, those variations are small compared to overall spectrum shape evolution as illustrated by \mmpt\ results in Fig.~\ref{mptss} (b) where hard-component variations cause relatively small deviations from {\em fixed} TCM trends (solid lines) that dominate.

The TCM description of PID spectrum evolution is quite simple regarding small collision systems, for example \pp\ collisions: With increasing event \nch\ an {\em approximately fixed} hard-component shape $\hat H_{0i}(y_t)$ increases in amplitude {\em linearly relative to} fixed soft-component shape $\hat S_0(m_t)$ per the universal \nn\ trend $\bar \rho_h / \bar \rho_s \approx \alpha \bar \rho_s$. That trend is especially obvious for higher-mass baryons as exhibited in Fig.~\ref{omegarats} (right) demonstrating that above $y_t \approx 2$ (0.5 GeV/c) PID spectra for more-massive hadrons are dominated by MB jet fragments. Also notable is the direct quantitative connection between \mmpt\ evolution and the high-\pt\ portions of spectra, specifically the TCM hard component as illustrated in  Sec.~\ref{directcomp}.

Those results are of particular interest in part because of the dramatic contrast to the blast-wave (BW) model and interpretation of spectrum ``hardening'' as indicating radial flow. Simultaneous BW model fits to multiple hadron species typically fall off rapidly with increasing \yt\ relative to spectrum data~\cite{tommodeltests,tombw,pppid}, and thus cannot represent jet-related hard components that dominate \mmpt\ trends, especially for more-massive hadrons. 

\subsection{Conventional descriptions of spectrum evolution}

The TCM describes PID \pt\ spectrum data across multiple collision systems and collision energies within statistical uncertainties~\cite{ppbpid,pidpart1,pidpart2,pppid} and has been related to basic QCD processes~\cite{eeprd,fragevo,jetspec2}. It is of interest to examine interpretations of spectrum evolution in Refs.~\cite{alippss,alippbss}.

Reference~\cite{alippss}, referring to its Figs.~2-4, addresses systematic variation of spectrum {\em ratios} (lower panels in those plots): In the results section ``The $p_T$ spectra become harder as the multiplicity increases, as also shown in Fig. 5, which shows [ensemble-mean \mmpt] as a function of the mid-rapidity charged particle multiplicity.'' And in the summary ``Hardening of the $p_T$ spectra with the increase of the multiplicity is observed, as already reported for pp...and p-Pb collisions...at lower energies.'' Reference~\cite{alippbss} observes that ``The spectra exhibit a progressive flattening with increasing multiplicity, which is qualitatively reminiscent of what is observed in Pb-Pb....'' In its conclusions it asserts that ``The multi-strange baryon spectra exhibit a progressive flattening with increasing multiplicity suggesting the presence of radial flow.''

Such qualitative references to ``hardening'' or ``flattening'' with increasing charge density $\bar \rho_0$ discard almost all of  the information content of PID spectra. The phrase ``...reminiscent of what is observed in Pb-Pb'' is an example of argument from analogy: Since we ``know'' that hydrodynamic flow occurs in more-central A-A collisions, and hydrodynamic flow causes ``hardening'' of \pt\ spectra, hardening of \pt\ spectra in small collision systems must also indicate flow. That is an example of {\em affirming the consequent} fallacy. What is missing from such argument is the reality of jet production as a {\em fundamental manifestation of QCD in high-energy nuclear collisions}. 

Jet  contributions to \pt\ spectra, predicted {\em quantitatively} by measured jet properties (jet energy spectra and fragmentation functions), peak near 1 GeV/c and dominate the evolution of hadron spectra with $\bar \rho_0$ as demonstrated for instance in Refs.~\cite{ppprd,hardspec,fragevo,alicetomspec,ppbpid,pidpart1,pidpart2,pppid}. 
Spectrum ``hardening'' is a manifestation of the quadratic relationship, first observed (2006) in 200 GeV \pp\ collisions~\cite{ppprd}, between jet production and the nonjet soft component in \nn\ collisions. One may also consider the detailed quantitative relation between features of PID \mmpt\ evolution with event class (multiplicity for \pp\ or centrality for \ppb) and the details of PID spectrum evolution within a TCM context that are {\em dominated by jet contributions} as demonstrated in Sec.~\ref{mptsec}. Most jet fragments appear at {\em low} \pt. Collision models that do not account {\em quantitatively} for jet fragments, especially those near 1 GeV/c, cannot properly describe high-energy nuclear collisions.

\subsection{PID integrated-yield $\bf \bar \rho_{0i}$ trends relative to $\bf \bar \rho_0$}

A major issue for Refs.~\cite{alippss,alippbss} is variation of strange-hadron integrated yields $\bar \rho_{0i}$ (for hadron species $i$) vs total charge density $\bar \rho_0$. Several strategies are utilized to seek insight for interpretation of such trends. Some common interpretations are considered here, accompanied by responses expressed in the context of the PID TCM. 

``When comparing to lower energy results, the yields of strange hadrons are found to depend only on [at least are correlated with] the mid-rapidity charged particle multiplicity [$\bar \rho_0$]''~\cite{alippss}. It cannot be otherwise since event classes are in effect defined based on $\bar \rho_0$. That such an inevitable correlation might exist does not exclude dependence on other factors also varying with event class and \nch.

``The production rates [yields] of [four hadron species] increase with the multiplicity faster than what is reported for inclusive charged particles [$\bar \rho_0$].%
\footnote{That observation is  interesting: ``multiplicity'' is presumably at least linearly proportional to ``inclusive charged particles.'' The statement translates to A increases faster with X than with X.} %
The increase is found to be more pronounced for hadrons with a larger strangeness content''~\cite{alippss}. The expression ``faster than'' is ambiguous, likely referring to deviation from {\em linear} proportionality $\bar \rho_{0i} \propto \bar \rho_0$. As addressed in the present study manifestations of ``nonlinearity'' (i.e.\ significant variation of fractional coefficient $z_{0i}(n_s)$) has two aspects: (a) reduction of $z_{0i}(n_s)$ below the NSD value for \nch\ that may be due to {\em jet suppression} and (b)  increase of $z_{0i}(n_s)$ linear with hard/soft ratio $x(n_s)\nu(n_s)$. The latter trend depends strongly on \pt\ suggesting that the effect is related to jet production (see Sec.~\ref{highpt}). That such trends are related {\em at all} to strangeness {\em per se} is not supported.

``As can be seen in Fig.\ 11, the yields of strange hadrons increase with the charged particle multiplicity following a power law behaviour, and the trend is the same at...7 and 13 TeV. ... This result indicates that the abundance of strange hadrons depends on the local charged particle density and turns out to be invariant with the collision energy.... It should also be noted that the yields of particles with larger strange quark content increase faster as a function of multiplicity....'' As noted, the power in ``power law'' is in this case simply 1. Deviations from linear proportionality (varying $z_{0i}$ in $\bar \rho_{0i} \approx z_{0i} \bar \rho_0$) are subtle as indicated by Fig.~\ref{ratios} above. Again, ``increase faster'' presumably refers to small deviations of $z_{0i}(n_s)$ from a constant value in that figure. That such effects might relate to strangeness, not jet production, is unlikely.

In reference to Fig.~13 of Ref.~\cite{alippss} ``This behaviour [effectively, variation of $z_{0i}(n_s)$ with $\bar \rho_0$] is even more pronounced for particles with higher strangeness content.\ This leads to a multiplicity-dependent increase of [$\Xi / K_\text{S}^0$ and $\Omega / K_\text{S}^0$] while [$\Lambda / K_\text{S}^0$] {\em turns out to be constant within uncertainties} [emphasis added].'' Given the systematic uncertainties in that figure the Lambda trend shows decreases for {\em lower} \nch\ similar to those for Cascades.

Reference~\cite{alippbss} reports that ``Both the [$\Xi/\pi$ and $\Omega/\pi$] ratios are observed to increase as a function of multiplicity, as seen in [its] Fig. 4. The relative increase is more pronounced for the [$\Omega$] than for [$\Xi$], being approximately 100\% for the former and 60\% for the latter. These relative increases are larger than the 30\% increase observed for the [$\Lambda/\pi$] ratio, indicating that strangeness content may control the rate of increase with multiplicity.'' The statement about $\Lambda/\pi$ (``the 30\% increase'') contradicts a conclusion of Ref.~\cite{alippss} emphasized above. Also, {\em linear increase} of $z_{0i}(n_s)$ with hard/soft ratio $x\nu$ corresponds to 30\% for Cascades and 60\% for Omegas. Those estimates are derived from variation of $x\nu$ by 0.3 in Fig.~\ref{ratiosx} and $\delta z_{0i}^*\approx 1$ for Cascades and 2 for Omegas from Tables~\ref{coefparams} and \ref{coefparams2}. Any Lambda variation is within data uncertainties. Hadron mass and/or jet fragment production $\propto x\nu$ are not excluded as control parameters. 

\subsection{Integrated high-$\bf p_t$ yields $\bf \bar \rho_{0i}(p_t)$ relative to $\bf \bar \rho_0$}

In reference to high-\pt\ yields Ref.~\cite{alippss} asserts that ``The trend at high-$p_T$ is highlighted in Fig.~6, which shows the integrated yields for $p_T >$ 4 GeV/c as a function of the mid-rapidity multiplicity. ... The high-$p_T$ yields of strange hadrons increase faster than the charged particle multiplicity. Despite the large uncertainties, the data also hint at the increase being non-linear.''%
\footnote{Given Fig.~\ref{highptx} (a) this may be a reference to quadratic factor $\bar \rho_h \propto \bar \rho_s^2 \sim \rho_0^2$ in Eq.~(\ref{ratint}) above.} %
It is not clear how ``large uncertainties'' play a role. Figure~\ref{highptx} (a) above shows the same data (solid points) with their published uncertainties. With data and curves transformed to panel (b) it is evident that the uncertainties are small compared to {\em relevant and informative} data trends. The choice of plotting format in Ref.~\cite{alippss} tends to obscure important information.
The phrase ``increase faster than the charged particle multiplicity'' is ambiguous. As demonstrated in Sec.~\ref{highpt} the dominant trend at high \pt\ is {\em quadratic} increase of high-\pt\ $\bar \rho_{0i} \sim \bar \rho_h \propto \bar \rho_s^2$ vs $\bar \rho_0 \sim \bar \rho_s$ reflecting {\em jet production}. Relative to that trend are variations due to hard-component mode $\bar y_t$ shifting on \yt\ as illustrated in Fig.~\ref{highptx} (b) (high-\pt\ yields) and Fig.~\ref{ybarcomp} (\mmpt\ hard components). Those {\em jet-related} effects are then manifested in integrated {\em total} yields~\cite{transport}. There is no significant strangeness dependence apparent in the figures, whereas meson trends {\em are} distinct from baryon trends.

The results and interpretations reported in Refs.~\cite{alippss,alippbss} concerning variation of PID integrated yields with event class present a confused picture. The phrases ``depend on,'' ``depend only on,'' ``increase with'' and ``increase faster than'' are not well defined. What one should want to know is the {\em specific} algebraic relation between variable $x$ and variable $y$, but that information is not forthcoming. Contrast the various and confusing plots of the same basic quantities in Ref.~\cite{alippss} with Fig.~\ref{ratios} of the present study that efficiently summarizes all structure in PID integrated {\em total} yields in a format (right panel) that demonstrates {\em near equivalence} of \pp\ and \ppb\ collision systems in that context. Again, whether those trends have anything to do with strangeness {\em per se} is not demonstrated.

\subsection{Thermalization and canonical suppression}

As noted in the introduction, inferred strangeness enhancement in more-central \aa\ collisions compared to reference \pp\ collisions has been interpreted as one indication of QGP formation in large collision systems: a thermalized medium of deconfined quarks and gluons. Thermalization is associated with hadron species abundances distributed according to a statistical model assuming a grand canonical ensemble. ``Canonical suppression'' of strangeness in smaller systems was expressed in terms of collision volumes: ``The suppression of the relative abundance of strange hadrons with respect to lighter flavours was suggested to be...a consequence of the finite volume, which makes the application of a grand-canonical ensemble not valid in hadron-hadron and hadron-nucleus interactions (canonical suppression)''~\cite{alippss}. That narrative however has been challenged by observation of (multi)strange hadron abundances in \pp\ and \ppb\ collisions comparable to those in more-central \aa\ collisions, albeit with greater particle multiplicities than minimum bias. 

Also as noted in the introduction, Ref.~\cite{alippss} cautions that ``...strangeness enhancement is no longer considered an unambiguous signature for deconfinement [i.e.\ QGP],'' and notes that ``These are surprising observations, because thermal strangeness production was considered to be a defining feature of heavy ion collisions....''
Reference~\cite{alippss} concludes that ``In the context of a canonical thermal model, an increase of the relative strangeness abundance depending on the strange quark content can be understood as a consequence of an increase in the system volume leading to a progressive removal of canonical suppression.'' That may well be true, but it is not demonstrated that such models are relevant to available data, especially compared to data models that accurately describe and {\em predict} data and explain their trends simply.

In the language of their proponents such arguments assume that almost all hadrons emitted in \aa\ collisions emerge from a large-volume ``fireball'' in thermal equilibrium. Strangeness abundances in that case should be consistent with a grand-canonical ensemble (GCE). Relative to those abundances strangeness abundances from small collision systems should then be reduced via canonical suppression. Increase of strangeness abundances from low-\nch\ \pp\ through increasing \nch\ for \pp\ and \ppb\ to central \pbpb\ must be due to increasing fireball volumes and restoration of GCE conditions. It is then inferred that strangeness production in \pp\ and \ppb\ collisions is different {\em only in degree} (emitter volume) from that in \aa\ collisions arising from QGP formation.

There are several problems with such scenarios: 

(a) Jet production in high-\nch\ \pp\ and \ppb\ collisions dominates hadron production, as illustrated in Secs.~\ref{omegajets} and \ref{strangemmpt} above. PID spectrum and \mmpt\ systematics are dramatically inconsistent with a thermalized emitting system resulting from partonic or hadronic {\em rescattering} (see Ref.~\cite{stock}). In high-\nch\ 13 TeV \pp\ collisions for instance 75\% of Omegas arise from jet production, with a unique \pt\ distribution characteristic of jet fragments (see Sec.~\ref{omegajets}). Copious production of strange and multistrange baryons is expected from gluons: ``We found that predominantly the gluon component in the QGP produces strange quark pairs rapidly...''~\cite{kochqgp}. But large-angle scattered gluons near midrapidity may, by the same argument, be copious sources of strange-quark pairs emerging then as elements of (multi)strange jet fragments. Such a result can also be seen for example in Sec.~\ref{omegajets}.

(b) Arguments related to canonical suppression based on varying emitter or fireball volume are problematic. Reference~\cite{cleymanssuppress} states that ``This strongly supports that the yields are directly proportional to the volume of the fireball....'' But ``this'' is an assertion that the inferred system volume is proportional to charge density $\bar \rho_0$ based on fits of a resonance gas model to strangeness abundances (from \pp, \ppb\ and \pbpb\ collisions) with volume as a fit parameter. The fitted data are in the form $\bar \rho_{0i}$ vs $\bar \rho_0$ (their Fig.~5) that increase on the log-log plot with slope near 1. What would test such models are ratios $z_{0i}(n_s) = \bar \rho_{0i}(n_s)/\bar \rho_{0}$ as in Fig.~\ref{ratios} above (exhibiting relevant information). Measured \pp\ charge multiplicities typically vary by at least a factor ten, and jet production (with most-probable fragment momentum 1 GeV/c) thus {\em increases 50-100 fold}~\cite{ppprd,tomnewppspec,pppid}. Which component (hard or soft) corresponds to a varying fireball volume?

Figure~\ref{ratios} (right) demonstrates that variation of strangeness fractional abundances $z_{0i}(n_s)$ for 5 TeV \ppb\ collisions and  13 TeV \pp\ collisions are {\em essentially equivalent} when compared on the basis of {\em jet-related} hard/soft ratio $x(n_s)\nu(n_s)$. Yet in ``central'' $n=1$ \ppb\ collisions (as defined in Ref.~\cite{aliceglauber}) the particle source consists of more than four nucleon participants and three binary collisions (see Table~\ref{rppbdata} unprimed values) while in \pp\ collisions there are by definition two and one respectively. Volume-based arguments also conflict with observation of \ppb\ collisions as {\em linear superpositions} of individual \pn\ collisions~\cite{pidpart1} and the principle of {\em exclusivity}, that each \pn\ collision proceeds independently~\cite{tomexclude}.

(c) It is concluded that certain data features are ``...compatible with thermal equilibrium,'' but correspondence of some data with statistical-model predictions does not demonstrate thermalization by multiple scattering. An interesting alternative approach to \ee\ collision data based on a statistical model is reported in Ref.~\cite{bec}. The analysis assumes that ``each jet [of an \ee\ hadronic final state] represents an independent phase in complete thermodynamic equilibrium....'' That assumption is taken to imply that ``one can describe a jet as an object defined by thermodynamic and mechanical quantities such as temperature, volume....'' The basis for analysis is ``...the canonical partition functions of systems with internal symmetries.'' The study concludes that ``...this model [thermal equilibrium] is able to fit impressively well the average multiplicities of light hadrons....''

It can be argued that ``an independent phase in complete thermodynamic equilibrium'' is not demonstrated by model agreement with hadron species abundances. In discussing the statistical properties of hadrons emerging from ``freezeout'' of A-A collisions at the SPS, Ref.~\cite{stock} warns that ``...this apparent `thermal' equilibrium is a result of the decay process, the nature of which lies well beyond the statistical model which `merely' captures the apparent statistical order, prevailing right after decay. ... The observed equilibrium is, thus, {\em not achieved by inelastic transmutation of the various hadronic species densities, in final hadron gas rescattering cascades}, i.e.\ not by hadron rescattering approaching a dynamical equilibrium [emphasis added].'' One may argue by analogy that the emerging hadrons within a jet do not rescatter so as to achieve thermodynamic equilibrium during hadronization. In fact, the process of parton hadronization to jets via splitting cascade, for instance within \ee\ collisions, was intensely studied over thirty years~\cite{dok}. It is the hadronization process itself, as a {\em quantum transition following a least-action principle}, that leads to approximate agreement with the statistical model of Refs.~\cite{bec,statmodel}. While systems in thermodynamic equilibrium may exhibit certain data features, observation of similar features in data doesn't demonstrate thermodynamic equilibrium.

\section{Summary}\label{summ}

This article reports analysis of identified-hadron (PID) \pt\ spectra for strange and multistrange hadrons from 5 TeV \ppb\ collisions and 13 TeV \pp\ collisions based on a two-component (soft+hard) model (TCM) of hadron production in high-energy nuclear collisions. A major motivation is claims in recent years that certain data features associated with particle spectra from small collision systems are similar to features in more-central \aa\ collisions that are conventionally interpreted to demonstrate formation of a thermally equilibrated quark-gluon plasma or QGP. Relevant data features include: 
(a) so-called ``hardening'' or ``flattening'' of PID \pt\ spectra increasing with charged-particle multiplicity \nch\ and with hadron mass commonly interpreted to indicate radial flow, 
(b) ensemble-mean \mmpt\ increasing with \nch, more so for higher mass and more strangeness, also commonly associated with radial flow and
(c) variation of total hadron species yields {\em in ratio to} charge density $\bar \rho_0$ for strange and multistrange hadrons apparently increasing with $\bar \rho_0$ and with strangeness toward values observed in central \aa\ collisions. Such trends are interpreted to indicate strangeness enhancement and formation of a thermalized QGP. 

The present study is based on previous TCM analysis of PID spectra for lower-mass hadrons from the same \ppb\ and \pp\ collision systems wherein spectrum data were described within their statistical uncertainties. 
Major new insights included (a) discovery of a simple mass dependence of individual hadron species fractions of total charge densities that permit predictions for multistrange hadrons, (b) discovery of transport from soft to hard components via jet production that conserves {\em total} species yields consistent with a statistical model, (c) systematic evolution of jet-related hard-component shapes varying linearly with hard/soft or jet/nonjet ratios, (d) detailed explanation of PID spectrum {\em ratio} variation with charge density $\bar \rho_0$ in terms of spectrum hard components shifting on transverse rapidity \yt\ differently for mesons vs baryons, and (e) detailed explanation of PID ensemble-mean \mmpt\ trends again in terms of jet production.

Given those developments a TCM for multistrange hadron spectra has been quantitatively predicted and the predictions confirmed in the present study via PID spectrum data for Cascades and Omegas. Reduction of total hadron  yields with charge density $\bar \rho_0$ {\em below} the non-single-diffractive (NSD) value are substantial for all species. 
Variation above that point is significant only for Cascades and Omegas and only at the two-$\sigma$ level. The \pt\ intervals covered by available multistrange data are dominated by jet contributions. Substantial variation of hadron yields integrated above 4 GeV/c are quantitatively explained by systematic shifts on \yt\ of jet-related spectrum hard components, differently for mesons vs baryons but with no clear relation to strangeness. The jet contribution to {\em total} integrated yields depends strongly on hadron mass. 
Hard components of PID ensemble-mean \mmpt\ extracted via TCM  are observed  to vary linearly vs hard/soft (jet/nonjet) ratio. Again, variations for different species are quantitatively related to measured shifts on \yt\ of spectrum hard-component peak modes. Given those systematics, pairs of spectra from \pp\ and \ppb\ collisions are compared for those event classes where hard-component mode positions and/or \mmpt\ hard components coincide. Spectrum pairs for each species coincide within data uncertainties except for Cascades where the 13 TeV \pp\ {\em spectra} appear to be systematically biased by factor 2/3 compared to Cascade {\em yield} data which agree with 5 TeV \ppb\ data.

In summary, PID spectra for strange and multistrange hadrons from 5 TeV \ppb\ collisions and 13 TeV \pp\ collisions are accurately and completely described by a simple few-parameter TCM based on two related fragmentation processes that provide the {\em default} QCD description of high-energy nuclear collisions. TCM parametrizations do not depend on fits to individual spectra and arguably represent all significant information carried by particle spectrum data.
Some conventional interpretations of spectrum features appear to arise from a strong assumption that {\em any} deviation of particle spectra from a Boltzmann exponential must arise from ``collective'' motion of the particle source, hence reference to spectrum ``hardening'' or ``flattening'' as evidence for flow. That  assumption is consistent with the Bevalac discovery of collectivity (flows) in A-B collisions, and data from more-central \mbox{A-A} collisions seem compatible with that scenario. 

However, at RHIC and LHC energies jet production dominates spectrum evolution and is {\em quantitatively predictable}. The present study provides simple and detailed explanations for data features in terms of minimum-bias jet production.  Variation of fractional abundances interpreted as related to strangeness remains at the two-$\sigma$ level  and, based on a study of the \pt\ dependence of $z_{0i}(n_s)$ for 13 TeV \pp\ data, is more likely due to strong mass dependence of jet production. \pp\ and \ppb\ small collision systems remain essential references that can be understood within a QCD context. Small collision systems restored as a reference system then form a counter to claims of novel phenomena in more-central \aa\ collisions that may lead to improved understanding of QCD.


\end{document}